\documentclass[mnsc,nonblindrev]{informs3_homepage}
\usepackage{xcolor}

\OneAndAHalfSpacedXI

\definecolor{darkblue}{rgb}{0.47, 0.62, 0.80}
\definecolor{darkgreen}{rgb}{0.34, 0.65, 0.34}
\definecolor{darkcoral}{rgb}{0.76, 0.42, 0.48}
\definecolor{lightblue}{rgb}{0.58, 0.76, 0.80}
\definecolor{lightcoral}{rgb}{0.90, 0.61, 0.66}
\definecolor{linecolor}{rgb}{0.65, 0.65, 1}

\definecolor{pastelBlue}{rgb}{0.0,0.4,0.7}

\usepackage{hyperref}
\hypersetup{
colorlinks=true,
allcolors = pastelBlue  
}

\usepackage{natbib}
 \bibpunct[, ]{(}{)}{,}{a}{}{,}%
 \def\bibfont{\small}%
 \def\newblock{\ }%
\TheoremsNumberedThrough     
\ECRepeatTheorems

\EquationsNumberedThrough    

\usepackage{float}
\usepackage[caption=false]{subfig}
\usepackage{amsmath}
\usepackage{algorithm2e}
\RestyleAlgo{ruled}
\LinesNumbered
\let\oldnl\nl
\newcommand{\nonl}{\renewcommand{\nl}{\let\nl\oldnl}}

\usepackage{algpseudocode}
\usepackage{ifthen}
\usepackage{enumitem}
\usepackage{cases}
\usepackage{mathtools}
\usepackage{amssymb}
\usepackage{graphicx}
\usepackage{empheq}
\usepackage{tikz}
\usetikzlibrary{shapes.geometric}
\usetikzlibrary{shapes}
\usetikzlibrary{arrows}
\usetikzlibrary{calc,positioning}
\usepackage{bm}
\usepackage{dsfont}
\usepackage{changepage}
\usepackage{Macros}

\def \subsecDirectEffect{\ref{subsec:direct-effect}}
\def \subsecGeneralSpecification{\ref{subsec:general-specification}} 
\def \secApndxNumericals{8}
\def \NIasMPHD{10}
\def \subsecGenDistrMatrix{\ref{subsec:gen-distr-int-matrix}}
\def \secServerTimetrend{\ref{sec:server_with_timetrend}}
\def \subsecTteTquilibrium{\ref{subsec:tte-at-equilibrium}}
\def \secDetailedProofs{\ref{sec:detailed_proofs}}
\def \secBOMWBI{\ref{sec:BOM_WBI}}
\def \thmSLLN{\ref{thm:SLLN-2}}

\begin{document}

\RUNTITLE{Causal Message-passing}

\TITLE{Causal Message Passing for Experiments with Unknown and General Network Interference}

\ARTICLEAUTHORS{%
\AUTHOR{Sadegh Shirani~~~~~~~~~~~~~Mohsen Bayati}
\AFF{Graduate School of Business, Stanford University}
}
\ABSTRACT{%
Randomized experiments are a powerful methodology for data-driven evaluation of decisions or interventions. Yet, their validity may be undermined by \emph{network interference}. This occurs when the treatment of one unit impacts not only its outcome but also that of connected units, biasing traditional treatment effect estimations. Our study introduces a new framework to accommodate complex and unknown network interference, moving beyond specialized models in the existing literature. Our framework, termed causal message-passing, is grounded in high-dimensional \emph{approximate message passing} methodology. It is tailored for multi-period experiments and is particularly effective in settings with many units and prevalent network interference. The framework models causal effects as a dynamic process where a treated unit’s impact propagates through the network via neighboring units until equilibrium is reached. This approach allows us to approximate the dynamics of potential outcomes over time, enabling the extraction of valuable information before treatment effects reach equilibrium. Utilizing causal message-passing,  we introduce a practical algorithm to estimate the total treatment effect, defined as the impact observed when all units are treated compared to the scenario where no unit receives treatment. We demonstrate the effectiveness of this approach across five numerical scenarios, each characterized by a distinct interference structure.}


\maketitle

\section{Introduction}
\label{sec:Intro}
Randomized experiments are crucial for establishing causal relationships by randomly assigning units (experiment subjects) to treatment and control groups and comparing outcomes. Their effectiveness relies on the Stable Unit Treatment Value Assumption (SUTVA)\citep{cox1958planning,rubin1978bayesian}, which assumes a unit’s outcome is unaffected by other units’ treatments. However, SUTVA often fails in real-world contexts with unit interactions \citep{manski2013identification}. For instance, in studies of medication for contagious diseases, interactions between units can lead to network effects, where treated units influence control units, potentially biasing results. Addressing such network interference is vital for accurate estimation of treatment effect.

When SUTVA is violated, new methods are needed to measure causal effects with minimal assumptions. Arbitrary interference models make estimation infeasible due to non-identifiability and exponential growth in cardinality of potential outcomes~\citep{forastiere2022estimating, yu2022estimating, basse2018limitations, karwa2018systematic, aronow2017estimating, manski2013identification, sussman2017elements}. Estimation results are also sensitive to model misspecifications~\citep{karwa2018systematic}. Therefore, it is important to relax assumptions and ensure tractable estimation to effectively study causal inference under network interference.

The present study introduces a new framework for modeling and analyzing causal effects in the presence of network interference. Specifically, when a unit is treated, the intervention impacts outcomes of its neighboring units. These changes, in turn, impact the outcomes of units interacting with those neighbors, creating a \emph{dynamic} process that propagates through the network of unit interactions until equilibrium is reached. Drawing inspiration from statistical physics \citep{mezard1986spin,mezard2009information}, this dynamics can be seen as the dissemination of information through a network of units via message exchanges. We then employ the approximate message-passing (AMP) methodology \citep{donoho2009message,bayati2011dynamics} to show that the dynamics of potential outcomes over time can be approximated by one-dimensional \emph{state evolution} equations. In light of this, we refer to our approach as \emph{Causal Message-Passing} (Causal-MP).

The proposed framework has also a potential outcome interpretation \citep{imbens2015causal}. 
Specifically, at each time instant, the model represents the outcome for each unit as a weighted combination of non-linear functions applied to the outcomes of units in previous time periods, their treatment assignments, and their covariates. This structure, reminiscent of neural networks, enables the model to adapt to broad families of network interference patterns.

The rigorous theoretical investigation of the proposed model introduces a new toolkit for designing and analyzing methods to study unobserved counterfactuals in the presence of time-dependent network interference. To the best of our knowledge, no existing research addresses time-dependent network interference. However, further assumptions, particularly regarding identifiability, are necessary to estimate causal effects. We illustrate this in scenarios where the interference pattern has a time-invariant mean and in multi-period Bernoulli randomized designs under a nonlinear parametric assumption on the interference structure. This leads to a straightforward algorithm with strong consistency guarantees for estimating the total treatment effect (TTE), which we validate through extensive simulations. Notably, our estimation procedure operates without any prior knowledge of the interference structure. We defer the adaptation of this algorithm to settings with more systematic time-dependent interference, using our time-dependent state evolution results, to future studies.

Unlike mean-field models in the network interference literature, which typically assume that each unit interacts with a collective average field, Causal-MP focuses on units and their unique interactions within a network, accounting for heterogeneous unit behaviors, and enabling analysis of treatment effects before they reach equilibrium.

\subsection*{Example and Challenges in Network Interference}
\label{subsec:example}

\begin{figure}
  \centering
  \includegraphics[width=0.8\linewidth]{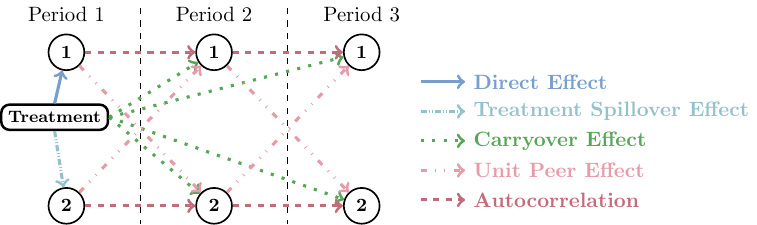}
  \caption{Illustration of different causal relationships; circles represent unit outcomes.}
  \label{fig:TEs}
\end{figure}

Assume we aim to evaluate a new medical treatment for a contagious disease by observing two individuals over three time periods (Figure~\ref{fig:TEs}). The treatment triggers several effects: a \emph{direct effect} on the treated unit (unit 1), a \emph{treatment spillover effect} impacting the control unit (unit 2), and a \emph{carryover effect} influencing future outcomes~\citep{athey2018exact,forastiere2022estimating,xiong2019optimal}. Additionally, interactions between units lead to \emph{unit peer effects} and \emph{autocorrelation}, adding complexity~\citep{yu2022estimating, imai2019identification}. An \emph{anticipation effect}, where units’ behavior is influenced by expected future events, also exists but is not shown in Figure~\ref{fig:TEs}.

A causal model must encompass these diverse effects while tackling several challenges. Identifying causal relationships is complex due to intricate interactions, especially when interference structures are unknown. Interference patterns evolve, requiring models that account for temporal changes. Treatment efficacy may also change over time, demanding models that span multiple periods. Additionally, budgetary constraints limit the treatment group size, necessitating methods that provide reliable inferences even with fewer treated units. Noise and unobserved covariates further complicate the analysis, requiring models that can handle incomplete data effectively.

Our proposed framework aims to address these complexities by simplifying evolving high-dimensional complexities into tractable one-dimensional dynamics at any point in time. This approach opens the door to utilizing all observed data during the experiments, even before the effects stabilize.


\section{Other Related Literature}
\label{sec:Relev_Lit}
The assumption of no interference, known as SUTVA, is foundational in causal inference \citep{cox1958planning, rubin1978bayesian, manski1990nonparametric, sussman2017elements}. However, recent research increasingly aims to relax this assumption. Some studies focus on testing for interference \citep{aronow2012general,bowers2013reasoning,saveski2017detecting,athey2018exact,pouget2019testing, hu2022average,han2022detecting}, while others propose new assumptions and methods for estimating causal effects without SUTVA \citep{leung2020treatment,viviano2020experimental,leung2022causal,yu2022estimating,cortez2022exploiting,cortez2022staggered,agarwal2022network,belloni2022neighborhood,li2022network,li2022random}. We survey these emerging developments and proceed by a brief discussion of the literature on approximate message-passing algorithms.

\paragraph{Neighborhood Interference Assumption (NIA)} The NIA is a prominent approach in the literature for relaxing SUTVA, positing that an individual's outcome is influenced solely by the treatments of neighboring units in the network~\citep{sussman2017elements}. Studies have expanded NIA under known or partially observed interference patterns \citep{leung2020treatment, viviano2020experimental, agarwal2022network, belloni2022neighborhood, li2022random}. In the absence of known network structures, \cite{cortez2022staggered, cortez2022exploiting, yu2022estimating} have proposed unbiased estimators under various designs and constraints. Recently, under a more flexible version of NIA, the consistency of inverse-probability weighting estimators has been derived \cite{leung2022causal}. Unlike these studies, this work does not assume NIA.

\paragraph{Partial Interference} Partial interference is another route for relaxing SUTVA, where the population is partitioned into non-overlapping clusters without any interference between them \citep{rosenbaum2007interference, candogan2021correlated}. Extending this concept to complex networks, the bias in standard estimators increases by the number of inter-cluster edges \citep{yu2022estimating}. Examples of approaches to mitigate this bias include variants of cluster-randomized designs \cite{eckles2016design,ugander2013graph,ugander2023randomized}, all necessitate knowledge of the interference network, a prerequisite not essential for the current study.

\paragraph{Restrictions on Interference Network} In the literature, various constraints have been placed on interference structures beyond the NIA and partial interference. Examples include bounding the largest node degree and proving the asymptotic normality of certain estimators \citep{li2022network,chin2018central}, introducing methods based on unique observation patterns or localized interference \citep{agarwal2022network,jagadeesan2020designs,wang2020design}, and restricting network topologies \citep{viviano2020experimental,belloni2022neighborhood,cai2015social,leung2022causal}.

\paragraph{Application-specific Interference Structures}
The literature has also explored interference patterns customized for specific applications \citep{holtz2020reducing,wager2021experimenting,munro2021treatment,johari2022experimental,bright2022reducing,farias2022markovian,farias2023correcting}. These studies cover a wide range of contexts, from marketplace dynamics to complex systems, offering tailored approaches to managing interference across diverse experimental settings.

\paragraph{Single-time Point Observation} The majority of the literature on network interference has concentrated on single-time point observations, thereby observing each unit's outcome once \citep{hudgens2008toward,aronow2017estimating,basse2019randomization,jackson2020adjusting,leung2020treatment,savje2021average,li2022random,yu2022estimating,cortez2022exploiting,leung2022causal}.
Recently, there is a pivot toward multi-time point observations \citep{ni2023design}. In this evolving context, \cite{li2022network} discuss challenges and the utility of additional data, while \cite{boyarsky2023modeling} utilize temporal treatment variations to model interference. This work also studies multi-period experiments.

\paragraph{Deterministic versus Stochastic Models} The literature typically considers outcomes and network structures as deterministic, with randomness only in treatment assignments \citep{aronow2017estimating,aronow2012general,basse2019randomization, leung2020treatment, savje2021average, yu2022estimating, harshaw2022design, leung2022causal}. Recent studies, however, have incorporated stochastic elements, using random graph models \citep{li2022random}, Bernoulli-distributed outcomes \citep{li2022network}, and analyzing network noise impacts on bias and variance \cite{li2021causal}. This work considers a stochastic model with randomness in outcomes, treatments, covariates, and interference patterns.

\paragraph{Model-based Approaches} An alternative approach uses structured outcome models, typically linear, to account for interference \citep{goldsmith2013social,toulis2013estimation, blume2015linear, cai2015social,basse2018model,chin2019regression, yu2022estimating, belloni2022neighborhood, jiang2023causal}. This method requires prior knowledge of network topology to choose relevant statistics. A key issue is that these models can oversimplify complex social interactions \citep{angrist2014perils}. The model in this paper aims to capture more complex patterns of interference.

\paragraph{Approximate Message Passing (AMP)} The origins and motivations for AMP trace back to \cite{thouless1977solution,kabashima2003cdma,donoho2009message}. The theoretical foundation was established by \cite{bolthausen2014iterative,bayati2011dynamics} and further expanded with various degrees of generality by \cite{javanmard2013state,bayati2015universality,berthier2020state,chen2020universality,xinyi2021approximate,wang2022universality,dudeja2023universality,rush2018finite,li2022non}. For a comprehensive overview, see \cite{zdeborova2016statistical,montanari2018mean,feng2022unifying}. AMP traditionally studies high-dimensional estimation problems via iterative dynamical systems involving nonlinear transformations and a mixing matrix. The goal is to use state evolution to refine the nonlinear functions and improve estimation accuracy. Our approach diverges as we observe the outcomes, \emph{not the mixing matrix or nonlinear functions}, and use state evolution to infer statistics for estimating causal effects. AMP is related to tree approximation methods from statistical physics \citep{bethe1935statistical}, information theory \citep{gallager1962low}, and computer science \citep{pearl1982reverend}, with its insights primarily rooted in statistical physics \citep{thouless1977solution}.


\section{Problem Formulation and Main Results}
\label{sec:PF}
In this section, we present an outcome specification that captures a general interference structure, followed by an informal description of our theoretical results\footnote{Rigorous statements are provided in \S \ref{sec:Tech_result}.}, which form the basis for a practical algorithm to estimate causal effects and an algorithm for estimating confidence intervals. Finally, we illustrate an application of Causal-MP in the context of the Bernoulli randomized design.

\subsection{Potential Outcome Specification}
\label{sec:POM}
Before introducing the setup, we define the notations $[M]$ and $[M]_0$ to refer to the sets ${1, \ldots, M}$ and ${0, 1, \ldots, M}$, respectively, for any positive integer $M$. Now, consider a setting with $N$ units indexed by $n$ in $[N]$ over a time horizon of $T+1$ periods, indexed by $t$ in $[T]_0$. At time instant $t>0$, each unit~$n$ is randomly assigned a treatment denoted by $\treatment{n}{t}$ which is distributed according to a probability distribution~$\pi_t$. We refer to the set $\expd = \{\pi_1,\ldots,\pi_T\}$ as the \emph{experimental design} \cite{sussman2017elements,bajari2021multiple} and assume that for all $t$, the support of $\pi_t$ is a subset of the real line that includes $0$ and at least one non-zero element. Whenever $\treatment{n}{t}=0$, we say that unit $n$ is under \emph{control}; otherwise, we say unit $n$ receives the \emph{treatment}. We let $\Mtreatment{}{} \in \R^{T\x N}$ be the treatment assignment matrix, where its $n^{th}$ column $\Vtreatment{n}{}$ represents the treatment assignment of unit~$n$ throughout the experiment: $\Vtreatment{n}{} = \big(\treatment{n}{1},\ldots,\treatment{n}{T}\big)^\top$. Additionally, $\OMtreatment{}{}$ denotes a specific realization of $\Mtreatment{}{}$. 

We adopt the Neyman-Rubin causal framework \citep{imbens2015causal} and denote by $\outcomeD{}{n}{t}(\Mtreatment{}{})$ the potential outcome of unit $n$ at time~$t$. In this context, the observed data consists of a treatment allocation~$\OMtreatment{}{}$ and outcomes $\outcomeD{}{n}{t}(\Mtreatment{}{}=\OMtreatment{}{})$, $n\in [N],\;t\in[T]_0$. In addition, for an integer $M$, we let the matrix $\covar \in \R^{M\x N}$ be the covariate matrix such that its $n^{th}$ column (denoted by $\Vcovar{n}$) gives the characteristics of unit~$n$ (e.g., age, gender).

To define the potential outcome specification, we let $\{\outcomeg{t}{}\}_{t\in[T]_0}$ be a family of unknown measurable functions such that $\outcomeg{t}{}: \R\x\R^T\x\R^M \mapsto \R$. 
For each $t$ and $n$, the output from $g_t$ is given as $\outcomeg{t}{}\left(\outcomeD{}{n}{t}(\Mtreatment{}{}) ,\Vtreatment{n}{},\Vcovar{n}\right)$. With a slight extension in notation, $\outcomeg{t}{}\big(\VoutcomeD{}{}{t}(\Mtreatment{}{}) ,\Mtreatment{}{}, \covar\big)$ represents a vector of size $N$, where the $n^{th}$ element is defined by $\outcomeg{t}{}\left(\outcomeD{}{n}{t}(\Mtreatment{}{}) ,\Vtreatment{n}{},\Vcovar{n}\right)$. Here, we denote $\VoutcomeD{}{}{t}(\Mtreatment{}{}) = \big(\outcomeD{}{1}{t}(\Mtreatment{}{}), \ldots,\outcomeD{}{N}{t}(\Mtreatment{}{})\big)^\top$ which is the column vector that contains the outcomes of all individuals at time~$t$. Then, given the vector of initial outcomes $\VoutcomeD{}{}{0}$, for $t=0,1,\ldots,T-1$, we define
\begin{align}
    \label{eq:outcome_function_matrix}
    \Voutcome{}{}{t+1}(\Mtreatment{}{}) =
    \big(\IM+\IMatT{t}\big)\outcomeg{t}{}\left(\Voutcome{}{}{t}(\Mtreatment{}{}) ,\Mtreatment{}{}, \covar\right) 
    +\Vnoise{}{t},
\end{align}
where $\IM$ and $\IMatT{t}$ are $N\times N$ matrices and $\Vnoise{}{t} = \big(\noise{1}{t},\ldots,\noise{N}{t}\big)^\top$. 
Here, the matrices $\IM$ and $\IMatT{t}$ are unknown and capture the interference structure. Let $\IMatl{ij}$ and $\IMatTl{ij}{t}$ denote the element in the $i^{th}$ row and $j^{th}$ column of $\IM$ and $\IMatT{t}$, respectively; $\IMatl{ij}+\IMatTl{ij}{t}$ quantifies the impact of unit $j$ on unit $i$ at time~$t$. We refer to $\IM$ and $\IMatT{t}$ as the \emph{fixed interference matrix} and \emph{time-dependent interference matrix}, respectively. Let $\IMatG{t} = \IM+\IMatT{t}$, the \emph{interference matrix} at time $t$. The function $\outcomeg{t}{}$ represents the impact of past outcomes, treatment assignments, and covariates on current outcomes. Finally, $\Vnoise{}{t}$ is the zero-mean noise term accounting for misspecifications and measurement errors. We eliminate the notation $\Mtreatment{}{}$ in $\Voutcome{}{}{t}(\Mtreatment{}{})$ and simply write $\Voutcome{}{}{t}$ whenever there is no ambiguity in the potential outcome concept.

\begin{remark}[Direct Effect]\label{rem:direct-effect}
In \eqref{eq:outcome_function_matrix}, the direct effect of the treatment of each unit $i$ on its outcome, $\outcome{}{i}{t+1}$, results from the combination of impact of $\treatment{i}{t+1}$ on the function $\outcomeg{t}{}$ followed by multiplication by $\IMatl{ii}+\IMatTl{ii}{t}$. In Appendix \S \subsecDirectEffect, we discuss a setting involving a more explicit direct effect.
\end{remark}

\begin{remark}[Non-additive Noise]\label{rem:gen-random-outcome}
In Appendix \S \subsecGeneralSpecification, we consider a more general setting where a non-linear random operator is applied to \eqref{eq:outcome_function_matrix}, allowing for a broader range of outcome distributions, including binary outcomes.
\end{remark}

One can interpret the specification in \eqref{eq:outcome_function_matrix} as follows:
If we exclude the noise term $\Vnoise{}{t}$, the potential outcome for unit $n$ at time $t+1$ is a weighted combination of ``messages'' it receives from all other units. Each message is a (nonlinear) function of the outcome of the sending unit at time~$t$, its entire treatment assignment, and its covariate vector. This is expressed as
$\outcomeD{}{n}{t+1}=\sum_{i\in [N]} \IMatGl{ni}{t} \outcomeg{t}{}(\outcomeD{}{i}{t},\Vtreatment{i}{},\Vcovar{i})
$. 
The potential outcome $\outcomeD{}{n}{t+1}$ will subsequently be used in the message that unit $n$ sends to other units in future periods. Tracing back in time and interpreting each $\outcomeD{}{i}{t}$ as a weighted combination of the messages it receives from other units at time $t$, the impact of unit $i$ on unit $n$ is shaped by a combination of the impacts it receives from other units, combined with the treatment assignment and personal characteristics of unit $i$. This message exchange captures dynamics of the interference effect within the network of units. A more detailed discussion of this message-passing interpretation, including the absence of the well-known Onsager term present in AMP formulations, is deferred to Appendix \S\NIasMPHD.

The specification in \eqref{eq:outcome_function_matrix} captures key interference effects depicted in Figure~\ref{fig:TEs}, such as treatment spillover, unit peer effects, and autocorrelation. Furthermore, the treatments and outcomes of all units can affect the outcome of any specific unit, thus relaxing the commonly employed neighborhood interference assumption in the literature~\citep{sussman2017elements}. Additionally, \eqref{eq:outcome_function_matrix} accounts for the propagation of intervention effects across the network, enabling distant units to influence each other’s outcomes through intermediary units over multiple time periods. The presence of each unit’s entire treatment assignment over time on the right-hand side of \eqref{eq:outcome_function_matrix} also represents the anticipation effect.

In the remaining, we assume the interference matrices, as well as the noise vectors to be Gaussian random variables with \emph{unknown} parameters. 
\begin{assumption}
    \label{asmp:Gaussian Interference Matrice}
    Entries of $\IM$ are i.i.d. Gaussian random variables with mean $\mu/N$ and variance~$\sigma^2/N$, independent from anything in the model. Similarly, for any $t\in[T]_0$, entries of $\IMatT{t}$ are i.i.d. Gaussian random variables with mean $\mu_t/N$ and variance~$\sigma^2_t/N$, independent from all other sources of the randomness in the model.
\end{assumption}
\begin{assumption}
    \label{asmp:Gaussian noise}
    Elements of noise vector $\Vnoise{}{t}$ are i.i.d Gaussian random variables with mean zero and finite variance~$\sigma_e^2$, independent from anything in the model.
\end{assumption}
We note that Assumption \ref{asmp:Gaussian Interference Matrice}  simplifies the theoretical analysis and supports rich interference patterns observed in real data. In Appendix \S \subsecGenDistrMatrix, we show that our results extend to settings where entries of $\IM$ and $\IMatT{t}$ are binary under neighborhood or partial interference assumptions. Leveraging insights from the \emph{universality} results from the AMP literature, we discuss how our results are expected to hold when these entries are not necessarily i.i.d. or Gaussian. Moreover, numerical results in \S \ref{sec:Numerical} validate our theoretical predictions in settings with binary entries under the neighborhood interference assumption, across various networks such as Facebook friends data, random geometric graphs, Erd\"{o}s–R\'{e}nyi graphs, and clustered networks. To provide some intuition for the richness of Assumption \ref{asmp:Gaussian Interference Matrice}, we note that the standard deviations of the entries of $\IM$ and $\IMatT{t}$ are assumed to be of order $O(1/\sqrt{N})$, while their means are of order $O(1/N)$. This allows the entries to be heterogeneous, even though they share a common mean and distribution. Such heterogeneity permits diverse \emph{local dependencies} between outcomes and treatments that cannot be captured with a \emph{mean-field} assumption requiring outcomes to be \emph{independent}, conditional on a \emph{global} average quantity.

In the following sections, we will rigorously analyze data generated by \eqref{eq:outcome_function_matrix} under Assumptions~\ref{asmp:Gaussian Interference Matrice} and~\ref{asmp:Gaussian noise}. Our main objective is to develop practical algorithms for identifying and estimating the underlying causal effects.

\subsection{Causal Estimands}
\label{sec:estimands}
The literature has explored several estimands for causal effects, with a focus on average effects rather than individual-level effects \citep{han2023model,yu2022estimating,li2022network, leung2022causal,savje2021average,hu2022average}. Here, we aim to estimate the Total Treatment Effect (TTE), also known as Global Treatment Effect (GTE). This estimand measures the average effect of altering the treatment for the entire community. For instance, in the case of a contagious disease, TTE quantifies the reduction in the number of infections or the impact on healthcare costs whenever the new medication is administered to all individuals, as compared to a scenario where nobody receives it. Precisely, for any two treatment allocations $\OMtreatment{}{}'$ and $\OMtreatment{}{}''$, we define
\begin{align}
    \label{eq:TTE_def_fixed}
    \TTE{t}{\OMtreatment{}{}'',\OMtreatment{}{}'} = \lim_{N \rightarrow \infty} \frac{1}{N} \sum_{n=1}^N
    \big(\outcomeD{}{n}{t}(\OMtreatment{}{}'')
    - \outcomeD{}{n}{t}(\OMtreatment{}{}')\big)\,.
\end{align}
A notable special case arises when entries of $\OMtreatment{}{}''$ and $\OMtreatment{}{}'$ are drawn from Bernoulli distributions with means $\pi''$ and $\pi'$, respectively \cite{eckles2016design,li2022network}. Then, with a slight abuse of notation, we write $\TTE{t}{\pi'',\pi'}$. For example, the case with $\pi''=1$ and $\pi'=0$ represent scenarios in which all treatment assignments are set to 1 or 0, respectively. In problems where only a single observation of the data is accessible, the literature typically assumes that $t$ is sufficiently large for the effect to have stabilized and then estimates the TTE at the equilibrium \citep{leung2022causal,yu2022estimating,savje2021average,sussman2017elements}. However, in this paper, we aim to estimate $\TTE{t}{\desired,0}$ for any desired $\desired \in [0,1]$ and all values of $t$, which will allow us to trace the evolution of $\TTE{t}{\desired,0}$, even before stabilization. $\TTE{t}{\desired,0}$ is particularly relevant when it is impractical to deliver the treatment to the entire population. In general, estimating the TTE is insightful in scenarios where a decision maker aims to use the result of the experiment to determine whether the treatment should be expanded to everyone or not \citep{eckles2016design}.

\paragraph{Direct and indirect effects.} TTE is generally a combination of indirect (network) and direct effects. In Appendix \S \subsecDirectEffect, we also discuss how to estimate these two components.

\subsection{State Evolution of the Experiment}
\label{subsec:SE}
Let the potential outcomes $\left\{\outcomeD{}{n}{t} \right\}_{n\in[N],t\in[T]_0}$ follow \eqref{eq:outcome_function_matrix}.  The goal is to derive efficient estimators for the TTE defined in \eqref{eq:TTE_def_fixed}. To this end, we will show that
\begin{align}
    \label{eq:average of outcomes}
    \lim_{N\rightarrow\infty}  \sum_{n=1}^N \frac{\outcomeD{}{n}{t}}{N} \eqasWOS \AVO{\expd}{}{t},
    ~
    \lim_{N\rightarrow\infty} \sum_{n=1}^N \frac{(\outcomeD{}{n}{t})^2}{N} -  \AVO{\expd}{}{t}^2\eqasWOS \VVO{\expd}{}{t}^2 ,
\end{align}
where deterministic quantities $\AVO{\expd}{}{t}$ and $\VVO{\expd}{}{t}^2$ are defined by the recursions,
\begin{equation}
    \label{eq:state evolution}
    \begin{aligned}
        \AVO{\expd}{}{t+1} &=
        (\mu+\mu_t) \E\left[
        \outcomeg{t}{}\big(\AVO{\expd}{}{t} + \VVO{\expd}{}{t} Z, \Vtreatment{}{},\Vcovar{}\big)
        \right],
        \\
        \VVO{\expd}{}{t+1}^2 &=
        (\sigma^2+\sigma_t^2) \E\left[
        \outcomeg{t}{}\big(\AVO{\expd}{}{t} + \VVO{\expd}{}{t} Z, \Vtreatment{}{},\Vcovar{}\big)^2
        \right] + \sigma_e^2\,,
    \end{aligned}
\end{equation}
that are initialized by $\AVO{\expd}{}{1}$ and $\VVO{\expd}{}{1}^2$, representing the expected sample mean and variance of outcomes at $t=1$, respectively. In addition, $Z \sim \Nc(0,1)$ is independent from $(\Vtreatment{}{},\Vcovar{}) \sim \Pi \x p_{\covar}$, where $\Pi := \pi_1 \times \ldots \times \pi_T$ and $p_{\covar}$ denotes the (large $N$) limit of the empirical distribution of columns of the covariate matrix $\covar$. Thus, $p_{\covar}$ defines a probability distribution over~$\R^M$ and $\Pi$ denotes the probability distribution of the treatments determined by~$\expd$.

Note that \eqref{eq:state evolution} allows tracking the dynamics of  $\AVO{\expd}{}{t}$ and $\VVO{\expd}{}{t}^2$, which represent the large $N$ sample mean and variance of the potential outcomes at time $t$, respectively. To simplify the notation, we drop the reference to $\expd$ and simply write $\AVO{}{}{t}$ and $\VVO{}{}{t}^2$, whenever there is no ambiguity.

We also show a more general result than \eqref{eq:state evolution}, under some moment conditions concerning $p_{\covar}$ and $\Pi$. For any continuous\footnote{The result still holds true if the function $\psi$ is almost everywhere continuous in the first argument and continuous in the other arguments.} function~$\psi$ with at most polynomial growth, we show:
\begin{align}
    \label{eq:general convergence-approiximation}
    \lim_{N\rightarrow\infty} \frac{1}{N} \sum_{n=1}^N \psi \big(\outcomeD{}{n}{t},\Vtreatment{n}{},\Vcovar{n}\big)
    \eqas
    \E \left[\psi \big( \AVO{}{}{t} + \VVO{}{}{t} Z, \Vtreatment{}{},\Vcovar{}\big) \right]\,.
\end{align}
Then, inspired by the literature on AMP algorithms, we refer to \eqref{eq:state evolution} as \emph{state evolution} equations of the experiment. We provide the rigorous statements and related details in~\S\ref{sec:Tech_result}. Here, we discuss the intuition and implications derived from \eqref{eq:general convergence-approiximation} and \eqref{eq:state evolution}.

Consider an experimental data $\left\{\outcomeD{}{n}{t}(\OMtreatment{}{}) \right\}_{n\in[N],t\in[T]_0}$ from a given experimental design $\expd$.
Note that we can utilize this data to estimate the limiting sample means $\left\{\AVO{}{}{t}\right\}_{t \in [T]_0}$ and the sample standard deviations $\left\{\VVO{}{}{t}\right\}_{t \in [T]_0}$ using \eqref{eq:average of outcomes}. These statistics serve as the basis for estimating complex functions involving units' outcomes, treatment assignments, and covariates using \eqref{eq:general convergence-approiximation}. Having access to $\expd$ and $p_{\covar}$, the unknown parameters in the state evolution equations are denoted as $\Uc := \left(\mu,\sigma,\{\mu_t\}_{t \in [T]_0},\{\sigma_t\}_{t \in [T]_0},\{\outcomeg{t}{}\}_{t\in[T]_0},\sigma_e\right)$. By estimating $\Uc$, we can estimate any desired average counterfactual, including the TTE defined in \eqref{eq:TTE_def_fixed}.

Note that our approach does not impose any constraints on the experimental design and does not require specific knowledge about interference structure. As a result, the state evolution equations, besides \eqref{eq:general convergence-approiximation}, offer a versatile framework for the causal analysis of data. Notably, \eqref{eq:general convergence-approiximation} suggests that we can effectively compute various functions, relevant to understanding the average behavior of individuals.

Overall, the framework presented in this section addresses the challenges of analyzing high-dimensional interconnected data by reducing it to the study of one-dimensional state evolution equations. However, the problem of estimating the unknown parameters $\Uc$ still needs to be tackled.

\subsection{Causal Effects Estimation}

In this section, we propose a meta algorithm for estimating causal effects, building on Eqs. \eqref{eq:general convergence-approiximation} and \eqref{eq:state evolution}. This is done by estimating the parameters $\Uc := \left(\mu,\sigma,\{\mu_t\}_{t \in [T]_0},\{\sigma_t\}_{t \in [T]_0},\{\outcomeg{t}{}\}_{t\in[T]_0}, \sigma_e\right)$ using the available data. We then utilize \eqref{eq:general convergence-approiximation} and \eqref{eq:state evolution} for a second time to compute the desired counterfactuals and estimands. In this regard, according to the available data and our specific research objectives, we can introduce constraints on the parameters of the outcome model \eqref{eq:outcome_function_matrix} to facilitate the estimation procedure of $\Uc$. Algorithm~\ref{alg:CE estimation} outlines a general scheme to estimate the total treatment effect when altering the treatment allocations from $\OMtreatment{}{}'$ to $\OMtreatment{}{}''$, corresponding to experimental designs $\expd'$ and $\expd''$, respectively.

\begin{algorithm}
\caption{Causal Message-passing: TTE estimation}
\label{alg:CE estimation}
\textbf{Input:} $\OMtreatment{}{}',\OMtreatment{}{}''$,  $\left\{\outcomeD{}{n}{t}(\OMtreatment{}{}) \right\}_{n\in[N],t\in[T]_0}$, $\expd$, and $p_{\covar}$.

\textbf{Step 1: Data processing}

\quad Compute $\left\{\AVO{}{}{t},\VVO{}{}{t}\right\}_{t\in[T]_0}$ by \eqref{eq:average of outcomes}.

\textbf{Step 2: Parameters estimation}

\quad Estimate $\Uc$, guided by \eqref{eq:state evolution}.

\textbf{Step 3: TTE estimation}

\quad \textbf{for} $t \gets 0$ \textbf{to} $T$:

\quad\quad Compute $\AVO{\expd'}{}{t},\VVO{\expd'}{}{t}$ and $\AVO{\expd''}{}{t},\VVO{\expd''}{}{t}$, by \eqref{eq:state evolution}.

$\quad\quad\ETTE{t}{\OMtreatment{}{}'',\OMtreatment{}{}'} \gets \AVO{\expd''}{}{t} - \AVO{\expd'}{}{t}$

\textbf{Output:} $\ETTE{t}{\OMtreatment{}{}'', \OMtreatment{}{}'},\;t\in[T]_0.$

\end{algorithm}
\subsubsection*{Explanation of Algorithm \ref{alg:CE estimation}}
\label{subsub:alg-explanation}
Algorithm~\ref{alg:CE estimation} presents a systematic and flexible framework for conducting the estimation process, which comprises three main steps explained below.

\paragraph{Step 1: Data processing} The initial step of the algorithm focuses on data preprocessing, transforming the data into a suitable format for further analysis. The computational cost of this step scales linearly with the number of units and results in the generation of two vectors of size $T+1$.
    
\paragraph{Step 2: Parameters estimation} In the second step, we utilize the preprocessed data to estimate~$\Uc$. To accomplish this, we can make use of a parametric function class for $\{\outcomeg{t}{}\}_{t \in [T]_0}$, and introduce simplifying assumptions about the interference matrices. By leveraging the observed outcomes and \eqref{eq:state evolution}, we estimate the relevant parameters.

Note that we can consider more complex models by collecting richer data. For example, if we can divide units into distinct clusters or conduct the experiment in multiple stages with different treatments, we can incorporate additional data into the estimation. This allows using more sophisticated outcome models that capture the intricacies of the data.

\paragraph{Step 3: TTE estimation} In the final step, the algorithm estimates the TTE by estimating the counterfactual scenarios under experimental designs $\expd'$ and $\expd''$. The algorithm then computes the TTE over the desired time horizon. It is important to note that the time horizon for estimating the TTE can differ from the time horizon of collecting the data.

Overall, Algorithm~\ref{alg:CE estimation} is designed to estimate the total treatment effect by observing a specific scenario and estimating other counterfactual scenarios. It should be noted that the designs $\expd_1$ and $\expd_2$, and therefore $\OMtreatment{}{}'$ and $\OMtreatment{}{}''$, are arbitrary and at least one of the counterfactual scenarios cannot be directly observed. However, by leveraging the available data and estimation techniques, the algorithm provides an estimate of the TTE. It is worth mentioning that the same procedure can be adapted to estimate other estimands related to the system. In the next sections, we discuss a heuristic for the estimation of the confidence interval of TTE and then we give a more concrete version of Algorithm~\ref{alg:CE estimation} tailored to the context of Bernoulli randomized design.

\subsection{Confidence Intervals}
\label{subsec:CI}

To guarantee the reliability of our TTE estimates, obtaining confidence intervals is important.
While we defer a proper treatment of this topic to follow up studies, we describe a heuristic approach here that seems promising, based on the numerical studies in \S \ref{sec:Numerical}.
By utilizing the data $\left\{\outcomeD{}{n}{t} (\OMtreatment{}{}) \right\}_{n\in[N],t\in[T]_0}$, we employ the state evolution equations (Eq.~\eqref{eq:state evolution}) to devise a resampling-based heuristic for confidence interval estimation of the TTE. This method involves resampling unit outcomes and repeatedly using Algorithm~\ref{alg:CE estimation} to approximate the TTE distribution, considering the outcome dependencies due to network interference.

Specifically, confidence interval computation involves a two-step method. For a given $q \in (0,1)$ and positive integer $B$, the first step is resampling the units $B$ times, with each unit's inclusion relies on an independent Bernoulli random variable with probability $q$. Then, for each sample, Algorithm~\ref{alg:CE estimation} is applied to estimate the TTE over the time horizon, yielding $B$ values of TTE estimates for each time $t$. The second step involves calculating the confidence interval using the mean and standard deviation of these estimated TTEs across the $B$ samples. Selecting $q$ should balances a trade-off between the accuracy of each sample's estimates, which increases with higher $q$, against the correlation between samples, which favors lower $q$ values. While a thorough analysis of this method is reserved for future work, our numerical results in $\S$\ref{sec:Numerical} suggest that selecting smaller $q$ as $N$ grows provides reasonable confidence intervals.

\subsection{Application to Bernoulli Randomized Design}
\label{sec:BerRD}

We consider a two-stage Bernoulli randomized experiment as a specific case of the experimental design. This approach aligns with the prevailing practice in many firms, where a dynamic phase release of the new treatment is carried out through a sequence of randomized experiments \citep{kohavi2020trustworthy,han2022detecting}.
Subsequently, the time horizon is divided into two intervals: $\{0,1,\ldots,T_1\}$ and $\{T_1+1, \ldots, T=T_1+T_2\}$. Each unit receives the treatment with probabilities $\pi_1$ and $\pi_2$ in the first and second intervals, respectively, where $\pi_1\neq \pi_2$. The main objective is to estimate the $\TTE{t}{\desired,0}$, as defined in \S \ref{sec:estimands}, throughout the time horizon $[T]_0 = \{0,1,\ldots,T\}$. Here, letting $\pi_1 = 0$ is equivalent to considering historical data with no experiment in the first stage.

Additionally, we consider approximation of the function~$\outcomeg{t}{}$, described as follows:
\begin{equation}
\begin{aligned}
    \label{eq:function_structure}
    \outcomeg{t}{}\left(\outcomeD{}{n}{t} ,\Vtreatment{n}{},\Vcovar{n}\right) &= \CE^n + \PE^n \outcomeD{}{n}{t} + \DE^n \treatment{n}{t+1} 
    + \ME^n \outcomeD{}{n}{t} \treatment{n}{t+1} + (\CC^n)^\top \Vcovar{n}.
\end{aligned}
\end{equation}
In \eqref{eq:function_structure}, $\CE^n,\PE^n,\DE^n,\ME^n \in \R$ and $\CC^n \in \R^M$ are unknown random objects independent of everything else.  \eqref{eq:function_structure} is a first-order approximation of the function~$\outcomeg{t}{}$ plus the second-order term $\ME^n \outcomeD{}{n}{t} \treatment{n}{t+1}$. In this context, $\CE^n$ represents the baseline effect, the coefficients $\PE^n, \DE^n, \ME^n$ correspond to the specific effects of the current outcome and treatment, and the random vector $\CC^n$ captures the influence of the covariates of the unit $n$. Note that all these coefficients are treated as random objects, specific to each unit. This differs from the existing literature, where both the outcomes and network structure are assumed to be deterministic \citep{aronow2017estimating, aronow2012general, athey2018exact, basse2019randomization}.

Considering the contagious disease example in Figure~\ref{fig:TEs}, the expression $\CE^n + \PE^n \outcomeD{}{n}{t} + (\CC^n)^\top \Vcovar{n}$ represents the severity of symptoms in the absence of medication. It combines several components: $\CE^n$ captures the inherent severity level, $\PE^n \outcomeD{}{n}{t}$ reflects the influence of the current health condition on future health outcomes, and $(\CC^n)^\top \Vcovar{n}$ incorporates the impact of the individual specific covariates such as age, gender, etc., on their health condition. The term $\DE^n \treatment{n}{t+1} + \ME^n \outcomeD{}{n}{t} \treatment{n}{t+1}$ accounts for the effect of administering the new medication to individual~$n$. The inclusion of the multiplicative term $\ME^n \outcomeD{}{n}{t} \treatment{n}{t+1}$ allows the consideration that the efficacy of treatment can vary according to the severity of symptoms. This flexible modeling approach recognizes the potential heterogeneity in treatment effects, allowing for a more nuanced understanding of the impact of the new medication.

We restrict ourselves to the setting where $\mu_t$ is constant. To simplify the notation, without loss of generality, we modify the coefficients $\CE^n$, $\PE^n$, $\DE^n$, $\ME^n$, and $\CC^n$, assuming $\mu_t+\mu=1$. This implies that while the interference pattern can vary over time, i.e., the matrix $\IMatT{t}$ can change completely over time, its mean remains the same throughout the experiment. In the context of our example, this means that individuals may interact with different intensity on different days, but the overall interaction level remains constant over time. We will show in the numerical simulations of Appendix \S \secServerTimetrend~that the above model is robust even when the data contain seasonal trends (i.e., when $\mu_t$ varies). Meanwhile, we anticipate that one approach to handling the case of varying $\mu_t$, similar to our parametric assumption on $\outcomeg{t}{}$, would be to assume a parametric form for $\mu_t$ by expressing it as a linear combination of basis functions of $t$, which we defer to future work.

Algorithm~\ref{alg:causal-mp-original} outlines a the procedure for estimating the TTE in the specified context. It employs data segmented into two parts of lengths $T_1$ and $T_2$, with treatment probabilities $\pi_1$ and $\pi_2$, respectively. Similar to Algorithm~\ref{alg:CE estimation}, it begins with a preprocessing step, followed by two linear regressions to estimate the necessary parameters. The algorithm then uses the sample mean of observed outcomes to estimate the counterfactual scenario for the desired treatment level $\desired$ and calculates the TTE in the final step. The consistency of this estimator is demonstrated in \S \ref{sec:Tech_result}.

\begin{algorithm}
\caption{TTE estimation: two-stage Bernoulli design}
\label{alg:causal-mp-original}
\textbf{Input:}{ $\desired$,  $\left\{\outcomeD{}{n}{t} (\OMtreatment{}{}) \right\}_{n\in[N],t\in[T]_0}$, $\pi_1,\pi_2$, $p_{\covar}$, $T_1$, and $T_2$.}

\textbf{Step 1: Data processing}

\quad \textbf{for} $t \gets 0$ \textbf{to} $T$:

\quad\quad $\HAVO{}{}{t} \gets \frac{1}{N} \sum_{n=1}^N \outcomeD{}{n}{t} (\OMtreatment{}{})$

\textbf{Step 2: Parameters estimation (by linear regression)}

\quad Regress $\big(\HAVO{}{}{1},\ldots, \HAVO{}{}{T_1}\big)$ on
$\big(\HAVO{}{}{0},\ldots, \HAVO{}{}{T_1-1}\big)$ to get $b_1$ and intercept~$a_1$

\quad  Regress $\big(\HAVO{}{}{T_1+1}, \ldots, \HAVO{}{}{T}\big)$ on
$\big(\HAVO{}{}{T_1}, \ldots, \HAVO{}{}{T-1}\big)$ to get $b_2$ and intercept~$a_2$

\quad $(\widehat{\PE},\widehat{\ME},\widehat{\DE}) \gets \left(\frac{1}{2} \left(b_2+b_1-\frac{(b_2-b_1)(\pi_2+\pi_1)}{\pi_2-\pi_1}\right),\frac{b_2-b_1}{\pi_2-\pi_1},\frac{a_2-a_1}{\pi_2-\pi_1}\right)$

\textbf{Step 3: TTE estimation by $(\widehat{\PE},\widehat{\DE},\widehat{\ME})$}

\quad $\desiredHAVO{}{}{0} \gets \HAVO{}{}{0}$ and $\ETTE{0}{\desired,0} \gets 0$


\quad \textbf{for} $t \gets 0$ \textbf{to} $T-1$:

\quad\quad \textbf{if} $t \leq T_1-1$, \textbf{then} $\pi \gets \pi_1$, \textbf{else} $\pi \gets \pi_2$

\quad\quad $\desiredHAVO{}{}{t+1} \gets \HAVO{}{}{t+1} + \widehat{\PE}\left(\desiredHAVO{}{}{t}-\HAVO{}{}{t}\right) + \widehat{\DE} (\desired-\pi) + \widehat{\ME} \left(\desired\desiredHAVO{}{}{t}-\pi\HAVO{}{}{t}\right)$

\quad\quad $\ETTE{t+1}{\desired,0} \gets \widehat{\PE} \ETTE{t}{\desired,0} + \widehat{\DE} \desired + \widehat{\ME} \desired \desiredHAVO{}{}{t}$

\textbf{Output:} $\ETTE{t}{\desired,0},\; t \in [T]_0$.
\end{algorithm}

Note that Algorithm~\ref{alg:causal-mp-original} operates without requiring covariate estimation. This feature is particularly advantageous in scenarios with unobserved covariates. By bypassing the need for covariate inclusion, Algorithm~\ref{alg:causal-mp-original} streamlines the analysis, reducing potential biases and complexities associated with covariate observation or estimation.

While we impose certain restrictions on the structure of the functions $\outcomeg{t}{}$ in \eqref{eq:function_structure}, note that these restrictions still encompass a wide range of linear and non-linear models that have been extensively studied in the existing literature \citep{sussman2017elements,belloni2022neighborhood,li2022network,yu2022estimating}. In addition, in \S \ref{sec:Numerical}, we provide comprehensive numerical analysis to support the flexibility and adaptability of Algorithm~\ref{alg:causal-mp-original}. The data generation processes in these examples do not obey the specifications assumed by Algorithm~\ref{alg:causal-mp-original} and are designed to assess the applicability of the algorithm in estimating the TTE, even when the data is generated using more complex underlying structures. 

\begin{remark}[Incorporating Prior Information]\label{rem:prior}
In Algorithm~\ref{alg:causal-mp-original}, performance may improve by integrating prior information relevant to the data and setting at hand. For instance, during the update $\ETTE{t+1}{\desired,0} \gets \widehat{\PE} \ETTE{t}{\desired,0} + \widehat{\DE} \desired + \widehat{\ME} \desired \HAVO{\desired}{}{t}$, one might project the right-hand side to specific sub-intervals of $\mathbb{R}$, informed by prior knowledge.
\end{remark}

\begin{remark}[TTE Estimation at Equilibrium]
\label{rem:TTE_at_EQ}
    We can also use state evolution equations \eqref{eq:state evolution} to establish an estimator for the TTE at equilibrium. This stands in contrast to Algorithm~\ref{alg:causal-mp-original}, which is tailored to estimate the TTE over the entire time horizon. The formal statement of this finding is laid out in Appendix \S \subsecTteTquilibrium.
\end{remark}


\section{Technical Results and Proofs Overview}
\label{sec:Tech_result}

In this section, we first outline and discuss the technical assumptions required for the proofs. We then present the rigorous statements of the theoretical contributions. We conclude by analyzing Algorithm~\ref{alg:causal-mp-original} and assessing the consistency of the resulting estimator.

\subsection{Main Result}
\label{subsec:main-result}
We proceed by considering a sequence of systems indexed by $N$, representing the population size. Accordingly, we use $\VoutcomeD{N}{}{t}$, $\Mtreatment{}{N}$, and $\covar(N)$ to denote the vector of potential outcomes at time $t$, treatments, and covariates corresponding to the $n^{th}$ system. Specifically, we added the notation $N$ to different terms to emphasize the dimension of the quantities. Then, for any fixed $t>0$, we analyze the behavior of the elements in $\VoutcomeD{N}{}{t}$ as the system size~$N$ approaches infinity. This investigation provides us with valuable insights into the evolution of the outcomes and how they are influenced by the design of the experiment.

To present the main results, we introduce several notations. For any vector $\Vec{v} \in \R^n$, we denote its Euclidean norm as $\norm{\Vec{v}}$. For a fixed $k \geq 1$, we define $\poly{k}$ as the class of functions $f: \R^\ell \rightarrow \R$, for some $\ell\geq 1$, that are continuous and exhibit polynomial growth of order $k$. That is, there exists a constant $c$ such that $|f(\Vcovar{})| \leq c(1+\normWO{\Vcovar{}}^k)$. Moreover, we consider a probability space $(\Omega, \F, \P)$, where $\Omega$ represents the sample space, $\F$ is the sigma-algebra of events, and $\P$ is the probability measure. We denote the expectation with respect to $\P$ as $\E$. Additionally, for any other probability measure~$p$, we use $\E_p$ to denote the expectation with respect to $p$. 

Next, we state an assumption that is standard in the AMP literature and then discuss it in the context of our experimental design problem.
\begin{assumption}
    \label{asmp:BL}
    Fix $k \geq 2$. We assume that
    \begin{enumerate}[label=(\roman*)]
        \item \label{asmp:BL-pl functions} For any $t\in[T]_0$, the function $\outcomeg{t}{}:\R^{1+T+M}\rightarrow \R$ is a $\poly{\frac{k}{2}}$ function.

        \item \label{asmp:BL-covariates distribution}
        Let $p_{N}$ be the empirical distribution of columns of $\covar(N)$, then $p_{N}$ converges weakly to a probability measure $p_{\covar}$ on $\R^M$ such that $\E_{p_{\covar}}\big[\normWO{\Vcovar{}}^k\big]< \infty$ and, as $N$ grows to $\infty$, $\E_{p_{N}}\big[\normWO{\Vcovar{}}^k\big] \rightarrow \E_{p_{\covar}}\big[\normWO{\Vcovar{}}^k\big]$. 

        \item \label{asmp:BL-bound on treatment moment} If $\Vtreatment{}{} \sim \Pi$, then $\E\big[\normWO{\Vtreatment{}{}}^{k}\big] < \infty$.

        \item \label{asmp:BL-bound on initials} $\VoutcomeD{N}{}{0}$, $\Mtreatment{}{N}$, $\covar(N)$, and the function $\outcomeg{0}{}$ are such that for some deterministic values $\AVO{}{}{1}$ and $\VVO{}{}{1}$, we have
        \begin{align*}
            \AVO{}{}{1} &= 
            \lim_{N\rightarrow \infty}
            \frac{\mu+\mu_0}{N} \sum_{n=1}^N
            \outcomeg{0}{}\big(
            \outcomeD{}{n}{0},\Vtreatment{n}{},\Vcovar{n}
            \big) < \infty\,,
            \\
            0
            <
            \VVO{}{2}{1}
            - \sigma_e^2
            &=
            \lim_{N\rightarrow \infty}
            \frac{\sigma^2+\sigma_0^2}{N} \sum_{n=1}^N
            \outcomeg{0}{}\big(
            \outcomeD{}{n}{0},\Vtreatment{n}{},\Vcovar{n}
            \big)^2 < \infty\,.
        \end{align*}

        \item \label{asmp:BL-empirical dist of inits} There exist a $\poly{\frac{k}{2}}$ function $\bar{g}_0: \R^{T+M} \mapsto \R$ such that for all $t$ and $\poly{\frac{k}{2}}$ functions $\varphi: \R^{T+M} \mapsto \R$, we have
        \begin{align*}
            \lim_{N\rightarrow \infty}
            \frac{1}{N} \sum_{n=1}^N
            \outcomeg{0}{}\big(\outcomeD{}{n}{0},\Vtreatment{n}{},\Vcovar{n}\big) \varphi(\Vtreatment{n}{},\Vcovar{n})
            \eqas \E \left[\bar{g}_0\big(\Vtreatment{}{},\Vcovar{}\big) \varphi(\Vtreatment{}{},\Vcovar{})\right],
        \end{align*}
        and $\E\left[\bar{g}_0(\Vtreatment{}{},\Vcovar{})^2\right] \leq \frac{\rho_1^2-\sigma_e^2}{\sigma^2+\sigma_0^2}$
        where $(\Vtreatment{}{},\Vcovar{}) \sim \Pi \x p_{\covar}$.
    \end{enumerate}
\end{assumption}

Assumption~\ref{asmp:BL}  encompasses a set of regularity conditions on the system parameters and model attributes. Specifically, Part~\ref{asmp:BL-pl functions} ensures that the functions $\outcomeg{t}{}$ do not exhibit fast explosive behavior, guaranteeing the well-posedness of the large system asymptotics. Part~\ref{asmp:BL-covariates distribution} ensures that the empirical distribution $p_N$ remains stable and does not diverge as the sample size increases. This assumption holds, for instance, when unit covariates (the columns of the covariate matrix~$\covar$) are i.i.d. with distribution $p_{\covar}$ with finite moments of order~$k$. Moreover, Assumption~\ref{asmp:BL}-\ref{asmp:BL-bound on treatment moment} holds for a wide range of treatment assignments, including cases where the support of $\Pi$ is bounded, such as the Bernoulli design. 

Assumptions~\ref{asmp:BL}-\ref{asmp:BL-bound on initials} and \ref{asmp:BL}-\ref{asmp:BL-empirical dist of inits} are required for the proofs and rule out restrictive initial conditions. For example, both are satisfied if $\outcomeg{0}{}$ is a non-zero function and the sequence of initial outcomes $\VoutcomeD{N}{}{0}$ is drawn from a distribution that possesses finite moments of order $k$. Specifically, a straightforward application of the law of large numbers in Theorem~\thmSLLN, outlined in the appendices, yields \ref{asmp:BL}-\ref{asmp:BL-bound on initials}. Subsequently, reusing the same theorem along with \ref{asmp:BL}-\ref{asmp:BL-bound on initials} implies \ref{asmp:BL}-\ref{asmp:BL-empirical dist of inits}. Overall, Assumptions~\ref{asmp:BL}-\ref{asmp:BL-bound on initials} and \ref{asmp:BL}-\ref{asmp:BL-empirical dist of inits} ensure that the initial outcomes $\VoutcomeD{N}{}{0}$ and the function~$\outcomeg{0}{}$ are informative and contribute to the estimation process.

Regularity conditions related to the outcome functions are commonly found in the existing literature. Examples include assuming bounded moments of a certain degree for the potential outcome functions \citep{savje2021average}, considering bounded outcomes \citep{leung2022causal}, and assuming boundedness of the potential outcome function and its derivatives \citep{li2022random}. 

Given $\AVO{}{}{1}$ and $\VVO{}{}{1}$ as in Assumption~\ref{asmp:BL}, we proceed by considering the state evolution equations in \eqref{eq:state evolution} for $t\geq 1$.
Then, we present the following theorem that formalizes a more general version of the result that was previously stated in \S \ref{subsec:SE}. This theorem also characterizes the joint distribution of $\outcome{}{n}{1}, \ldots, \outcome{}{n}{t+1}$ within large sample asymptotics, providing a better understanding of the underlying statistical properties of a high-dimensional network data.
\begin{theorem}
    \label{thm:Big theorem}
    Fixing $k\geq 2$, assume the sequence of initial outcomes $\VoutcomeD{N}{}{0}$, the treatment assignment~$\Mtreatment{}{N}$, as well as the covariates $\covar(N)$ are given and suppose Assumption~\ref{asmp:BL} holds. Then, we have the following statements for all $t \geq 0$.
    \begin{enumerate}[label=(\alph*)]
        \item \label{part:BT-a} For any function $\psi: \R^{t+1+T+M} \mapsto \R$ that $\psi \in \poly{k}$, we have
        \begin{equation}
            \begin{aligned}
                \label{eq:BT-average limit}
                \lim_{N \rightarrow \infty}
                \frac{1}{N} \sum_{n=1}^N
                \psi\big(
                \outcomeD{}{n}{1}
                , \ldots,
                \outcomeD{}{n}{t+1}
                ,
                \Vtreatment{n}{},\Vcovar{n}
                \big)
                \eqas
                \E
                \Big[
                \psi\big(
                \AVO{}{}{1} + \VVO{}{}{1} Z_1,
                \ldots,
                \AVO{}{}{t+1} + \VVO{}{}{t+1} Z_{t+1},
                \Vtreatment{}{},\Vcovar{}
                \big)
                \Big],
            \end{aligned}
        \end{equation}
        where $Z_s \sim \Nc(0,1),\; s= 1,\ldots,t+1,$ are independent of $(\Vtreatment{}{},\Vcovar{}) \sim \Pi \x p_{\covar}$.

        \item \label{part:BT-c} Let $\VMoutcome{t}$ be a matrix with columns equal to $\outcomeD{}{}{s},\; s=1,\ldots,t$; that is $\VMoutcome{t} := \left[ \outcomeD{}{}{1} \big| \outcomeD{}{}{2} \big| \ldots \big| \outcomeD{}{}{t} \right]$. Then, the following matrix is positive definite almost surely:
        \begin{align}
            \label{eq:BT-positive definite covar matrix}
            \lim_{N \rightarrow \infty} \frac{\VMoutcome{t}^\top \VMoutcome{t}}{N} 
            -
            \lim_{N \rightarrow \infty} \frac{\VMoutcome{t}^\top\ones{N\times 1}}{N}
            \lim_{N \rightarrow \infty} \frac{\ones{1\times N}\VMoutcome{t}}{N}
            \succ 0,
        \end{align}
        where $\ones{l_1\times l_2}$ is matrix of size $l_1\times l_2$ with all entries equal to $1$.
    \end{enumerate}
\end{theorem}

Broadly speaking, Theorem~\ref{thm:Big theorem} yields concise one-dimensional dynamical equations that consolidate the analysis of high-dimensional network data in the large sample asymptotic. Specifically, \eqref{eq:BT-average limit} indicates that unit outcomes follow a Gaussian distribution at any given time. Moreover, when considering the entire time horizon, the outcomes exhibit a non-degenerate multivariate normal distribution (\eqref{eq:BT-positive definite covar matrix}). This implies that the joint distribution of outcomes can be well-characterized and analyzed. In summary, Theorem~\ref{thm:Big theorem} provides a framework for analyzing high-dimensional network data. It establishes the Gaussian nature of outcomes, facilitating the computation of various statistics and functions that capture the average behavior of the system.

To obtain the results of Theorem~\ref{thm:Big theorem}, 
the main challenge arises from the dependence between the fixed interference matrix $\IM$ and potential outcomes $\left\{\outcome{}{n}{t}\right\}_{n\in[N]}$ for any time $t > 0$. Indeed, the observed outcomes reveal some information about $\IM$ that we need to incorporate into any calculation concerning future observations. 
To overcome this obstacle, we leverage the ``conditioning technique" introduced by \cite{bolthausen2014iterative} and developed further by \cite{bayati2011dynamics}. This technique is commonly employed in the literature on Approximate Message Passing (AMP) algorithms, e.g., \citep{javanmard2013state,rush2018finite,berthier2020state,feng2022unifying}. They typically consider a fixed symmetric coefficient matrix or a Wishart matrix, with entries of order $O(1/\sqrt{N})$, and also assume a pseudo-Lipschitz non-linearity. We adapt the analysis of \cite{bayati2011dynamics} to the current setting, 
assuming that the coefficient matrix is non-symmetric, dropping the Lipschitz assumption of non-linearity, and consider an additional noise term in each round. As a result, the analysis becomes easier in some sense (as there is no ``Onsager" or memory term), yet involving additional randomness structures. We provide the rigorous proofs in Appendix~\secDetailedProofs.

Recent literature on finite sample analysis of AMP \citep{rush2018finite,li2022non} shows that the distribution of unit outcomes at a finite time $t$ converges to a Gaussian distribution at a rate of $\sqrt{\log(N)/N}$ \citep{li2022non}. Accordingly, Algorithm \ref{alg:causal-mp-original} must grapple with an added $O(\sqrt{1/N})$ noise present in the sample means. To reduce the effect of this noise term, one can either increase the population size $N$ or extend the experiment duration $T$. In~\S \ref{sec:Numerical}, we demonstrate that Algorithm \ref{alg:causal-mp-original} remains effective even with values of $N$ as small as 500 and $T$ around 60. While deriving the precise impact of finite sample error on estimation results is an exciting research direction, it is worth noting that many tech companies conduct online experiments with thousands to millions of units \citep{gupta2019top}.

\subsection{Consistency of the TTE Estimator in Algorithm~\ref{alg:causal-mp-original}}
\label{subsec:tte-estimation-consistency}

We proceed by demonstrating that Algorithm~\ref{alg:causal-mp-original} yields a strongly consistent estimator for the total treatment effect defined in \eqref{eq:TTE_def_fixed}.

\begin{theorem}
    \label{thm:consistency}
    Suppose that there exist at least two distinct sample means $\HAVO{}{}{t'} \neq \HAVO{}{}{t''}$, for some $t' \neq t''$, and $\CE^n,\PE^n,\DE^n,\ME^n \in \R$ and $\CC^n$, $n \in [N]$, are random objects independent of everything else with bounded $k^{th}$ moments.
    Let $\ETTE{t}{\desired,0}$ be the estimator defined in Algorithm~\ref{alg:causal-mp-original}. Then, $\ETTE{t}{\desired,0}$ is a strongly consistent estimator for the total treatment effect; that is, for any $t \in [T]_0$, we have
    \begin{align}
        \label{eq:consistency}
        \lim_{N \rightarrow \infty}
        \ETTE{t}{\desired,0}
        \eqas
        \TTE{t}{\desired,0}.
    \end{align}
\end{theorem}
According to Theorem~\ref{thm:consistency}, as the number of individuals $N$ grows large, the estimator converges to the true value of the TTE almost surely.

We can establish a similar consistency result for Algorithm~\ref{alg:CE estimation} as long as the estimation of parameters in the second step remains consistent. A proof scheme analogous to Theorem~\ref{thm:consistency}, relying on the results of Theorem~\ref{thm:Big theorem} and the state evolution dynamics in \eqref{eq:state evolution}, would be sufficient to obtain these results. In summary, by appropriately choosing the model specifications and accurately estimating the set~$\Uc$ in the second step of Algorithm~\ref{alg:CE estimation}, we can leverage the proposed framework to design and analyze various desired causal effects in diverse models. This claim is supported in $\S$~\ref{sec:Numerical} through a comprehensive analysis of different systems with more general interference patterns that relax the assumptions of Theorem~\ref{thm:consistency}.

In addition to TTE estimation, Appendix \S \subsecDirectEffect details the estimation of direct and indirect effects. Using Theorem \ref{thm:Big theorem}, we show that the difference-in-means and Horvitz-Thompson estimators are strongly consistent for the direct effect. This is similar to results of \cite{savje2021average,munro2021treatment} but in our distinct setting, with prevalent and locally heterogeneous interference,  as discussed in \S \ref{sec:PF}. This result, combined with Theorem \ref{thm:consistency}, is then used to estimate the indirect effect.


\section{Numerical Illustrations}
\label{sec:Numerical}

In this section, we investigate five experimental scenarios utilizing a Bernoulli design and employ Algorithm~\ref{alg:causal-mp-original} to estimate the total treatment effect. We also use the resampling idea outlined in \S \ref{subsec:CI} to estimate confidence intervals. 
The first scenario involves a linear-in-means model under a staggered roll-out design with three different interference structures. 
The second scenario considers a binary potential outcome model, and the third scenario focuses on estimating the total effect of speeding up servers in a parallel server system.

The design employed in each case is detailed subsequently. Unless otherwise specified, each experiment begins from a near-equilibrium state with all units in the control condition. This state is obtained by running the data-generating process with all treatment variables set to zero for 10 “burn-in” periods before the experimentation phase begins. Each simulation is replicated 5,000 times, with each replication involving a new network, a new realization of the randomized treatment assignment, and new noise.

In \S \ref{sec:SBM}, we aim to compare Algorithm~\ref{alg:causal-mp-original} with several benchmarks. Therefore, we use a stochastic block model that includes a clustered interference pattern, enabling the use of clustered randomized experiments. The remaining scenarios are aimed at showcasing the capability of Algorithm~\ref{alg:causal-mp-original} to estimate the entire dynamics of TTE and its robustness to varying interference structures or outcome dynamics. In all scenarios, we compare the estimates with the ground truth value of the treatment effect, which is feasible because we have access to the data-generating process and can recreate the necessary counterfactual outcomes. Additional numerical results are provided in Appendix \S \secApndxNumericals.

\subsection{Linear-in-Means Model with Staggered Roll-out Design}\label{subsec-LiM-sythetic}

We begin by adapting and replicating the linear-in-means model  \citep{cai2015social,leung2022causal}. However, similar to \cite{eckles2016design}, we make slight modifications to the original static model to capture dynamic settings. Specifically, for $i \in [N]$, we define the following dynamic outcome model:
\begin{equation}
\begin{aligned}
    \label{eq:linear-in-mean}
    \outcome{}{i}{t+1}
    =
    \alpha_1
    +
    \alpha_2 \frac{\sum_{j=1}^N \adjMe^{ij}\outcome{}{j}{t}}{\sum_{j=1}^N \adjMe^{ij}}
    +
    \alpha_3 \frac{\sum_{j=1}^N \adjMe^{ij}\treatment{j}{t+1}}{\sum_{j=1}^N \adjMe^{ij}}
    +
    \alpha_4 \treatment{i}{t+1}
    +
    \eps_i.
\end{aligned}
\end{equation}
In \eqref{eq:linear-in-mean}, the matrix $\adjM$ $= [\adjMe^{ij}]_{i,j}$ defines the adjacency matrix of a graph and $\eps_n \sim \Nc(0,1)$. Following \cite{leung2022causal}, we set $(\alpha_1,\alpha_2,\alpha_3,\alpha_4) = (-1,0.8,1,1)$ as parameters. Here, $\alpha_1$ signifies the baseline effect, $\alpha_2$ indicates the autocorrelation and peer effect, $\alpha_3$ reflects the spillover effect, and $\alpha_4$ corresponds to the direct effect, as shown in Figure~\ref{fig:TEs}. Additionally, to deliver the treatments, we have employed a staggered roll-out design with $(\pi_1, \pi_2) = (0.2, 0.5)$. Under this design, each unit that receives treatment in the first stage remains under treatment in the second stage as well. This assumption takes into account practical constraints that might prevent units from switching between control and treatment groups. For example, in the context of a contagious disease, the new treatment can induce long-lasting effects on the treated individuals, making it impractical or ethically challenging to reverse the treatment once it has been applied.

Next, we investigate the robustness of the proposed method against misspecification in the interference structure by considering three different graphs as detailed below.

\subsubsection{Linear-in-means model with stochastic block network}
\label{sec:SBM}
In the first scenario, inspired by \cite{eckles2016design}, we consider an interference network captured through a directed stochastic block model with $10$ blocks, each representing a different cluster. We examine populations with $N=$ 500, 1,000, and 10,000, where each unit is connected to an average of $5$ units within the same cluster and an average of $0.9$ units from other clusters.

We benchmark the performance of Algorithm~\ref{alg:causal-mp-original} as follows. Considering a Randomized Controlled Trial (RCT) where we randomize the treatment over units, when the interference structure is completely unknown, a potential approach is to ignore the interference and rely on simple estimators \citep{eckles2016design}. Accordingly, we consider Difference-in-Means (DM) and Horvitz-Thompson (HT) estimators, outlined in Appendix \S \subsecDirectEffect. In this setting, both DM and HT indeed estimate the direct effect, as shown in Appendix \S \subsecDirectEffect. 
However, if the network structure is partially known up to clusters, the literature suggests randomizing the treatments over the clusters, referred to as Cluster RCT \citep{puffer2005cluster}. In Cluster RCTs, and in the ideal case where clusters have no inter-cluster connections, both DM and HT are expected to yield unbiased estimates of the TTE \citep{ugander2013graph,eckles2016design,aronow2017estimating}.

\begin{figure}
    \centering
    \includegraphics[width=0.95\linewidth]{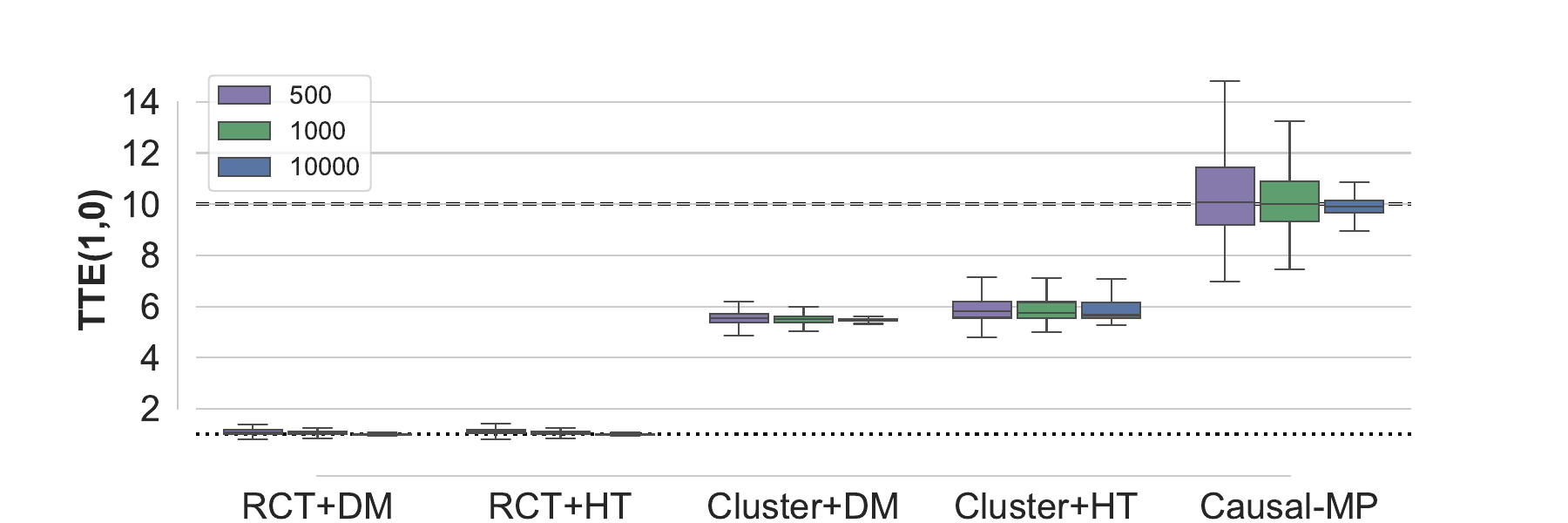}
    \caption{Linear-in-means model with stochastic block network; Causal-MP vs. Difference-in-Means (DM) and Horvitz-Thompson (HT) estimator in Randomized Controlled Trial (RCT) and Cluster RCT. The ground truth value for the TTE is 10 (the dashed line) and for the direct effect is 1 (the dotted line).}
    \label{fig:SBM}
\end{figure}

Figure~\ref{fig:SBM} depicts the results, highlighting the accuracy of Causal-MP compared to other benchmarks, which exhibit significant biases. Specifically, in the absence of perfect network observation, other estimators might overlook unit connections (whether within clusters or between clusters), resulting in biased estimations of the treatment effect. In contrast, applying Algorithm~\ref{alg:causal-mp-original} without any knowledge of the interference structure provides an accurate estimation of the TTE, underscoring the relevance of the proposed framework. 

\subsubsection{Linear-in-means model with Facebook friends lists network}
\begin{figure}
    \centering
    \includegraphics[width=0.95\linewidth]{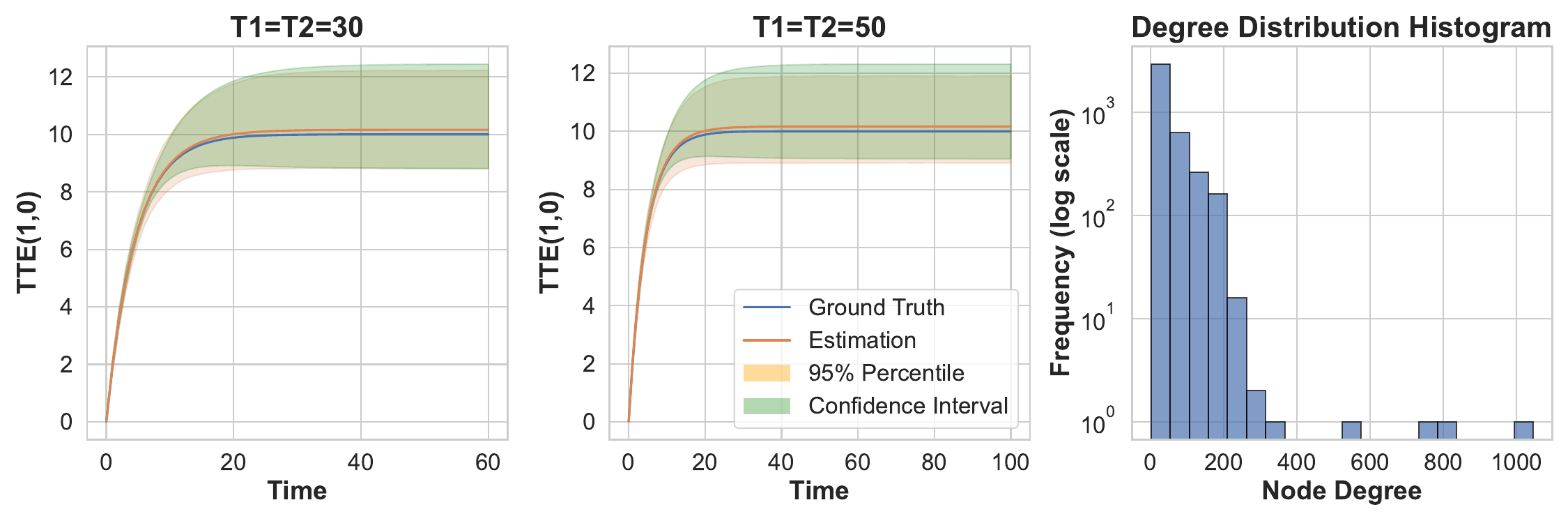}
    \caption{Left: Linear-in-means model with Facebook network; 95\% confidence interval for the total treatment effect estimation when $(\pi_1,\pi_2)=(0.2,0.5)$. Right: degree distribution of the Facebook network.}
    \label{fig:Facebook}
\end{figure}

Next, we explore the linear-in-means model as described in \eqref{eq:linear-in-mean}, but utilize a Facebook network dataset with 4,039 nodes and 88,234 edges, obtained from \cite{leskovec2012learning}. Then, Figure~\ref{fig:Facebook} displays the results. The right plot in this figure illustrates the degree distribution of the nodes, highlighting significant variation in node degrees. 
Additionally, Figure~\ref{fig:Facebook} includes the (estimated) 95\% confidence intervals for the output of Algorithm~\ref{alg:causal-mp-original}, calculated via the heuristic detailed in \S \ref{subsec:CI} using $B=500$ and and $q=0.15$.
It is worth noting that the slight bias observed is potentially due to the relatively small network size ($N =$ 4,039) or misspecifications between the assumptions of our proposed methodology and the real network data. But the overall performance demonstrates the versatility of the proposed approach.

\subsubsection{Linear-in-means model with random geometric graph}
In this setting, proposed by \cite{leung2022causal}, we generate the graph with adjacency matrix $\adjM$ using random geometric graph models. Specifically, we create a graph with vertex set $[N]$, such that for each pair of distinct units $i$ and $j$ in $[N]$, we define $\adjMe^{ij} = \1_{\|\Vec{V}_i-\Vec{V}_j\|\leq r_N}$, where for each unit $n$ in $[N]$, $\Vec{V}_n=[V_{n}^1,V_{n}^2]^\top\in \R^2$ is determined by independently sampling each coordinate from the uniform distribution over the interval $[0,1]$, and $r_N = \sqrt{8(\pi N)^{-1}}$. Importantly, this model implies an average connectivity where each individual is linked to 8 other units, which implies a moderate level of interference. Moreover, we let $\eps_n = V_{n}^1 - 0.5 + \Nc(0,1)$ in \eqref{eq:linear-in-mean}; consequently, as \cite{leung2022causal} notes, the noise term generates unobserved homophily, and units with closer $V_{n}^1$ values have similar outcomes.

\begin{figure}
    \centering
    \includegraphics[width=1\linewidth]{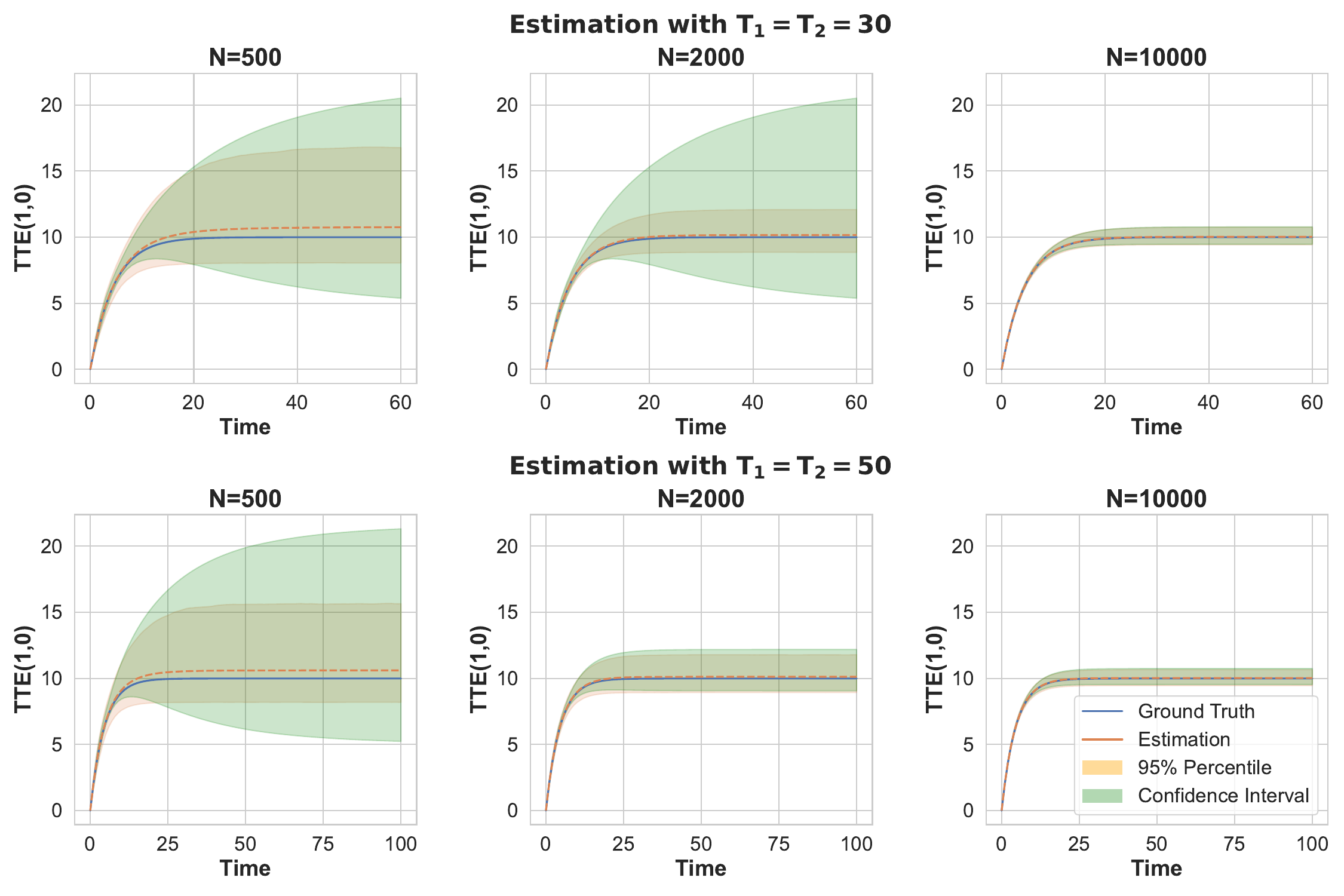}
    \caption{Linear-in-means model with random geometric graph;   
    95\% confidence intervals for the total treatment effect estimation when $(\pi_1,\pi_2)=(0.2,0.5)$.}
    \label{fig:LinM_pi}
\end{figure}

Figure~\ref{fig:LinM_pi} displays the results, while the first row depicts experiments with $T=60$ and the second row illustrates longer experiments with $T=100$. This figure presents both the average estimated TTE, obtained using Algorithm~\ref{alg:causal-mp-original} along with its (true) 95\% confidence interval, and the ground-truth TTE derived from the replications. Additionally, Figure~\ref{fig:LinM_pi} includes the (estimated) 95\% confidence intervals for the output of Algorithm~\ref{alg:causal-mp-original}, calculated via the heuristic detailed in \S \ref{subsec:CI} using $B=500$. The parameter $q$ is set to $0.4$, $0.3$, and $0.25$ corresponding to $N=500$, $2,000$, and $10,000$, respectively. Finally, in light of Remark~\ref{rem:prior}, we incorporate a ``prior-knowledge'' that the magnitude of the TTE is at most $100$.

As $N$ or $T$ grow in Figure~\ref{fig:LinM_pi}, we observe more accurate estimation. The former directly aligns with the consistency result in Theorem~\ref{thm:consistency}. The latter is also (but indirectly) related to Theorem~\ref{thm:consistency}. Particularly, the inputs to Algorithm~\ref{alg:causal-mp-original} contain error terms, that go to zero with $N$. For any finite~$N$, their impact on the estimation error is compensated with increasing $T$.
This indicates that as we gather more data over an extended period, the accuracy of the estimates further improves, as reflected in the lower width of the confidence intervals in the figure.

\subsection{Binary Outcome Model with Micro-Randomized Trial}\label{subsec:MRT}
The setting we examine next is a binary potential outcome model \citep{li2022network}. Precisely, we consider $\outcome{}{i}{t+1} \sim \text{Bernoulli}(O_t)$, such that,
\begin{align}
    \label{eq:MRT}
    O_t
    =
    \alpha_1
    +
    \alpha_2
    \treatment{i}{t+1} Z^i_t
    +
    \alpha_3
    \outcome{}{i}{t} Z^i_t
    +
    \alpha_4 \treatment{i}{t+1} \outcome{}{i}{t} Z^i_t
\end{align}
where $Z^i_t = \sum_{j=1}^N \adjMe^{ij}\outcome{}{j}{t}$ represents the number of neighbors of individual $i$ with an outcome of 1.
Following Example 1 in \cite{li2022network}, we let $\adjM$ to be the adjacency matrix of an Erd\"os-R\'enyi graph where each pairs of vertices are connected, independently, with probability~$p_{\text{edge}}$. We set the parameter values as $(\alpha_1,\alpha_2,\alpha_3,\alpha_4,p_{\text{edge}}) = (0.5,0.15,0.20,0.01,3/N)$. Consequently, on average, each individual is connected to three other units, resulting in a low interference level in the network.

We adopt a Micro-Randomized Trial (MRT) with a Bernoulli design to determine the treatment group in each period. Specifically, we set $(\pi_1,\pi_2)=(0.25,0.75)$, and for each period in stage $j\in\{1,2\}$, we generate a new treatment vector $\Vtreatment{}{t}$ such that $\treatment{i}{t} \simiid \text{Bernoulli}(\pi_j)$, where $t \in [T_j]$. MRTs were initially introduced as an experimental design for developing just-in-time adaptive interventions by \cite{liao2016sample} and \cite{klasnja2015microrandomized}. Since then, they have gained popularity in various research areas, particularly in studying mobile health interventions aimed at increasing physical activity among sedentary individuals \cite{klasnja2019efficacy}.

Here, we consider the binary outcome model without the initial “burn-in” periods and defer the case with the “burn-in” periods to Appendix~\S \secBOMWBI. Indeed, not all experiments begin with the population under study in a stable equilibrium. For instance, consider the evaluation of a new medication for a recently emerged contagious disease. In this situation, the health status of experimental units might be subject to rapid changes due to the disease's spread and the dynamic interactions among individuals. This instability complicates the estimation of treatment effects because baseline conditions are not consistent. Therefore, it is crucial to develop methods capable of accurately tracking the dynamics of the treatment.

\begin{figure}
    \centering
    \includegraphics[width=1\linewidth]{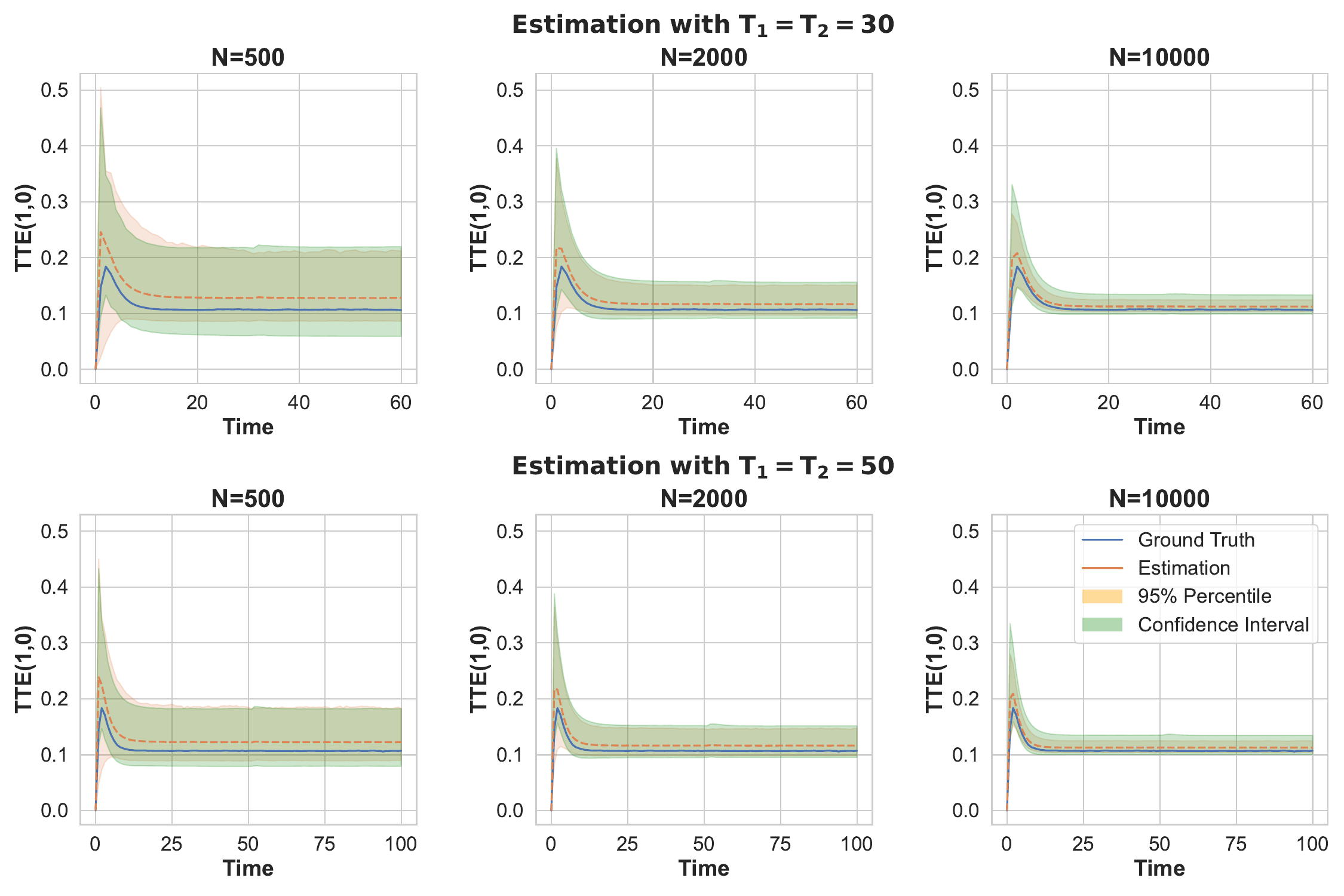}
    \caption{Non-equilibrium binary outcome model: 95\% confidence interval for the total treatment effect estimation when $(\pi_1,\pi_2)=(0.25,0.75)$.}
    \label{fig:MRT_pi_WOWU}
\end{figure}

Figure~\ref{fig:MRT_pi_WOWU} shows the results, highlighting a quick jump to the ground truth value of the TTE in the early stages of the experiment. This jump is successfully captured by Algorithm~\ref{alg:causal-mp-original}, demonstrating its reliable performance in tracking the dynamics of the TTE throughout the experiment. Similar to Figure~\ref{fig:LinM_pi}, increasing the sample size and extending the time horizon leads to significantly improved precision in the estimates.
Here, $B=500$ and $q$ is set to $0.6$, $0.5$, and $0.3$ corresponding to $N=500$, $2,000$, and $10,000$, respectively. Given the binary outcome model, we incorporate the prior knowledge that the TTE's magnitude is capped at 1, as discussed in Remark~\ref{rem:prior}.

\subsection{Parallel Server System Speed-up Effect}
We next consider a parallel server system with $N$ servers that operates under the join-the-shortest queue policy, a widely adopted routing strategy for server farms \cite{gupta2007analysis}. This policy assigns each incoming job to one of the shortest queues. Motivated by
\cite{kuang2024detecting}, our primary focus is on understanding the interference impact of change in the service rate (speeding up the servers in our case) on the overall ``server utilization". 
To study this, we conduct a randomized experiment by speeding up selected servers. However, it is crucial to account for the interference effect among servers, which arises due to the join-the-shortest queue policy. This interference can cause servers in the control group, maintaining their original service rates, to experience reduced demand as a result of the allocation policy. Consequently, the treatment assignment impacts the control group.

We generate the data by simulating a parallel server system characterized by a Poisson arrival process with a rate of $0.95N$ and an exponential service time, where each server has a service rate of 1. The treatment we consider involves doubling the speed of randomly selected servers. At the beginning of stage 1 ($t=1$), we increase the service rate of each server to 2 with a probability $\pi_1 = 0.15$, continuing this setting until the end of the first stage ($t=T_1$). Subsequently, at the beginning of stage 2 ($t=T_1+1$), we increase the service rate of each server to 2 with a probability $\pi_2 = 0.5$. Therefore, the treatment status of each server remains unchanged within each stage. Then, the observed data from server $n$ at time period $t$ (i.e., the outcome $\outcome{}{n}{t}$) represents the total time that server $n$ is busy during the time interval $[t,t+1)$. This setup allows us to evaluate the performance of Algorithm~\ref{alg:causal-mp-original} in a setting with implicit interference introduced by the join-the-shortest queue policy, versus the more explicit specifications as in \eqref{eq:linear-in-mean}-\eqref{eq:MRT}.

\begin{figure}
    \centering
    \includegraphics[width=1\linewidth]{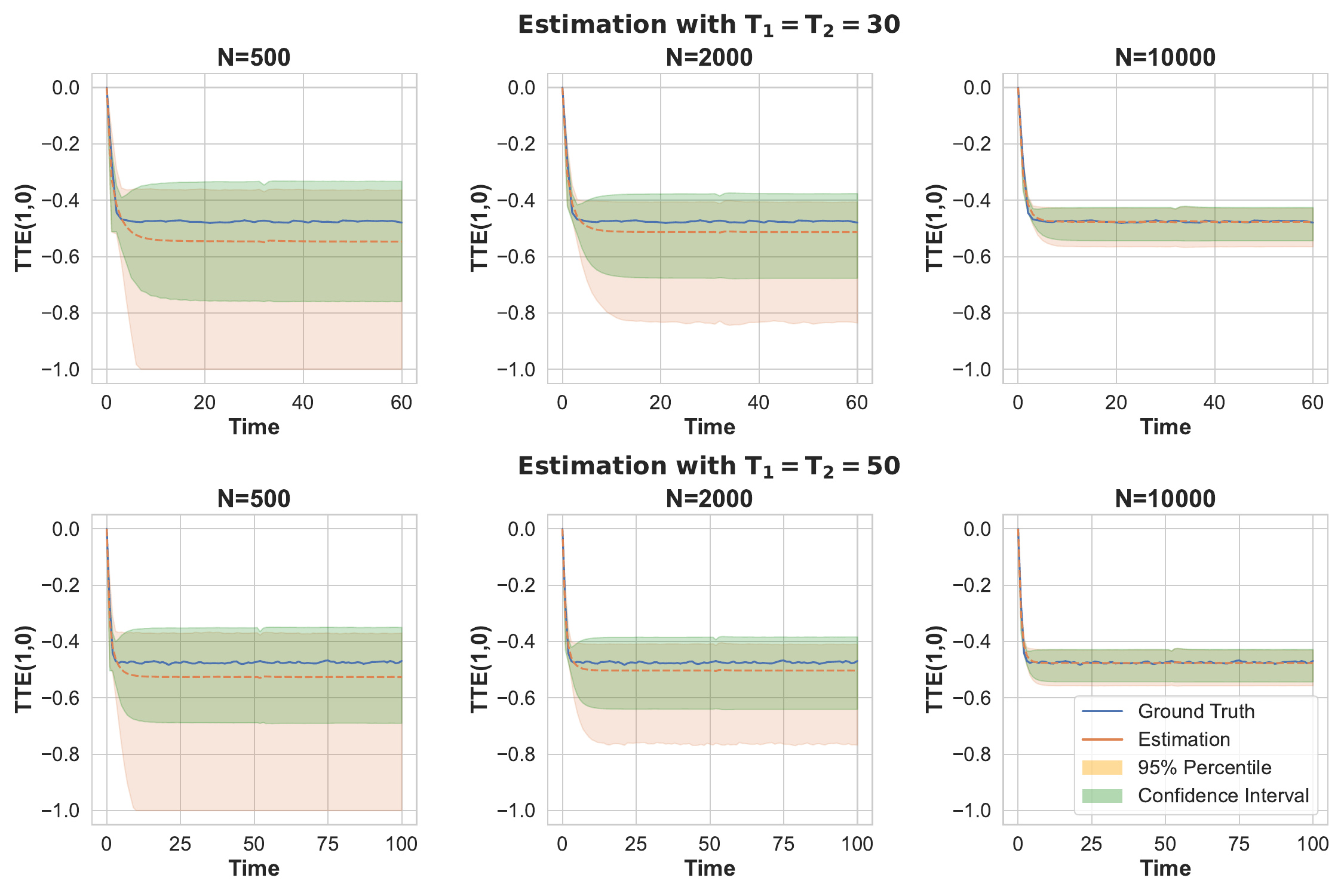}
    \caption{Server speed-up problem with implicit interference due to the join-the-shortest queue policy; 95\%~confidence interval for the total treatment effect estimation when $(\pi_1,\pi_2)=(0.15,0.5)$.}
    \label{fig:Servers_pi}
\end{figure}

Figure~\ref{fig:Servers_pi} presents the results of estimating the TTE in systems with different numbers of servers and time horizons. Algorithm~\ref{alg:causal-mp-original} successfully estimates the small treatment effect, even when treating at most half of the units in the system ($\pi_2=0.5$). This demonstrates the effectiveness of our proposed framework for real-world applications, as it accounts for the complex interference effects and yields reliable estimates of treatment effects in diverse scenarios. For the confidence interval heuristic, we set $B=500$ as before and $q$ is selected to be $0.2$, $0.15$, $0.1$ for $N=500$, $N=$2,000, and $N=$10,000, respectively. Consistent with earlier scenarios, we utilize the prior knowledge that the TTE must be between $-1$ and $0$, as noted in Remark~\ref{rem:prior}.

\subsection{TTE estimation at other treatment levels}
\begin{figure}
    \centering
    \includegraphics[width=1\linewidth]{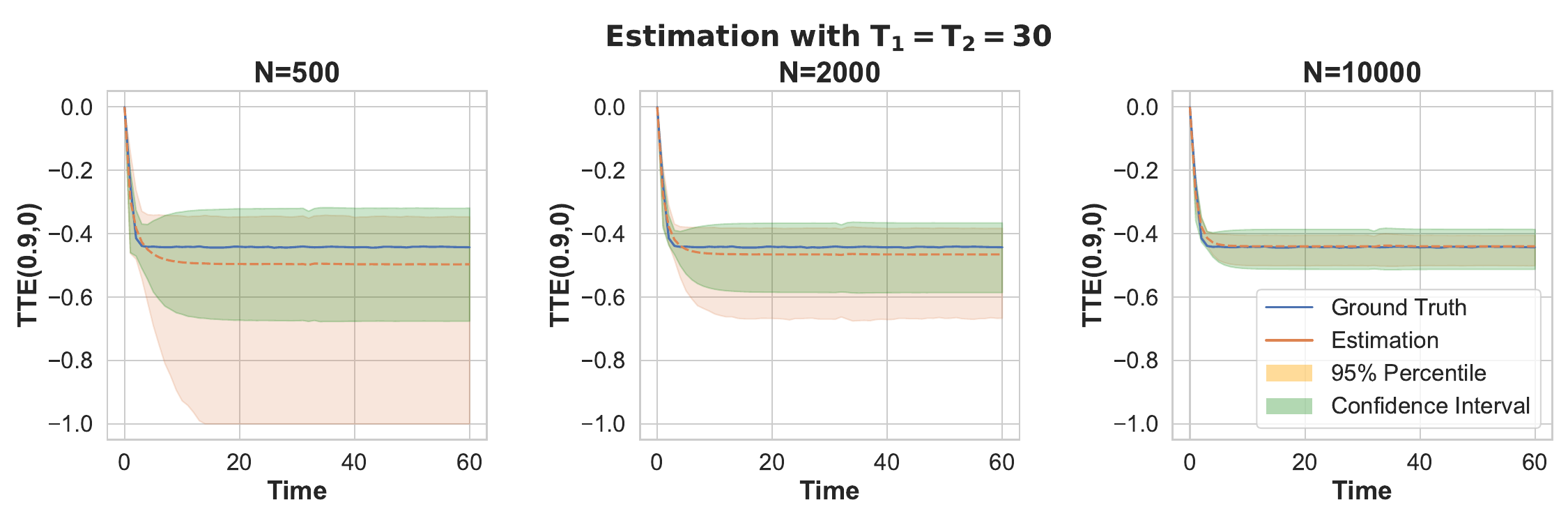}
    \caption{Server speed-up problem under a Bernoulli design: 95\% confidence interval for $\TTE{t}{0.9,0}$ estimation when $(\pi_1,\pi_2)=(0.15,0.5)$.}
    \label{fig:Servers_pi09}
\end{figure}

In many practical scenarios, it is not possible to deliver the treatment to the entire population, either due to failure in delivery by the experimenter or due to resistance in adoption by the units. For instance, in the contagious disease example, some individuals may avoid using the new medication. In these settings, a broader family of estimands is necessary, as covered by Algorithm~\ref{alg:causal-mp-original}. Figure~\ref{fig:Servers_pi09} presents the results of estimating $\TTE{t}{0.9,0}$ in the parallel server system speed-up, while
$\TTE{t}{0.9,0}$ for linear-in-means and binary outcome settings are deferred to Appendix \secApndxNumericals. $\TTE{t}{0.9,0}$ compares the scenario of treating 90\% of the population on average (with each experimental unit adopting the treatment with a probability of 0.9) to the scenario of treating no one. The results in Figure~\ref{fig:Servers_pi09} demonstrate the effectiveness of the proposed method for estimating a broader range of treatment effects, beyond just $\TTE{t}{1,0}$.


\section{Conclusion}
\label{sec:conclusion}

Estimating the total treatment effect in presence of network interference is an important scientific and practical question. Previous research has tackled this challenge by limiting interference to immediate neighbors, imposing structural constraints on interference, specializing outcome models, or innovating new estimands to overcome inherent difficulties. This study introduces the Causal Message-passing framework as a new methodology for experimental design under unknown network interference. We provide a theoretical analysis of the framework and formulate one-dimensional representations of the high-dimensional outcomes.

In the context of Bernoulli randomized designs, we propose a strongly consistent estimator for the total treatment effect. To exhibit the adaptability of the proposed framework, we present several distinct case studies, with varying interference patterns and outcome specifications. These studies serve as a ``proof-of-concept," affirming the utility of Caussal Message-passing in experiment design and analysis where interference patterns are unknown.

There are several directions left for future explorations. One immediate area is assessing applicability and robustness of the proposed methodology in more realistic contexts, such as real-world experimental design settings, with increased risks of model misspecification. On the theoretical side, an analysis of the confidence interval estimation method presents an important avenue for exploration.


\newpage

\onecolumn
\begin{center}
{\Large {\bf Supplementary Appendices for}\\
\vspace{5mm}

Causal Message Passing for Experiments with Unknown and General Network Interference}
\vspace{5mm}

\end{center}


\section{Further Extensions and Interpretations}
\label{sec:generalizations}

In this section, we explore various ways to generalize and interpret the outcome specification \eqref{eq:outcome_function_matrix}. In \S \ref{subsec:direct-effect}, we begin by extending the results to a setting with direct effects and discuss how both direct and indirect effects can be estimated. This is followed by a different generalization where a non-linear random operator is applied to the right-hand side of \eqref{eq:outcome_function_matrix}. Next, we examine a neural network interpretation of \eqref{eq:outcome_function_matrix} and briefly discuss extensions to more general interference patterns.

	\subsection{Direct and the Indirect Effects}\label{subsec:direct-effect} 
	
	In this section, we first start by discussing a slightly more general version of the outcome specification \eqref{eq:outcome_function_matrix}. Recall that the notation $\Vtreatment{}{t}$ refers to the treatment vector applied to all units at time $t$.
 
 Specifically, we consider:
	\begin{align}
		\label{eq:outcome_function_matrix_with_direct}
		\Voutcome{}{}{t+1} =
		\big(\IM+\IMatT{t}\big)\outcomeg{t}{}\left(\Voutcome{}{}{t} ,\Mtreatment{}{}, \covar\right) +\DTEP{}{} \Vtreatment{}{t+1}
		+\Vnoise{}{t}\,,
		\quad\quad\quad
		t=0,1,\ldots,T-1,
	\end{align}
	where a new term involving the unknown constant $\DTEP$, which captures the \emph{direct treatment effect}, is added. We will show that our results from the settings with $\DTEP=0$ can be applied here, through a change of variable. In addition, we will demonstrate that Algorithm \ref{alg:causal-mp-original}, without any modification can be used to estimate TTE, i.e., Theorem \ref{thm:consistency} continues to hold. However, to decompose TTE into its two components, the direct effect $\DTEP$ and the indirect effect, we would need to introduce a new pre-processing step.
	\begin{definition}
		If the treatment variables are binary, for any $t \in [T]$, the difference-in-means and Horvitz-Thompson estimators, denoted by $\DIME{t}$ and $\HTE{t}$ respectively, are defined as follows:
		\begin{align}
			\DIME{t} & := \frac{\sum_{n=1}^N\outcome{}{n}{t}\treatment{n}{t}}{\sum_{n=1}^N\treatment{n}{t}} - \frac{\sum_{n=1}^N\outcome{}{n}{t}(1-\treatment{n}{t})}{\sum_{n=1}^N(1-\treatment{n}{t})}\label{eq:DM-estimator}\\
			\HTE{t} &:= \frac{1}{N} \sum_{n=1}^N \left( \frac{\outcome{}{n}{t} \treatment{n}{t}}{\E[\treatment{n}{t}]} - \frac{\outcome{}{n}{t} (1 - \treatment{n}{t})}{\E[1 - \treatment{n}{t}]} \right)\,.\label{eq:HT-estimator}
		\end{align}
	\end{definition}
	Next, we show that we can utilize our theoretical results from \S \ref{sec:Tech_result} to recover a version of the known result in the network interference literature \citep{savje2021average,li2022random}, which states that the difference-in-means and Horvitz-Thompson estimators estimate the direct effect.
	\begin{theorem}[Direct effect via difference-in-means and Horvitz Thompson]\label{thm:DM-and-HT-estimate-direct-effect}
		Consider the same assumptions as in Theorem \ref{thm:Big theorem}, along with the additional assumption that treatment variables are binary with means strictly away from $0$ or $1$, i.e., $\E[\treatment{}{t}]\notin \{0,1\}$ for any $t \ge 1$. Then, $\DIME{t}$ and $\HTE{t}$ are strongly consistent estimators for $\DTEP$., i.e.,
		\[
		\lim_{N \rightarrow \infty} \DIME{t} \eqas \DTEP~~~~\text{and}~~~~\lim_{N \rightarrow \infty} \HTE{t} \eqas \DTEP\,.
		\]
	\end{theorem}
	{Proof.}
		Let us define, for all $t$, a new outcome process $\{\VCoVoutcome{}{}{t}\}_{t\ge 0}$ involving the same treatment matrix $\Mtreatment{}{}$, covariates $\covar$, and interference matrices $\IM+\IMatT{t}$, defined by:
		\[
		\VCoVoutcome{}{}{t}:=\Voutcome{}{}{0},~~~~\VCoVoutcome{}{}{t}:=\Voutcome{}{}{t}-\DTEP{}{}\Vtreatment{}{t},~~~t=1,\ldots,T\,,
		\]
		but using the updated function $\CoVoutcomeg{}{}$, defined as
		\[
		\CoVoutcomeg{t}{}\left(\VCoVoutcome{}{}{t} ,\Mtreatment{}{}, \covar\right):=\outcomeg{t}{}\left(\VCoVoutcome{}{}{t}+\DTEP{}{}\Vtreatment{}{t} ,\Mtreatment{}{}, \covar\right)
		\,.
		\]
		It is easy to see that \eqref{eq:outcome_function_matrix_with_direct} is equivalent to
		\begin{align}
			\label{eq:outcome_function_matrix_with_direct_change_of_var}
			\VCoVoutcome{}{}{t+1} =
			\big(\IM+\IMatT{t}\big)\CoVoutcomeg{t}{}\left(\VCoVoutcome{}{}{t},\Mtreatment{}{}, \covar\right)
			+\Vnoise{}{t}\,,
			\quad\quad\quad
			t=0,1,\ldots,T-1\,,
		\end{align}
		which is in the same format as the original \eqref{eq:outcome_function_matrix}. Moreover, Assumptions \ref{asmp:Gaussian Interference Matrice}-\ref{asmp:BL} hold for this updated process which means we can apply Theorem \ref{thm:Big theorem} and the updated state evolution equations,
		\begin{equation}
			\label{eq:state evolution_change_of_var}
			\begin{aligned}
				\CoVAVO{}{}{t+1} :=
				(\mu+\mu_t) \E\left[
				\CoVoutcomeg{t}{}\big(\CoVAVO{}{}{t} + \CoVVVO{}{}{t} Z, \Vtreatment{}{},\Vec{X}\big)
				\right],
				\quad\quad
				\CoVVVO{}{2}{t+1} :=
				(\sigma^2+\sigma_t^2) \E\left[
				\CoVoutcomeg{t}{}\big(\CoVAVO{}{}{t} + \CoVVVO{}{}{t} Z, \Vtreatment{}{},\Vec{X}\big)^2
				\right] + \sigma_e^2\,,
			\end{aligned}
		\end{equation}
		where $\lim_{N \rightarrow \infty}\sum_{n=1}^{N}\CoVoutcome{}{n}{t}/N \eqas \CoVAVO{}{}{t}$. 
		
		By setting the function $\psi$ in Theorem \ref{thm:Big theorem}(a), once equal to $(\CoVoutcome{}{n}{t} +\DTEP \treatment{n}{t})\treatment{n}{t}$, and once equal to $\treatment{n}{t}$, we obtain 
		\[
		\lim_{N \rightarrow \infty}\frac{1}{N}\sum_{n=1}^{N}\outcome{}{n}{t}\treatment{n}{t} \eqas (\CoVAVO{}{}{t}+\DTEP)\E[\treatment{}{t}]
		~~~\text{and} ~~~
		\lim_{N \rightarrow \infty}\frac{1}{N}\sum_{n=1}^{N}\treatment{n}{t} \eqas \E[\treatment{}{t}]\,,
		\]
		respectively, where $\treatment{}{t}\sim\pi_t$.	Similarly, we can obtain,
		\[
		\lim_{N \rightarrow \infty}\frac{1}{N}\sum_{n=1}^{N}\outcome{}{n}{t}(1-\treatment{n}{t}) \eqas \CoVAVO{}{}{t}\E[(1-\treatment{}{t})]
		~~~\text{and} ~~~
		\lim_{N \rightarrow \infty}\frac{1}{N}\sum_{n=1}^{N}(1-\treatment{n}{t}) \eqas \E[(1-\treatment{}{t})]\,.
		\]
		Therefore, since $\E[\treatment{}{t}]\notin\{0,1\}$,
		\[
		\lim_{N\rightarrow\infty}\DIME{t}\eqas\CoVAVO{}{}{t}+\DTEP-\CoVAVO{}{}{t}=\DTEP\,.
		\]
		which finishes the proof for the difference-in-means estimator. Proof for the Horvitz-Thompson estimator uses the same argument, plus recalling that for all $n \in [N]$, $\treatment{n}{t} \sim \pi_t$, which means $\E[\treatment{n}{t}] = \E[\treatment{}{t}]$.
		\ep

	\subsubsection{Estimating the indirect effect}\label{eq:indirect-effect}
	
	Here, we first define the indirect (or network) effect, in light of the specification \eqref{eq:outcome_function_matrix_with_direct}, and utilize Theorem \ref{thm:DM-and-HT-estimate-direct-effect}, combined with a more general version of Theorem \ref{thm:consistency}, to estimate the indirect effect.
	
	We define the indirect treatment effect (ITE) by
	\begin{align}
		\label{eq:ITE-definition}
		\ITE{t}{1,0} := \TTE{t}{1,0} - \DTEP\,.
	\end{align}

	Next, we show that Theorem \ref{thm:consistency} remains valid under the specification \eqref{eq:outcome_function_matrix_with_direct} when the treatment assignment is re-randomized in each time period. 
	\begin{theorem}
		\label{thm:consistency-with-direct-effect}
		Under the assumptions of Theorem \ref{thm:consistency},  
		consider a two-stage Bernoulli experiment as in Algorithm~\ref{alg:causal-mp-original} with outcome specification \eqref{eq:outcome_function_matrix_with_direct}, with the added assumption that the treatment assignments at each time period are selected independently of the past treatments, i.e. the vectors $\Vtreatment{}{t}$ are independent for all $t\in[T]$. 
		Then, for any $\desired$, the output $\ETTE{t}{\desired,0}$ of the algorithm is a strongly consistent estimator for the total treatment effect; that is, for any $t \in [T]_0$, we have
		\begin{align} 
			\label{eq:consistency-with-direct-effect}
			\lim_{N \rightarrow \infty}
			\ETTE{t}{\desired,0}
			\eqas
			\TTE{t}{\desired,0}.
		\end{align}
	\end{theorem}
	{Proof.}
		Following the same argument as in the proof of Theorem \ref{thm:consistency}, we can assume Algorithm \ref{alg:causal-mp-original} is applied to the limiting objects $\CoVAVO{\expd}{}{t}$, rather than their finite sample variants. Next,  utilizing the same outcome process, $\CoVoutcome{}{}{t}$, as in proof of Theorem \ref{thm:DM-and-HT-estimate-direct-effect} and expanding the state evolution equation \eqref{eq:state evolution_change_of_var} for the averages and the fact that treatment at time $t$ is independent of the one at time $t+1$, we obtain
		\begin{equation}
			\label{eq:consistency_proof_SE_CoV}
			\begin{aligned}
				\CoVAVO{\expd}{}{t+1}
				&\eqas
				\ACE + \APE [\CoVAVO{\expd}{}{t}+\DTEP\pi_1] + \ADE \pi_1 + \AME [\CoVAVO{\expd}{}{t}+\DTEP\pi_1] \pi_1 + \ACC^{\;\top} \Acovar,
				\quad\quad\quad
				&&t = 0,\ldots,T_1-1,
				\\
				\CoVAVO{\expd}{}{t+1}
				&\eqas
				\ACE + \APE [\CoVAVO{\expd}{}{t}+\DTEP\pi_2] + \ADE \pi_2 + \AME [\CoVAVO{\expd}{}{t}+\DTEP\pi_2] \pi_2 + \ACC^{\;\top} \Acovar,
				\quad\quad\quad
				&&t = T_1,\ldots,T_1+T_2-1.
			\end{aligned}
		\end{equation}
		Moreover,  by using the function $\psi$, equal to $\CoVoutcome{}{n}{t} +\DTEP \treatment{n}{t}$, in Theorem \ref{thm:Big theorem}(a), we get 
		\[
		\AVO{\expd}{}{t}:=\lim_{N \rightarrow \infty}\frac{1}{N}\sum_{n=1}^{N} \outcome{}{n}{t}
		\eqas
		\left\{
		\begin{array}{ll}
			\CoVAVO{\expd}{}{t}+\DTEP\pi_1&~~~~t = 0,\ldots,T_1\,,\\
			\CoVAVO{\expd}{}{t}+\DTEP\pi_2&~~~~t = T_1,\ldots,T_1+T_2\,.
		\end{array}
		\right.
		\]
		Combining this with \eqref{eq:consistency_proof_SE_CoV}, we obtain
		\begin{equation}
			\begin{aligned}
				\AVO{\expd}{}{t+1}
				&\eqas
				\ACE + \APE \AVO{\expd}{}{t} + (\ADE+\DTEP) \pi_1 + \AME \AVO{\expd}{}{t} \pi_1 + \ACC^{\;\top} \Acovar,
				\quad\quad\quad
				&&t = 0,\ldots,T_1-1,
				\\
				\AVO{\expd}{}{t+1}
				&\eqas
				\ACE + \APE \AVO{\expd}{}{t} +(\ADE +\DTEP)\pi_2 + \AME  \AVO{\expd}{}{t} \pi_2 + \ACC^{\;\top} \Acovar,
				\quad\quad\quad
				&&t = T_1,\ldots,T_1+T_2-1.
			\end{aligned}
		\end{equation}
		This means that, using an induction argument on $t$, the limits of sample averages for the experiment under specification \eqref{eq:outcome_function_matrix_with_direct}, i.e., $\AVO{\expd}{}{t}$, are equal to the corresponding values if the outcomes followed specification \eqref{eq:outcome_function_matrix}, but with the mean coefficient $\lambda$ in function $g$ replaced by $\lambda + \DTEP$. This statement holds for any values of $\pi_1$ and $\pi_2$, which means $\TTE{t}{\desired,0}$ would be equal in both scenarios. This also implies that the inputs to Algorithm \ref{alg:causal-mp-original} are the same under both scenarios. Therefore, we can invoke Theorem \ref{thm:consistency} in the latter scenario to show that the output of Algorithm \ref{alg:causal-mp-original}, under either scenario, is a strongly consistent estimator for $\TTE{t}{\desired,0}$. This completes the proof.
		\ep

	\begin{remark}\label{rem:dir-effect-equiv-combined-effect}
Proof of Theorem \ref{thm:consistency-with-direct-effect} reveals that the addition of the direct effect in the outcome specification, i.e., the outcome sequence $\Voutcome{}{}{t}$ with functions $\outcomeg{t}{}$, has the same impact on the state evolution for the average of outcomes $\AVO{\expd}{}{t}$ and $\TTE{t}{\desired,0}$ as increasing the mean coefficient $\lambda$ of the functions $\outcomeg{t}{}$ by the direct effect, i.e., the outcome sequence $\VCoVoutcome{}{}{t}$ with functions $\CoVoutcomeg{t}{}$. In other words, in outcome sequence $\VCoVoutcome{}{}{t}$,  the combined direct and indirect effects appear as an indirect effect, which does not affect the TTE.
	\end{remark}
	
	Now, we can revisit our initial aim of estimating the indirect effect. Specifically, we can first apply Algorithm \ref{alg:causal-mp-original} to obtain $\ETTE{t}{1,0}$, which is a strongly consistent estimator for $\TTE{t}{1,0}$. Then, we can obtain an estimate $\EDTEP$ for the direct effect $\DTEP$ using the difference-in-means or Horvitz-Thompson estimator, and use the following estimate for the indirect treatment effect (ITE).
	\begin{align}
		\label{eq:ITE-estimate}
		\EITE{t}{1,0} =
		\ETTE{t}{1,0} - \EDTEP ,.
	\end{align}
	Combining Theorems \ref{thm:DM-and-HT-estimate-direct-effect}–\ref{thm:consistency-with-direct-effect}, we obtain the following result.
	\begin{corollary}[Consistency of ITE estimate]\label{cor:indirect-estimate}
		Under the conditions of Theorems \ref{thm:DM-and-HT-estimate-direct-effect}–\ref{thm:consistency-with-direct-effect}, the estimate in \eqref{eq:ITE-estimate} is strongly consistent for the indirect effect, defined in \eqref{eq:ITE-definition}.
	\end{corollary}

	\subsection{Generalized randomized outcome specifications}\label{subsec:general-specification} 
	
	To justify the empirical success of Algorithm \ref{alg:causal-mp-original} as shown in \S \ref{sec:Numerical}, in settings beyond those covered by specification \eqref{eq:outcome_function_matrix}, we consider a more general version of \eqref{eq:outcome_function_matrix}, where the right-hand side is further transformed through a random function. We then provide a formal theorem for the applicability of Algorithm \ref{alg:causal-mp-original} for consistent estimation of the TTE under certain conditions.
	
	For all $t \ge 0$, consider random functions $\outcomegeneralnoise{t}{}: \R \to \R$ such that, for all $x \in \R$, $\E_{\outcomegeneralnoise{t}{}}[\outcomegeneralnoise{t}{x}] = x$, where, with a slight abuse of notation, $\E_{\outcomegeneralnoise{t}{}}$ refers to taking the expectation with respect to the randomness in $\outcomegeneralnoise{t}{}$, and $\E_{\outcomegeneralnoise{\le t}{}}$ refers to taking the expectation with respect to the randomness of $\outcomegeneralnoise{0}{}, \ldots, \outcomegeneralnoise{t}{}$. 
	
	Two special cases are:
	\begin{enumerate}
		\item \emph{Additive noise:} When $\outcomegeneralnoise{t}{x} = x + \noise{}{t}$ for a centered random variable $\noise{}{t}$, similar to the one used in \eqref{eq:outcome_function_matrix}.
		
		\item \emph{Bernoulli noise:} In cases where $x \in [0,1]$, $\outcomegeneralnoise{t}{x}$ is a Bernoulli random variable with mean $x$.
	\end{enumerate}
	
	We also follow the same convention as in prior sections that for any vector $\vec{V}$, the function $\outcomegeneralnoise{t}{}$ operates coordinatewise.
	
	Now, we introduce the following general outcome specification.
	\begin{align}
		\label{eq:outcome_function_matrix_general}
		\Voutcome{}{}{t+1} = \outcomegeneralnoise{t}{}
		\left(
		\big(\IM + \IMatT{t}\big)\outcomeg{t}{}\left(\Voutcome{}{}{t}, \Mtreatment{}{}, \covar\right)
		\right)
		\quad\quad\quad
		t = 0, 1, \ldots, T-1,
	\end{align}
	where the randomness of $\outcomegeneralnoise{t}{}$ is independent across all $t$ as well as independent of all other sources of randomness in \eqref{eq:outcome_function_matrix_general}.  	
	
	It is easy to see that \eqref{eq:outcome_function_matrix} is a special case of \eqref{eq:outcome_function_matrix_general} where the function $\outcomegeneralnoise{t}{}$ is additive noise that is independent across all $n$ as well. Similarly, the case where $\outcomegeneralnoise{t}{}$ is independent Bernoulli noise across all $n$, bears similarity to the setting studied in \S \ref{subsec:MRT}.
	
	\begin{theorem}\label{thm:consistency-with-gen-random-functiont}
		Under the conditions of Theorem \ref{thm:consistency}, as well as assuming that for all $t$, $	\lim_{N \rightarrow \infty}\sum_{n=1}^{N}\left(\outcome{}{n}{t} - \E_{\outcomegeneralnoise{\le t-1}{}}[\outcome{}{n}{t}]\right)/N\eqas0$.
		Then, for any desired $\desired\in[0,1]$, Algorithm \ref{alg:causal-mp-original} provides a strongly consistent estimator for $\TTE{t}{\desired,0}$ for all $t$, when applied to data generated from the generalized outcome specification \eqref{eq:outcome_function_matrix_general}.
	\end{theorem}
	
	{Proof.}
		Similar to the proofs in \S \ref{subsec:direct-effect}, we  define a new outcome process $\CoVoutcome{}{}{t}$ as follows:
		\[
		\VCoVoutcome{}{}{t} := \E_{\outcomegeneralnoise{\le t-1}{}}[\Voutcome{}{}{t}]\,.
		\]
		We obtain:
		\begin{align*}
			\VCoVoutcome{}{}{t+1} &= \E_{\outcomegeneralnoise{\le t-1}{}}\left[\E_{\outcomegeneralnoise{t}{}}\left[
			\big(\IM+\IMatT{t}\big)\outcomeg{t}{}\left(\Voutcome{}{}{t}, \Mtreatment{}{}, \covar\right)\right]\right] \\
			&= \E_{\outcomegeneralnoise{\le t-1}{}}\left[\big(\IM+\IMatT{t}\big)\outcomeg{t}{}\left(\Voutcome{}{}{t}, \Mtreatment{}{}, \covar\right)\right] \\
			&\stackrel{(a)}{=} \E_{\outcomegeneralnoise{\le t-1}{}}
			\left[
			\big(\IM+\IMatT{t}\big)\outcomeg{t}{}\left(\VCoVoutcome{}{}{t}, \Mtreatment{}{}, \covar\right)
			+
			\big(\IM+\IMatT{t}\big)\outcomegddx{t}\left(\VCoVoutcome{}{}{t}, \Mtreatment{}{}, \covar\right)(\Voutcome{}{}{t}-\VCoVoutcome{}{}{t})
			\right] \\    
			&\stackrel{(b)}{=}
			\big(\IM+\IMatT{t}\big)\outcomeg{t}{}\left(\VCoVoutcome{}{}{t}, \Mtreatment{}{}, \covar\right)\,,
		\end{align*}
		where $(a)$ uses the fact that family of functions $\outcomeg{t}{}$, studied in Theorem \ref{thm:consistency} and here, are linear with respect to their first coordinate, and $(b)$ uses $\E_{\outcomegeneralnoise{\le t-1}{}}[\Voutcome{}{}{t}-\VCoVoutcome{}{}{t}]=0$, that holds by definition, and the fact that all the terms $\IM+\IMatT{t}$, $\outcomeg{t}{}\left(\VCoVoutcome{}{}{t}, \Mtreatment{}{}, \covar\right)$, and $\outcomegddx{t}\left(\VCoVoutcome{}{}{t}, \Mtreatment{}{}, \covar\right)$ are independent of all randomness in $\outcomegeneralnoise{0}{}, \ldots, \outcomegeneralnoise{t-1}{}$.
		
		Therefore, the outcome process $\VCoVoutcome{}{}{t}$ satisfies specification \eqref{eq:outcome_function_matrix}, allowing us to invoke Theorem \ref{thm:Big theorem} and obtain state evolution equations. Following the same steps as in the proof of Theorem \ref{thm:DM-and-HT-estimate-direct-effect}, and using the assumption that yields $\lim_{N \rightarrow \infty}\sum_{n=1}^{N}\outcome{}{n}{t}/N \eqas \lim_{N \rightarrow \infty}\sum_{n=1}^{N}\CoVoutcome{}{n}{t}/N$, for all $t$, 
		one can show that the output of Algorithm \ref{alg:causal-mp-original}, when applied to data generated from a two-stage Bernoulli experiment on outcomes of specification \eqref{eq:outcome_function_matrix_general}, provides a strongly consistent estimator for $\TTE{t}{\desired,0}$ for all $t$. \ep

\subsection{Neural network interpretation}\label{subsec:NN-interpretation}

\begin{figure}[h!]
	\centering
        \includegraphics[width=0.6\linewidth]{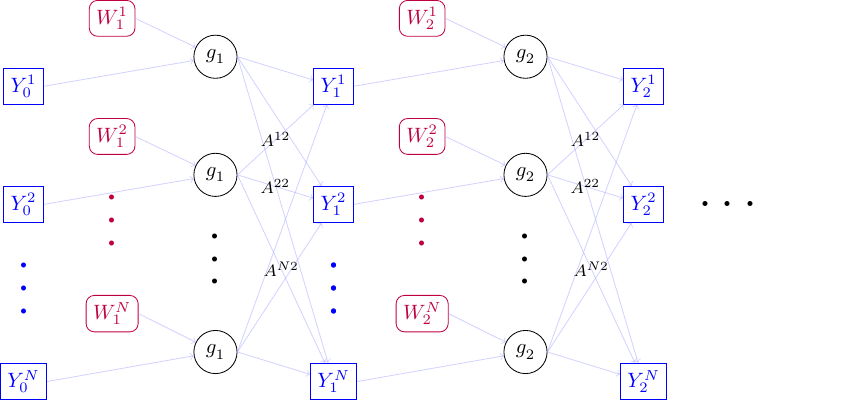}

	\caption{Illustration of the potential outcome specification~\eqref{eq:outcome_function_matrix} as a neural network model.}
	\label{fig:NNs}
\end{figure}

An alternative approach to interpret the specification~\eqref{eq:outcome_function_matrix} is motivated by the 
neural networks \citep{goodfellow2016deep} and may explain the empirical performance of Causal-MP
in \S \ref{sec:Numerical}, particularly its flexibility in capturing a wide range of outcome and interference patterns. Precisely, for a fixed $t>0$, the potential outcome of the $n^{th}$ unit at time~$t$, $\outcome{}{n}{t}$, is equal to the result of applying non-linear transformations to the outcomes of all units during previous time steps, weighted by linear operators. Figure~\ref{fig:NNs} illustrates this, in a simpler setting without covariates and static interference patterns. Mathematically, \eqref{eq:outcome_function_matrix} captures this interpretation and $\outcome{}{n}{t}$ is the output of a neural network with $t+1$ layers (consisting of 1 input layer, $t-1$ hidden layers, and 1 output layer), where the input data is $\Voutcome{}{}{0}$. In this context, the elements of the interference matrix correspond to the weights in the neural network.

The neural network depiction of \eqref{eq:outcome_function_matrix}, alongside the proven ability of neural networks to encapsulate complex and nonlinear interactions—especially as discussed in the context of randomly weighted neural networks \citep{rahimi2007random}—offers a promising avenue to study the Causal-MP framework. However, a distinct aspect of our approach, as informed by \eqref{eq:general convergence-approiximation}, is that as the number of units grows, our emphasis shifts to the estimation of the neural network's sufficient statistics regarding its weights, rather than the estimation of each individual weight.

\subsection{Distribution of the interference matrix} \label{subsec:gen-distr-int-matrix}

The numerical simulations in \S \ref{sec:Numerical} provide evidence that the predictions of Theorem~\ref{thm:consistency} regarding the consistency of the estimates from Algorithm \ref{alg:causal-mp-original} are applicable in settings that do not necessarily satisfy the assumptions in our theoretical analysis. Focusing on Assumption \ref{asmp:Gaussian Interference Matrice} regarding the distribution of the interference matrix, in this section, we discuss why such iid Gaussian  assumption can be relaxed to encompass a broader range of interference matrices.

As discussed in \S \ref{sec:Relev_Lit}, a large body network interference literature adopts the neighborhood interference assumption; 
outcome of a unit $i$ is obtained as a function of outcomes and treatments of all units in a neighborhood of $i$, where neighborhood is defined because of an explicit graph structure, e.g., \citep{eckles2016design,sussman2017elements,leung2020treatment, viviano2020experimental, agarwal2022network, belloni2022neighborhood, li2022random,cortez2022staggered, cortez2022exploiting, yu2022estimating,leung2022causal,jiang2023causal}. 
First, we show that these settings can be mapped to our specification \eqref{eq:outcome_function_matrix}. Taking the linear-in-means model from \S \ref{subsec-LiM-sythetic}, specifically \eqref{eq:linear-in-mean}, it can be writen as the following compact form
\begin{align}
	\label{eq:linear-in-mean-mat}
	\Voutcome{}{}{t+1}
	=
	\alpha_1
	+
	\IM{} (\alpha_2\Voutcome{}{}{t}+\alpha_3 \Vtreatment{}{t+1})+\alpha_4\Vtreatment{}{t+1}
	+
	\eps\,,
\end{align}
where $\IM{}$ is the row normalized adjacency matrix of the graph, representing the interference network, i.e., for units $i$ and $j$, $\IMatl{ij}:=\adjMe^{ij}/(\sum_{k=1}^N\adjMe^{ik})$. Following similar ideas as in \S \ref{subsec:direct-effect}, by a change of variable, $\VCoVoutcome{}{}{t}:=\Voutcome{}{}{t}-\alpha_1-\alpha_4\Vtreatment{}{t}$, we can consider the  specification
\begin{align}
	\label{eq:linear-in-mean-mat-cov}
	\VCoVoutcome{}{}{t+1}
	=
	\IM \left(\alpha_2[\VCoVoutcome{}{}{t}+\alpha_1+\alpha_4\Vtreatment{}{t}]+\alpha_3 \Vtreatment{}{t+1}\right)
	+
	\eps\,.
\end{align}
Further, by defining $\CoVoutcomeg{t}{}(y,w):=\alpha_2(y+\alpha_1+\alpha_4 w)+\alpha_3 w$, specification \eqref{eq:linear-in-mean-mat-cov} becomes a special case of \eqref{eq:outcome_function_matrix}, except that the interference matrix $\IM$ does not satisfy Assumption \ref{asmp:Gaussian Interference Matrice}. 
However, when the interference network is a random graph or has a stochastic block structure (e.g., \citep{eckles2016design, li2022random, li2022network}), $\IM$ becomes a random matrix. The literature on approximate message passing (AMP), in alignment with random matrix theory, provides evidence that one can adapt our analysis to obtain variants of Theorems \ref{thm:Big theorem}-\ref{thm:consistency} in settings when the entries of $\IM$ are neither necessarily i.i.d. nor Gaussian.

Specifically, there is a substantial body of literature demonstrating that such random matrices exhibit a certain \emph{universality property} in their asymptotic behavior. For instance, the literature on random matrix theory indicates that many asymptotic properties of the joint distribution of eigenvalues of these matrices are insensitive to the distribution of the matrix entries, under certain tail conditions, provided that the first two moments of the entries' distribution are preserved  \citep{bai2005spectral,anderson2009introduction, tao2012random}. Similarly, motivated by compressed sensing applications, the AMP literature highlights other asymptotic properties, such as state evolution of variants of \eqref{eq:outcome_function_matrix}, also exhibits a similar universality \citep{bayati2015universality, montanari2021estimation, berthier2020state, chen2020universality}. In their context, the entries of $\IM$ only need to be independent and are not required to follow the same distribution, and even allowing for a block structure in their moments. Further generalizations, such as when the interference matrix belongs to the family of rotationally invariant matrices \cite{xinyi2021approximate}, or other settings, have been studied by \cite{dudeja2023universality} and \cite{wang2022universality}.

Therefore, we expect generalizations of the state evolution and Theorems \ref{thm:Big theorem}-\ref{thm:consistency},  to a broader class of interference matrices beyond those considered in Assumption \ref{asmp:Gaussian Interference Matrice} to be possible. It is also important to note that for cases where there is structure among the entries of $\IM$, such as when the random graph potentially exhibits a clustered structure similar to the stochastic block model examined by \cite{eckles2016design} and used in \S \ref{sec:Numerical}, state evolution is expected to require modifications \citep{krzakala2012statistical, donoho2013information, javanmard2013state, bayati2015universality}. Exploring the implications of these modifications for estimating causal effects is an intriguing area for future research.


\section{Additional Numerical Results}
\label{sec:apndx_numericals}
In this section, we provide more details on the experimental settings studied in \S \ref{sec:Numerical}, followed by additional numerical results that support the applicability and relevance of Algorithm~\ref{alg:causal-mp-original} for estimating more general versions of the total treatment effect, defined in \eqref{eq:TTE_def_fixed}.

\subsection{More details on linear-in-Means model with stochastic block network}
To generate the stochastic block model with $N =$ 10,000, we consider the setting that units within each cluster are connected with a probability of $0.005$, while units from different clusters have a probability of $0.0001$ of being connected. These numbers for $N =$ 1,000 are $0.05$ and $0.001$, and for $N = 500$ are $0.1$ and $0.002$, respectively. Furthermore, we run Algorithm~\ref{alg:causal-mp-original} in two stages with $T_1 = T_2 = 30$ with $(\pi_1,\pi_2) = (0.2,0.5)$.

\subsection{Binary outcome model with a burn-in period}
\label{sec:BOM_WBI}
\begin{figure}
    \centering
    \includegraphics[width=0.9\linewidth]{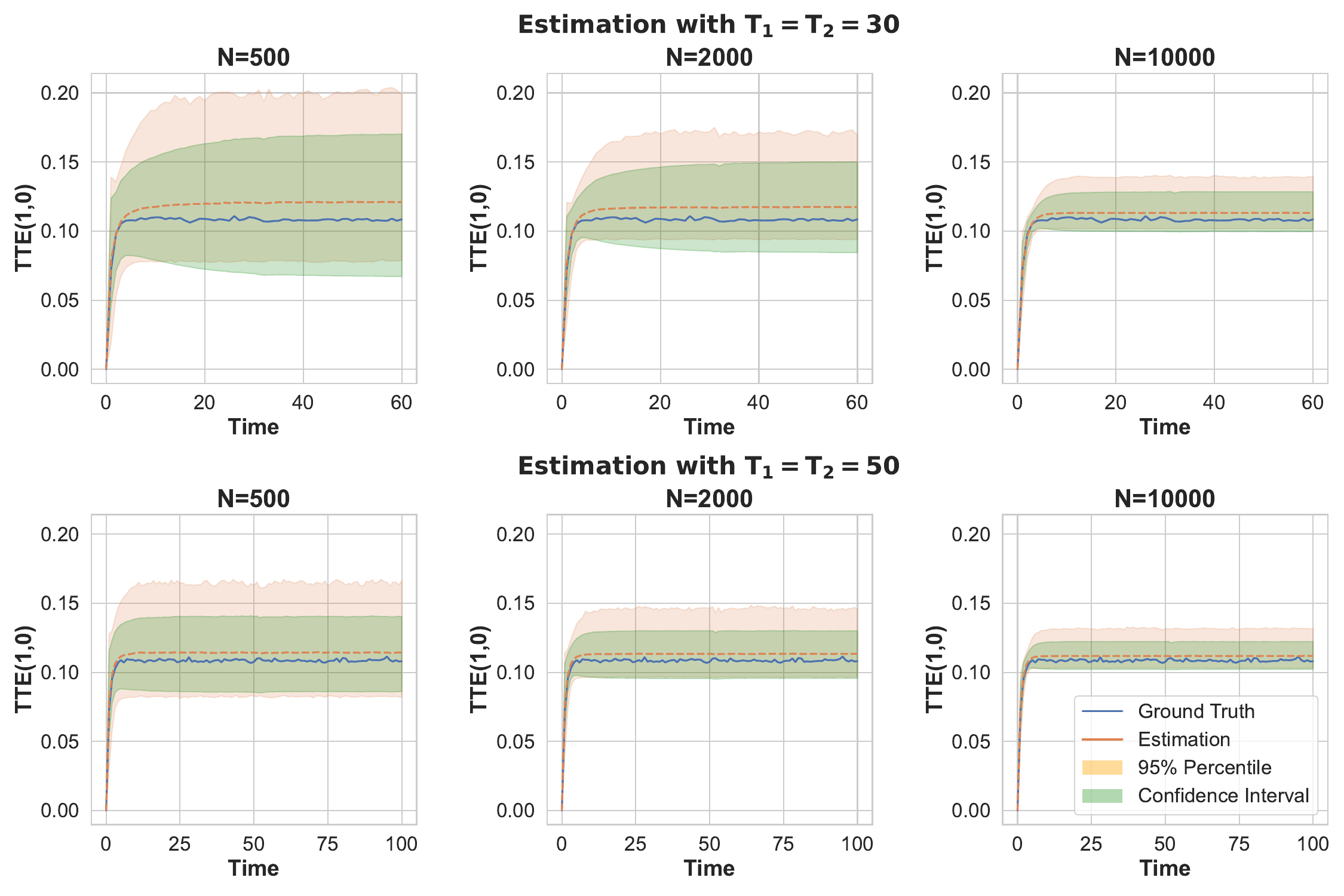}
    \caption{Binary outcome model with Erd\"os-R\'enyi graph; 95\% confidence interval for the total treatment effect estimation when $(\pi_1,\pi_2)=(0.25,0.75)$.}
    \label{fig:MRT_pi}
\end{figure}

Here, we consider another setup of the binary outcome model, in \S \ref{subsec:MRT}, with a burn-in period of length 10. Specifically, under Assumptions 3-5 in \cite{li2022network}, which ensure the uniqueness of the stationary distribution of the underlying system, we set the parameter values as $(\alpha_1,\alpha_2,\alpha_3,\alpha_4,p_{\text{edge}}) = (0.5,0.04,0.04,0.01,3/N)$. Figure~\ref{fig:MRT_pi} presents the results of estimating the TTE. For the confidence interval estimation, we set $B=500$ and $q=0.7$ for all $N$. The results demonstrate a high level of accuracy in estimating the TTE, showcasing the robustness of the proposed framework. 

\subsection{TTE estimation at other treatment levels}
\begin{figure}
    \centering
    \includegraphics[width=0.85\linewidth]{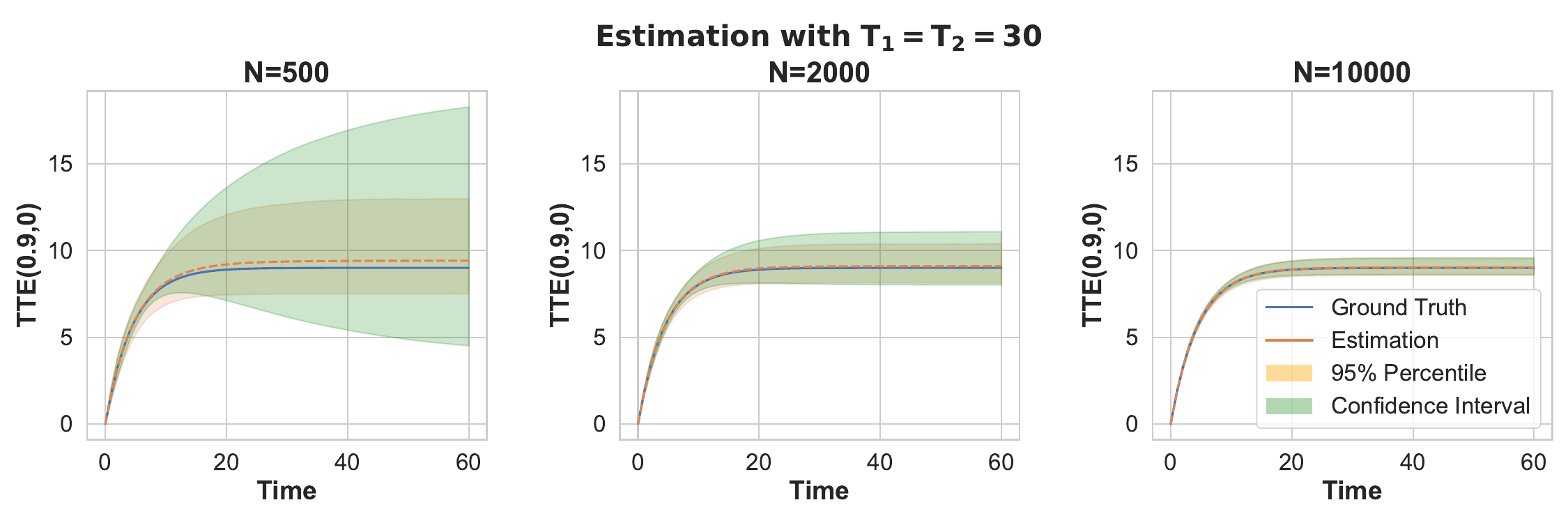}
    \caption{Linear-in-means model with random geometric graph;   
    95\% confidence intervals for $\TTE{t}{0.9,0}$ estimation when $(\pi_1,\pi_2)=(0.2,0.5)$.}
    \label{fig:LinM_pi09}

    \centering
    \includegraphics[width=0.85\linewidth]{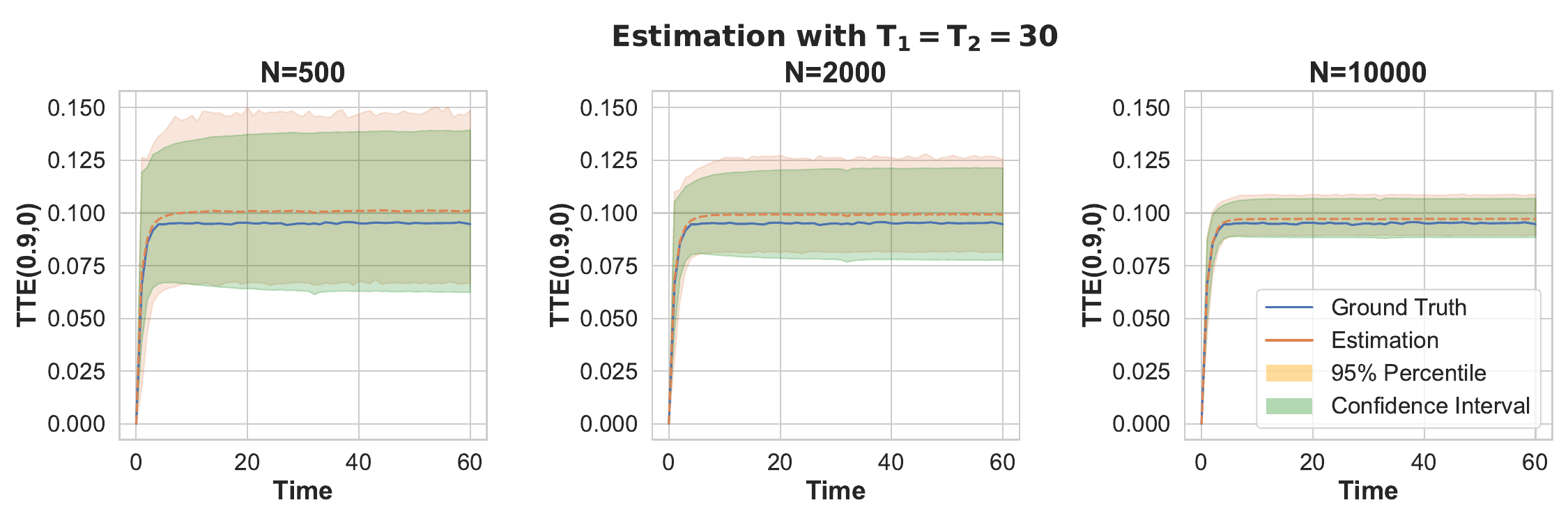}
    \caption{Binary outcome model with micro-randomized trial: 95\% confidence interval for $\TTE{t}{0.9,0}$ estimation when $(\pi_1,\pi_2)=(0.25,0.75)$.}
    \label{fig:MRT_pi09}
\end{figure}

Here, we present the results of estimating $\TTE{t}{0.9,0}$ in two other settings: the linear-in-means model with a random geometric graph (Figure~\ref{fig:LinM_pi09}) and the binary outcome model (Figure~\ref{fig:MRT_pi09}), both highlighting the robustness of the proposed method.

\subsection{TTE estimation with seasonal time trend}
\label{sec:server_with_timetrend}
\begin{figure}
    \centering
    \includegraphics[width=0.85\linewidth]{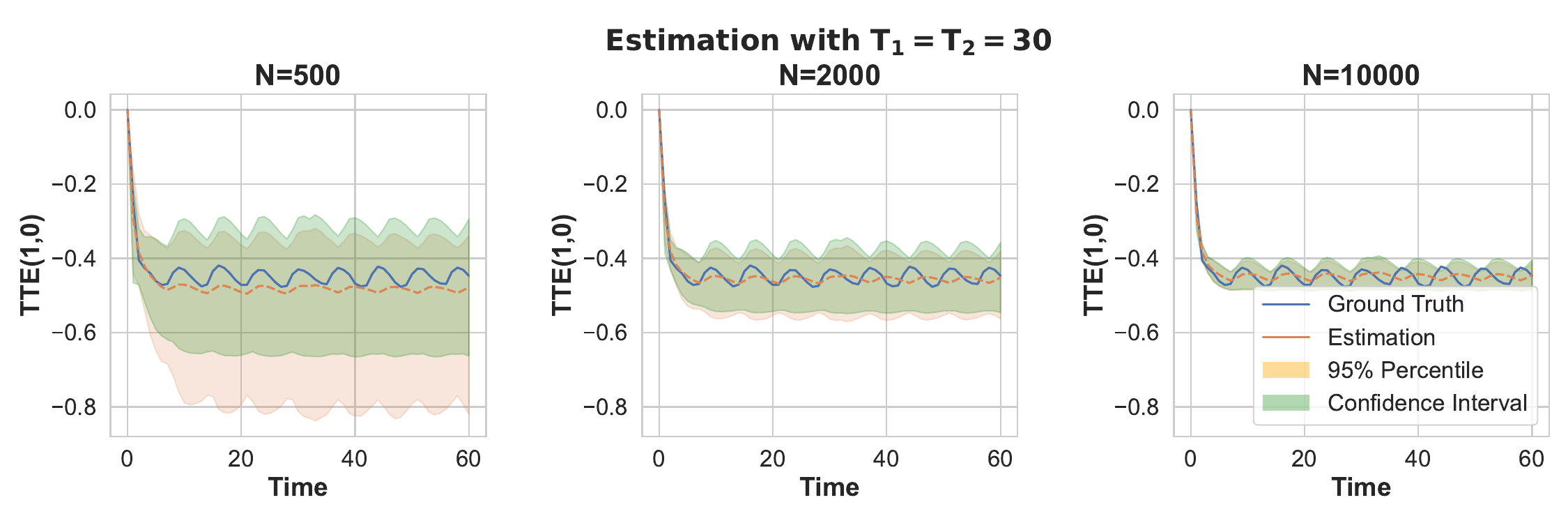}
    \caption{Server speed-up problem with a seasonal time trend: 95\% confidence interval for the TTE estimation when $(\pi_1,\pi_2)=(0.15,0.5)$.}
    \label{fig:Servers_time_dep}
\end{figure}

In the final setting, we consider the parallel server system and incorporate a weekly time trend, along with independent noise across time steps, in the demand rate for the servers. This means that the arrival rate to the queueing system varies randomly over time while following a specific trend. Figure~\ref{fig:Servers_time_dep} illustrates the results, showcasing the robustness of the proposed framework.

The inclusion of a time-varying demand rate adds complexity to the model, simulating more realistic scenarios where demand fluctuates. Despite these variations, the framework consistently provides accurate and reliable estimates of the treatment effect. Specifically, Algorithm~\ref{alg:causal-mp-original} yields a reliable estimate of the TTE even without accounting for heterogeneous variations over time. This further demonstrates the versatility and strength of our approach in handling dynamic and noisy environments, confirming its applicability to real-world situations where conditions are far from static.

\section{Detailed Explanation of Treatment Effects}
\label{Details_of_example}

\paragraph{Direct Effect.} The direct effect represents the immediate influence of the medication on the severity of symptoms for unit 1, following the first period of administration. This effect is specific to the treated units and stems directly from the treatment \citep{athey2018exact,forastiere2022estimating}.

\paragraph{Treatment Spillover Effect.}
Consider a hospital environment where the treatment for unit 1 involves the use of a highly effective air purifier designed to remove airborne pathogens. While the immediate goal is to alleviate unit 1's respiratory symptoms, the purifier also cleans the air in the shared space, indirectly benefiting unit 2. This is an example of how the treatment of one unit can influence the outcomes of another \citep{athey2018exact,forastiere2022estimating}.

\paragraph{Carryover Effect.}
A carryover effect implies that the effects of the medication continue to influence the health status of the units in subsequent periods, beyond the immediate treatment \citep{xiong2019optimal}.

\paragraph{Unit Peer Effect.}
The unit peer effect is especially significant in contagious diseases, where the severity of symptoms in one unit directly impacts the severity experienced by other units \citep{yu2022estimating, imai2019identification}.

\paragraph{Autocorrelation.}
Autocorrelation refers to the temporal interdependence of outcomes within the same unit. For example, severe symptoms today increase the likelihood of continued illness tomorrow.

\paragraph{Anticipation Effect.}
The anticipation effect describes how units might adjust their current behavior based on expected future treatments or outcomes, adding a layer of complexity to the network.

\section{Network Interference as a Massage Passing Model}
\label{NI_as_MP_HD}

Here we expand on the message-passing interpretation of the specification \eqref{eq:outcome_function_matrix} from \S \ref{sec:PF}. Specifically, we note that in the literature on message-passing algorithms, the message from unit $i$ to $n$ differs slightly from the description in \S \ref{sec:PF}. Specifically, the message from each unit $i$ to $n$ is as described, but it omits the message received from $n$ in the prior period. If we represent the message from unit $i$ to $n$ at time $t+1$ by $\FMessage{i}{n}{t+1}{}$, then
$\FMessage{i}{n}{t+1}{}=
\outcomeg{t}{}
(
\sum_{j\in[N]\setminus n}
\IMatGl{ij}{t}
\FMessage{j}{i}{t}{}
,
\Vtreatment{j}{}
,
\Vcovar{j}
)$.
However, in the large $N$ regime, this is approximately equivalent to the simpler specification 
\begin{align}
\label{eq:NI_as_MP_message_simple}
\outcomeD{}{n}{t+1}=\sum_{i\in [N]} \IMatGl{ni}{t} \underbrace{\outcomeg{t}{}(\outcomeD{}{i}{t},\Vtreatment{i}{},\Vcovar{i})}_{\textbf{Message of $i$ to $n$}}\,.
\end{align}
This approximation underpins the AMP algorithm. For completeness, we provide this heuristic argument, that has also appeared in AMP literature, below.  Additionally, in the AMP literature, this exercise often leads to a dynamic as in  \eqref{eq:outcome_function_matrix} that also incorporates a memory term involving $\outcomeg{t-1}{}(\Voutcome{}{}{t-1} ,\Mtreatment{}{}, \covar)$, also known as the Onsager term. However, in our context, this term disappears since the matrix $\IMatG{t}$ is not symmetric.

Next, we present a heuristic derivation of the potential outcome model. We proceed by writing the message-passing model as follows.
\begin{equation}
    \label{eq:NI_as_MP_dynamics}
    \FMessage{n}{m}{0}{}
    =
    \outcomeg{0}{}
    \left(
    \outcome{}{n}{0}
    ,
    \Vtreatment{n}{}
    ,
    \Vcovar{n}
    \right),
    \quad
    \FMessage{n}{m}{t+1}{}
    =
    \outcomeg{t}{}
    \bigg(
    \sum_{i\in[N]\setminus m}
    \IMG_{ni}
    \FMessage{i}{n}{t}{}
    ,
    \Vtreatment{n}{}
    ,
    \Vcovar{n}
    \bigg).
\end{equation}
Note that on the right-hand side of the second equation in \eqref{eq:NI_as_MP_dynamics}, we encounter a summation of order N terms. Furthermore, its dependence on $m$ is solely through the exclusion of the term $\FMessage{m}{n}{t}{}$. As a result, we can follow a similar line of reasoning as presented in Appendix A of \cite{bayati2011dynamics}, for $n,m \in[N]$, we write 
\begin{align}
    \label{eq:NI_as_MP_second_order_term}
    \FMessage{n}{m}{t}{}
    =
    \FMessage{n}{}{t}{}
    +
    \delta
    \FMessage{n}{m}{t}{},
\end{align}
where $\delta \FMessage{n}{m}{t}{} = O(N^{-1/2})$ and $\FMessage{n}{}{t}{}$ is a term independent of $m$. Considering \eqref{eq:NI_as_MP_dynamics} and \eqref{eq:NI_as_MP_second_order_term} together, we get
\begin{equation*}
    \begin{aligned}
        \FMessage{n}{}{t+1}{}
        +
        \delta
        \FMessage{n}{m}{t+1}{}
        =
        \outcomeg{t}{}
        \bigg(
        \sum_{i\in[N]}
        \IMG_{ni} 
        \big(\FMessage{i}{}{t}{}
        +
        \delta
        \FMessage{i}{n}{t}{}\big)
        -
        \IMG_{nm} 
        \big(\FMessage{m}{}{t}{}
        +
        \delta
        \FMessage{m}{n}{t}{}\big)
        ,
        \Vtreatment{n}{}
        ,
        \Vcovar{n}
        \bigg).
    \end{aligned}
\end{equation*}
Assuming that the function $\outcomeg{t}{}$ is continuous and differentiable in the first argument as well as $\IMG_{nm} \sim \Nc\big((\mu+\mu_t)/N,(\sigma+\sigma_t)/N\big)$, for large values of $N$, we can write
\begin{equation*}
    \begin{aligned}
        \FMessage{n}{}{t+1}{}
        +
        \delta
        \FMessage{n}{m}{t+1}{}
        \approx
        \outcomeg{t}{}
        \bigg(
        \sum_{i\in[N]}
        \IMG_{ni} 
        \big(\FMessage{i}{}{t}{}
        +
        \delta
        \FMessage{i}{n}{t}{}\big)
        -
        \IMG_{nm} 
        \FMessage{m}{}{t}{}
        ,
        \Vtreatment{n}{}
        ,
        \Vcovar{n}
        \bigg).
    \end{aligned}
\end{equation*}
Then, writing the first-order approximation, we have
\begin{equation}
    \label{eq:NI_as_MP_first_order_approx}
    \begin{aligned}
        \FMessage{n}{}{t+1}{}
        +
        \delta
        \FMessage{n}{m}{t+1}{}
        \approx
        &\;\outcomeg{t}{}
        \bigg(
        \sum_{i\in[N]}
        \IMG_{ni} 
        \big(\FMessage{i}{}{t}{}
        +
        \delta
        \FMessage{i}{n}{t}{}\big)
        ,
        \Vtreatment{n}{}
        ,
        \Vcovar{n}
        \bigg)
        \\
        &\;-
        \outcomeg{t}{'}
        \bigg(
        \sum_{i\in[N]}
        \IMG_{ni} 
        \big(\FMessage{i}{}{t}{}
        +
        \delta
        \FMessage{i}{n}{t}{}\big)
        ,
        \Vtreatment{n}{}
        ,
        \Vcovar{n}
        \bigg)
        \IMG_{nm} 
        \FMessage{m}{}{t}{},
    \end{aligned}
\end{equation}
where $\outcomeg{t}{'}$ denotes derivative with respect to the first argument.
Note that the last term is the only term involved with $m$ on the right-hand side; thus, we can argue that
\begin{align}
    \label{eq:NI_as_MP_second_order_term_approx}
    \delta \FMessage{n}{m}{t+1}{}
    \approx
    -
    \outcomeg{t}{'}
    \bigg(
    \sum_{i\in[N]}
    \IMG_{ni} 
    \big(\FMessage{i}{}{t}{}
    +
    \delta
    \FMessage{i}{n}{t}{}\big)
    ,
        \Vtreatment{n}{}
        ,
        \Vcovar{n}
    \bigg)
    \IMG_{nm} 
    \FMessage{m}{}{t}{}.
\end{align}
Substituting the result of \eqref{eq:NI_as_MP_second_order_term_approx} in \eqref{eq:NI_as_MP_first_order_approx} implies that
\begin{align*}
        \FMessage{n}{}{t+1}{}
        +
        \delta
        \FMessage{n}{m}{t+1}{}
        &\approx
        \outcomeg{t}{}
        \bigg(
        \sum_{i\in[N]}
        \IMG_{ni} 
        \FMessage{i}{}{t}{}
        -
        \sum_{i\in[N]}
        \outcomeg{t}{'}
        \Big(
        \sum_{j\in[N]}
        \IMG_{nj} 
        \big(\FMessage{j}{}{t}{}
        +
        \delta
        \FMessage{j}{i}{t}{}\big)
        \Big)
        \IMG_{in}
        \IMG_{ni} 
        \FMessage{n}{}{t}{}
        ,
        \Vtreatment{n}{}
        ,
        \Vcovar{n}
        \bigg)\\
        &~~~~~~~~~-
        \outcomeg{t}{'}
        \bigg(
        \sum_{i\in[N]}
        \IMG_{ni} 
        \big(\FMessage{i}{}{t}{}
        +
        \delta
        \FMessage{i}{n}{t}{}\big)
        ,
        \Vtreatment{n}{}
        ,
        \Vcovar{n}
        \bigg)
        \IMG_{nm} 
        \FMessage{m}{}{t}{}.
\end{align*}
Note that $\IMG_{in}$ and $\IMG_{ni}$ are two independent Gaussian random variables. Therefore, taking limit as $N \rightarrow \infty$, we get
\begin{align}
    \label{eq:NI_as_MP}
    \FMessage{n}{}{t+1}{}
    =
    \outcomeg{t}{}
    \bigg(
    \sum_{i\in[N]}
    \IMG_{ni} 
    \FMessage{i}{}{t}{}, \Vtreatment{n}{}, \Vcovar{n}
    \bigg).
\end{align}
Letting $\outcome{}{n}{t+1} = \sum_{i\in[N]} \IMG_{ni}  \FMessage{i}{}{t}{}$, we obtain the desired result.

\section{Total treatment effect estimation at the equilibrium}
\label{subsec:tte-at-equilibrium}

Here, we use the state evolution equations \eqref{eq:state evolution} to establish an estimator for the TTE at equilibrium. This stands in contrast to Algorithm~\ref{alg:causal-mp-original}, which is tailored to estimate the total treatment effect over the entire time horizon. This equilibrium estimand, denoted as $\TTE{}{1,0}$, characterizes the TTE as the time horizon extends towards infinity, formally expressed as $\lim_{T \rightarrow \infty} \TTE{T}{1,0}$. Subsequently, with access to two sets of observations of individuals' outcomes at the equilibrium, denoted as $\VoutcomeD{\pi_1}{}{}$ and $\VoutcomeD{\pi_2}{}{}$, we define the following estimator:
\begin{equation}
    \label{eq:Estimator at EQ}
    \ETTE{}{1,0}
    =
    \frac{1}{\pi_2-\pi_1} \sum_{n=1}^N 
    \frac{
    \outcomeD{\pi_2}{n}{}
    -
    \outcomeD{\pi_1}{n}{}
    }{N}.
\end{equation}
The estimator presented in \eqref{eq:Estimator at EQ} is inspired by the work of \cite{yu2022estimating} and extends a version analyzed by them. They demonstrated that with access to average baseline outcomes before an experiment, a special case of this estimator is unbiased under the neighborhood interference assumption and a linear outcome model. Our adaptation, which utilizes the function family $g$ from \eqref{eq:function_structure}, enables the examination of non-linear outcomes. We find that the estimator \eqref{eq:Estimator at EQ} is subject to bias, with the degree of bias linked to the expected values of $\E[\ME^n]$ and $\E[\PE^n]$. We formalize the statement below.

Note that we cannot let $T \rightarrow \infty$ as the results of Theorem~\ref{thm:Big theorem} hold true for finite values of $T$. Therefore, we formally assume that for some sufficiently large value of $T$, the quantity $\TTE{T}{1,0}$ stabilizes, and we denote its value as $\TTE{}{1,0}$. For any experimental design $\expd$, we also denote $\HAVO{\expd}{}{} := \frac{1}{N}\sum_{n=1}^N \outcomeD{}{n}{T}$; that is the sample mean at the equilibrium state under design $\expd$. We then have the following result.
\begin{theorem}
    \label{thm:Estimator at EQ}
    Let $\ETTE{}{1,0}$ be the estimator defined by \eqref{eq:Estimator at EQ}. Then, for any two values of $\pi_2 \neq \pi_1$, we have:
    \begin{align}
        \label{eq:consistency at EQ}
        \lim_{N \rightarrow \infty}
        \ETTE{}{1,0}
        \eqas
        \TTE{}{1,0} 
        +
        \frac{\AME}{1-\APE}
        \lim_{N \rightarrow \infty}
        \left(
        \frac{\pi_2\HAVO{\pi_2}{}{}-\pi_1\HAVO{\pi_1}{}{}}{\pi_2-\pi_1}-\HAVO{1}{}{}\right),
    \end{align}
    where $\APE = \E[\PE^n]$ and $\AME = \E[\ME^n]$.
\end{theorem}
The following corollary can be readily obtained as a special case of Theorem~\ref{thm:Estimator at EQ}.
\begin{corollary}
    \label{crl:consistency at EQ}
    If $\AME = 0$, then, $\ETTE{}{1,0}$ is a strongly consistent estimator for the total treatment effect at the equilibrium $\TTE{}{1,0}$.
\end{corollary}
Given access to historical data before the experiment, and by setting $\pi_1=0$ in Theorem~\ref{thm:Estimator at EQ}, we can simplify the bias term to $\frac{\AME}{1-\APE} \left(\HAVO{\pi_2}{}{}-\HAVO{1}{}{}\right)$.
Consequently, if we can establish bounds on the values of $\AME$, $\APE$, and $\HAVO{1}{}{}$ based on system characteristics, we can place bounds on the bias of the estimation using the estimator in \eqref{eq:Estimator at EQ}.

\section{Detailed Proofs of Technical Results}
\label{sec:detailed_proofs}
Below, we first introduce the necessary notations required for the proofs. Then, we focus on proving Theorem~\ref{thm:Big theorem} in multiple steps. We begin by stating Theorem~\ref{lm:Big lemma} as a generalization of Theorem~\ref{thm:Big theorem}. Subsequently, we describe the conditioning technique and present related results for adopting this technique for our specific purpose. This includes deriving the conditional distribution of the fixed interference matrix given observations of outcomes and treatments up to a certain point. 

Next, we utilize an induction argument to prove the main results in two major steps. Following that, we provide a detailed proof for the consistency statement of Algorithm~\ref{alg:causal-mp-original} in Theorem~\ref{thm:consistency}. Finally, we conclude this section by presenting two versions of the law of larger numbers that are frequently used in the proofs.

\subsection{Notations and Preliminaries}
For any set $S$, the indicator function $\1_S(\omega)$ evaluates to $1$ if $\omega$ belongs to $S$, and $0$ otherwise. We define $\R^{n\times m}$ as the set of matrices with $n$ rows and $m$ columns. Given a matrix $\bm{M}$, we denote its transpose as $\bm{M}^\top$, its Frobenius norm as $\fnorm{\bm{M}}$, and its trace as $\Tr(\bm{M})$. Additionally, we represent a matrix of ones with dimensions $n\times m$ as $\ones{n\times m} \in \R^{n\times m}$. The symbol $\eqd$ is used to denote equality in distribution, while $\eqas$ is used for equalities that hold almost surely. For $t\geq 1$, we define
\begin{equation}
    \label{eq:state evolution_fixed part}
    \begin{aligned}
        \BAVO{}{}{t+1} :=
        \mu \E\left[
        \outcomeg{t}{}\big(\AVO{}{}{t} + \VVO{}{}{t} Z, \Vtreatment{}{},\Vcovar{}\big)
        \right],
        \quad\quad
        \BVVO{}{2}{t+1} :=
        \sigma^2 \E\left[
        \outcomeg{t}{}\big(\AVO{}{}{t} + \VVO{}{}{t} Z, \Vtreatment{}{},\Vcovar{}\big)^2
        \right],
    \end{aligned}
\end{equation}
where $Z \sim \Nc(0,1)$ independent from $(\Vtreatment{}{},\Vcovar{}) \sim \Pi \x p_{\covar}$ and $\AVO{}{}{t}$ as well as $\VVO{}{}{t}$ are defined in \eqref{eq:state evolution}. Further, letting $\VUoutcome{}{}{t} = \outcomeg{t}{}\big(\Voutcome{}{}{t},\Mtreatment{}{},\covar\big)$ as well as $\VUoutcome{}{}{tn} = \outcomeg{t}{}\big(\outcome{}{n}{t},\Vtreatment{n}{},\Vcovar{n}\big)$, we denote
\begin{equation}
\label{eq:Q and R}
\begin{aligned}
    \bm{Q}_t
    :=
    \left[
    \VUoutcome{}{}{0}
    \Big|
    \VUoutcome{}{}{1}
    \Big|
    \ldots
    \Big|
    \VUoutcome{}{}{t-1}
    \right],
    \quad\quad
    \bm{R}_t
    :=
    \left[
    \Voutcome{}{}{1}
    -
    \IMatT{0} \VUoutcome{}{}{0}
    -
    \Vnoise{}{0}
    \Big|
    \ldots
    \Big|
    \Voutcome{}{}{t}
    -
    \IMatT{t-1} \VUoutcome{}{}{t-1}
    -
    \Vnoise{}{t-1}
    \right].
\end{aligned}
\end{equation}
That is, $\bm{Q}_t$ and $\bm{R}_t$ are matrices with columns of $\VUoutcome{}{}{s-1}$ and $\Voutcome{}{}{s} - \IMatT{s-1} \VUoutcome{}{}{s-1} - \Vnoise{}{s-1}$, when $s=1,\ldots,t$, respectively. Then, we use $\VUoutcome{\parallel}{}{t}$ to denote the projection of $\VUoutcome{}{}{t}$ onto the column space of $\bm{Q}_t$ and accordingly define $\VUoutcome{\perp}{}{t} = \VUoutcome{}{}{t} - \VUoutcome{\parallel}{}{t}$. Further, let $\Vec{\alpha}_t = (\alpha_0,\alpha_1,\ldots,\alpha_{t-1})^\top$ be such that
\begin{align}
    \label{eq:projection sum}
    \VUoutcome{\parallel}{}{t}
    =
    \sum_{s=0}^{t-1} \alpha_s \VUoutcome{}{}{s}
    =
    \sum_{s=0}^{t-1} \alpha_s
    \outcomeg{s}{}
    \big(
    \Voutcome{}{}{s}, \Mtreatment{}{}, \covar
    \big).
\end{align}
For vectors $\Vec{u},\Vec{v} \in \R^m$, we define the scalar product $\pdot{\Vec{u}}{\Vec{v}}:= \frac{1}{m} \sum_{i=1}^m u_i v_i$. Then, in \eqref{eq:projection sum}, we have
\begin{align}
    \label{eq:projection coefficients}
    \Vec{\alpha}_t
    =
    \left(
    \bm{Q}_t^\top \bm{Q}_t
    \right)^{-1}
    \bm{Q}_t^\top \VUoutcome{}{}{t}.
\end{align}

\subsection{General Result: Proof of Theorem~\ref{thm:Big theorem}}
Here, we state Theorem~\ref{lm:Big lemma} which is an expanded version of Theorem~\ref{thm:Big theorem}.

\begin{theorem}
    \label{lm:Big lemma}
    Fixing $k\geq 2$, assume the sequence of initial outcomes $\Voutcome{N}{}{0}$, the treatment assignments~$\Mtreatment{}{N}$, as well as the covariates $\covar(N)$ are given and suppose Assumption~\ref{asmp:BL} holds. Then, we have the following statements for all $t \geq 0$.
    \begin{enumerate}[label=(\alph*)]
        \item \label{part:BL-a} For any function $\psi: \R^{t+1+T+M} \mapsto \R$ that $\psi \in \poly{k}$, we have
        \begin{equation}
            \begin{aligned}
                \label{eq:BL-average limit}
                &\;\lim_{N \rightarrow \infty}
                \frac{1}{N} \sum_{n=1}^N
                \psi\big(
                \outcome{}{n}{1}
                , \ldots,
                \outcome{}{n}{t+1}
                ,
                \Vtreatment{n}{},\Vcovar{n}
                \big)
                \eqas
                \E
                \Big[
                \psi\big(
                \nu_1 + \rho_1 Z_1,
                \ldots,
                \AVO{}{}{t+1} + \rho_{t+1} Z_{t+1},
                \Vtreatment{}{},\Vcovar{}
                \big)
                \Big],
            \end{aligned}
        \end{equation}
        where $Z_s \sim \Nc(0,1),\; s= 1,\ldots,t+1,$ independent of $(\Vtreatment{}{},\Vcovar{}) \sim \Pi \x p_{\covar}$.

        \item \label{part:BL-b} For all $0 \leq r\neq s \leq t$, the following equations hold and all limits exist, are bounded, and have degenerate distribution (i.e. they are constant random variables)
        \begin{subequations}
            \label{eq:BL-b}
            \begin{align}
                \label{eq:BL-b-1}
                &\lim_{N \rightarrow \infty}
                \frac{1}{N}
                \sum_{n=1}^N
                \outcome{}{n}{s+1}
                \eqas
                \lim_{N \rightarrow \infty}
                \frac{\mu+\mu_s}{N}
                \sum_{n=1}^N
                \Uoutcome{n}{}{s}
                \eqas
                \AVO{}{}{s+1},
                \\
                \label{eq:BL-b-2}
                &\lim_{N \rightarrow \infty}
                \frac{1}{N}
                \sum_{n=1}^N
                \big(
                \outcome{}{n}{s+1}
                - \IMatTv{n\cdot}{s} \VUoutcome{}{}{s}
                - \noise{n}{s} \big)
                \eqas
                \lim_{N \rightarrow \infty}
                \frac{\mu}{N}
                \sum_{n=1}^N
                \Uoutcome{n}{}{s}\eqas
                \BAVO{}{}{s+1},
                \\
                \label{eq:BL-b-3}
                &\lim_{N \rightarrow \infty}
                \frac{1}{N}
                \sum_{n=1}^N
                (\outcome{}{n}{s+1})^2
                \eqas
                \AVO{}{2}{s+1}
                +
                \lim_{N \rightarrow \infty}
                \frac{\sigma^2+\sigma_s^2}{N}
                \sum_{n=1}^N
                (\Uoutcome{n}{}{s})^2
                + \sigma_e^2
                \eqas
                \AVO{}{2}{s+1}
                +
                \VVO{}{2}{s+1},
                \\
                \label{eq:BL-b-3-rs}
                &\lim_{N \rightarrow \infty}
                \frac{1}{N}
                \sum_{n=1}^N
                \outcome{}{n}{s+1}
                \outcome{}{n}{r+1}
                \eqas
                \AVO{}{}{s+1} \AVO{}{}{r+1}
                +
                \lim_{N \rightarrow \infty}
                \frac{\sigma^2}{N}
                \sum_{n=1}^N
                \Uoutcome{n}{}{s} \Uoutcome{n}{}{r},
                \\
                \label{eq:BL-b-4}
                &\lim_{N \rightarrow \infty}
                \frac{1}{N}
                \sum_{n=1}^N
                \big(
                \outcome{}{n}{s+1}
                - \IMatTv{n\cdot}{s} \VUoutcome{}{}{s}
                -
                \noise{n}{s} \big)
                \outcome{}{n}{r+1}
                \\
                \notag
                &\quad \eqas
                \lim_{N \rightarrow \infty}
                \frac{\mu(\mu+\mu_r)}{N^2} \left(\sum_{n=1}^N \Uoutcome{n}{}{s}\right)
                \left(\sum_{n=1}^N \Uoutcome{n}{}{r}\right)
                +
                \lim_{N \rightarrow \infty}
                \frac{\sigma^2}{N} \sum_{n=1}^N \Uoutcome{n}{}{s}\Uoutcome{n}{}{r},
                \\
                \label{eq:BL-b-5}
                &\lim_{N \rightarrow \infty}
                \frac{1}{N}
                \sum_{n=1}^N
                \big(
                \outcome{}{n}{s+1}
                - \IMatTv{n\cdot}{s} \VUoutcome{}{}{s}
                -
                \noise{n}{s} \big)
                \big(
                \outcome{}{n}{r+1}
                - \IMatTv{n\cdot}{r} \VUoutcome{}{}{r}
                -
                \noise{n}{r} \big)
                \\
                \notag
                &\quad \eqas
                \lim_{N \rightarrow \infty}
                \frac{\mu^2}{N^2} \left(\sum_{n=1}^N \Uoutcome{n}{}{s}\right)
                \left(\sum_{n=1}^N \Uoutcome{n}{}{r}\right)
                +
                \lim_{N \rightarrow \infty}
                \frac{\sigma^2}{N}
                \sum_{n=1}^N
                \Uoutcome{n}{}{s} \Uoutcome{n}{}{r}.
            \end{align}
        \end{subequations}
        
        \item \label{part:BL-c} For all $s=1,\ldots,t$, the following matrices are positive definite almost surely:
        \begin{align}
            \label{eq:BL-lower bound for perps}
            \lim_{N \rightarrow \infty} \frac{\bm{Q}_s^\top \bm{Q}_s}{N} \succ 0,
            \quad\quad\quad
            \lim_{N \rightarrow \infty} \frac{\bm{V}_s^\top \bm{V}_s}{N} 
            -
            \lim_{N \rightarrow \infty} \frac{\bm{V}_s^\top\ones{N\times 1}}{N}
            \lim_{N \rightarrow \infty} \frac{\ones{1\times N}\bm{V}_s}{N}
            \succ 0.
        \end{align}
    \end{enumerate}
\end{theorem}
In the following section, we will provide a comprehensive explanation of the conditioning technique, which will be employed to establish the results presented in Theorem~\ref{lm:Big lemma}.

\subsection{Conditioning Technique}
Let $\Gc_t$ denote the $\sigma$-algebra generated by $\Voutcome{}{}{0},\Voutcome{}{}{1},\ldots,\Voutcome{}{}{t}$, $\Mtreatment{}{}$, $\covar$, $\IMatT{0},\ldots,\IMatT{t-1}$, as well as $\Vnoise{}{0},\Vnoise{}{1},\ldots,\Vnoise{}{t-1}$. We calculate the conditional distribution of $\IM$ given $\Gc_t$. On the other hand, conditioning on $\Gc_t$ is equivalent to conditioning on the event
\begin{align}
    \label{eq:LOM-matrix form}
    \IM \bm{Q}_t = \bm{R}_t.
\end{align}
Note that, given $\Gc_t$, entries of both $\bm{Q}_t$ and $\bm{R}_t$ are deterministic known real numbers. Then, we need a generalization of Lemma 11 in \cite{bayati2011dynamics} which is based on the invariance property of the Gaussian distribution under rotations.
\begin{lemma}
    \label{lm:conditional dist-vector}
    Let $\bm{D} \in \R^{m \times n}$ be a full-row rank matrix
    and $\Vec{V} \in \R^n$ a vector with i.i.d. Gaussian entries with mean $\gamma$ and variance $\chi^2$. Then, for any vector $\Vec{b} \in \R^m$, we have
    \begin{align}
        \label{eq:conditional dist-vector}
        \Vec{V}|_{\bm{D}\Vec{V} = \Vec{b}}
        \eqd
        \gamma \ones{n} + \bm{D}^\top \left(\bm{D} \bm{D}^\top\right)^{-1} \Vec{d} +  P_{\{\bm{D}\Vec{J}=0\}}(\Vec{S}),
    \end{align}
    where $\Vec{d} = \Vec{b} - \gamma \bm{D} \ones{n}$ and $\Vec{J} = \Vec{V} - \gamma \ones{n}$. Further, $\Vec{S}$ is a random vector independent of $\Vec{J}$ with the same distribution, $P_{\{\bm{D}\Vec{J}=0\}}$ is the orthogonal projection onto the subspace ${\{\bm{D}\Vec{J}=0\}}$, and $\bm{D}^\top \left(\bm{D} \bm{D}^\top\right)^{-1} \Vec{d} = \argmin_{\Vec{J}} \left\{\norm{\Vec{J}}^2: \bm{D}\Vec{J}=\Vec{d}\right\}$.
\end{lemma}
Proof. By definition, $\Vec{J} \in \R^n$ is a random vector with i.i.d. Gaussian entries with zero mean and variance~$\chi^2$. We have
\begin{align*}
    \Vec{V}|_{\bm{D}\Vec{V} = \Vec{b}}
    \eqd
    \gamma \ones{n} + \Vec{J}|_{\bm{D}\Vec{V} = \Vec{b}}
    \eqd
    \gamma \ones{n} + \Vec{J}|_{\bm{D}\Vec{J} = \Vec{d}}.
\end{align*}
Then, we get the desired result by applying Lemma 11 in \cite{bayati2011dynamics} on the second term on the right-hand side. \ep

The next Lemma applies the result of Lemma~\ref{lm:conditional dist-vector} to obtain the conditional distribution of the fixed interference matrix $\IM$ given $\Gc_t$.
\begin{lemma}
    \label{lm:conditional dist of IM}
    Fix $t$ and assume that $\bm{Q}_t$ is a full-row rank matrix. Then, for the conditional distribution of the fixed interference matrix $\IM$ given $\IM \bm{Q}_t=\bm{R}_t$, we have
        \begin{align}
        \label{eq:conditional dist of IM}
        \IM|_{\IM \bm{Q}_t=\bm{R}_t}
        \eqd
        \frac{\mu}{N} \ones{N\times N}
        +
        \bar{\bm{R}}_t
        \left(
        \bm{Q}_t^\top \bm{Q}_t
        \right)^{-1}
        \bm{Q}_t^\top
        +
        \widetilde{\IM}_0 P^\perp.
        \end{align}
    where $\widetilde{\IM}_0 \eqd \IM - \frac{\mu}{N} \ones{N\times N}$ independent of $\IM$ and $P^\perp = (\I-P)$ that P denotes the orthogonal projector onto the column space of $\bm{Q}_t$ and
    \begin{align}
        \label{eq:LOM-matrix form-rewriting}
        \bar{\bm{R}}_t = \bm{R}_t - \frac{\mu}{N} \ones{N\times N} \bm{Q}_t
    \end{align}
\end{lemma}
Proof. To calculate the conditional distribution of $\IM$ given $\IM \bm{Q}_t = \bm{R}_t$, we proceed by rewriting the interference matrix $\IM$ as follows:
\begin{align*}
    \IM \eqd \frac{\mu}{N} \ones{N\times N} + \bar{\IM},
\end{align*}
where $\bar{\IM}$ is a matrix of i.i.d. Gaussian entries with zero mean and variance $\sigma^2$. Therefore, we rewrite~\eqref{eq:LOM-matrix form} as:
\begin{align*}
    \bar{\IM} \bm{Q}_t = \bar{\bm{R}}_t.
\end{align*}
Following Lemma~\ref{lm:conditional dist-vector}, we first solve the least square problem below:
\begin{align*}
    \bm{E}_t
    =
    \argmin_{\bar{\IM}}
    \left\{
    \fnorms{\bar{\IM}}:
    \bar{\IM} \bm{Q}_t = \bar{\bm{R}}_t
    \right\}.
\end{align*}
We write the Lagrangian
\begin{align*}
     \fnorms{\bar{\IM}}
     +
     \Tr
     \left(
     \bm{\Lambda}
     \left(
     \bar{\bm{R}}_t
     -
     \bar{\IM} \bm{Q}_t
     \right)^\top
     \right),
\end{align*}
where $\bm{\Lambda} \in \R^{N\times t}$ is the Lagrange multiplier. We get $2 \bar{\IM}=\bm{\Lambda} \bm{Q}_t^\top$, that implies
\begin{align*}
    \bm{E}_t
    =
    \bar{\bm{R}}_t
    \left(
    \bm{Q}_t^\top \bm{Q}_t
    \right)^{-1}
    \bm{Q}_t^\top.
\end{align*}
Next, we show that the orthogonal projection of $\widetilde{\IM}_0$ onto the subspace $\Ac:= \{\tilde{\IM}:\tilde{\IM} \bm{Q}_t = 0\}$ is equal to $\widetilde{\IM}_0\left(\I - P\right)$. For that purpose, we follow the same steps as the proof of Lemma~10 in \cite{bayati2011dynamics}. First, note that by definition $\widetilde{\IM}_0\left(\I - P\right) \bm{Q}_t = 0$; that is, $\widetilde{\IM}_0\left(\I - P\right) \in \Ac$. Second, the orthogonal projection of $\widetilde{\IM}_0\left(\I - P\right)$ onto the subspace $\Ac$ is equal to itself. That is,
\begin{align*}
    \widetilde{\IM}_0\left(\I - P\right)\left(\I - P\right)
    =
    \widetilde{\IM}_0\left(\I - P\right)
    -
    \bar{\IM} P
    +
    \bar{\IM} P P
    =
    \widetilde{\IM}_0\left(\I - P\right).
\end{align*}
Third, we show that if $\IM'\in\Ac$, then $\IM'(\I-P)=\IM'$. To this end, note that if $\IM'\in\Ac$, we have $\IM'\bm{Q}_t=0$; then, all the rows of the matrix $\IM'$ are perpendicular to the columns of $\bm{Q}_t$. This implies that $\IM'P=0$ and so $\IM'\left(\I - P\right) = \IM'$.
Finally, we need to show that the operator corresponding to this projection is symmetric. That is, for all matrices $\bm C$ and $\bm D$ it holds that $\Tr \left(\bm C\left(\I - P\right)\bm D^\top\right) = \Tr \left(\bm D\left(\I - P\right)\bm C^\top\right)$. We have,
\begin{align*}
    \Tr \left(\bm C\left(\I - P\right)\bm D^\top\right)
    =
    \Tr \left(\left(\bm C\left(\I - P\right)\bm D^\top\right)^\top\right)
    =
    \Tr \left(\bm D\left(\I - P\right)\bm C^\top\right).
\end{align*}
Applying Lemma~\ref{lm:conditional dist-vector} concludes the proof.\ep

Next lemma expresses the distribution of $\Voutcome{}{}{t+1}$ conditioning on the $\sigma$-algebra $\Gc_t$ or equivalently on the event $\IM \bm{Q}_t = \bm{R}_t$. 

\begin{lemma}
    \label{lm:conditional dist of outcome}
    Fix $t$ and assume that $\bm{Q}_t$ is a full-row rank matrix. The following holds for the conditional distribution of the outcome vector $\Voutcome{}{}{t+1}$:
            \begin{align}
                \label{eq:conditional dist of outcome_nonsym}
                \Voutcome{}{}{t+1}\big|_{\Gc_t}
                \eqd
                &\;
                \widetilde{\IM} 
                \VUoutcome{\perp}{}{t}
                + \bm{R}_t \Vec{\alpha}_t
                +
                \IMatT{t} \VUoutcome{}{}{t}
                +
                \Vnoise{}{t},
            \end{align}
    where the matrix $\widetilde{\IM}$ is independent of $\IM$ and has the same distribution.
\end{lemma}
Proof. By \eqref{eq:outcome_function_matrix}, we have
\begin{equation}
    \label{eq:CD_proof_1}
    \begin{aligned}
        \Voutcome{}{}{t+1}\big|_{\Gc_t}
        \eqd
        \;\big(
        \IM \VUoutcome{}{}{t}
        +
        \IMatT{t} \VUoutcome{}{}{t}
        +
        \Vnoise{}{t}
        \big)\big|_{\Gc_t}
        \eqd
        \;\IM\big|_{\Gc_t} \VUoutcome{}{}{t}
        +
        \IMatT{t} \VUoutcome{}{}{t}
        +
        \Vnoise{}{t},
    \end{aligned}
\end{equation}
where we used the fact that $\IMatT{t}$ and $\Vnoise{}{t}$ are independent of $\Gc_t$. Further, note that in the right-hand side of~\eqref{eq:CD_proof_1}, the matrix $\IMatT{t}$ and the vector $\Vnoise{}{t}$ are random objects while $\VUoutcome{}{}{t}$ is a deterministic known vector according to~$\Gc_t$. Now, we use the result of Lemma~\ref{lm:conditional dist of IM}. We get
\begin{equation}
    \label{eq:CD_proof_nonsym_1}
    \begin{aligned}
        \Voutcome{}{}{t+1}\big|_{\Gc_t}
        \eqd
        \frac{\mu}{N} \ones{N\times N}
        \VUoutcome{}{}{t}
        +
        \bar{\bm{R}}_t
        \left(
        \bm{Q}_t^\top \bm{Q}_t
        \right)^{-1}
        \bm{Q}_t^\top
        \VUoutcome{}{}{t} 
        +
        \widetilde{\IM}_0 P^\perp
        \VUoutcome{}{}{t}
        +
        \IMatT{t} \VUoutcome{}{}{t}
        +
        \Vnoise{}{t}.
    \end{aligned}
\end{equation}
By $\VUoutcome{}{}{t} = \VUoutcome{\parallel}{}{t} + \VUoutcome{\perp}{}{t}$ and the fact that $\bm{Q}_t^\top \VUoutcome{\perp}{}{t} = 0$ and using \eqref{eq:LOM-matrix form-rewriting}, we can write
\begin{equation}
    \label{eq:CD_proof_nonsym_2}
    \begin{aligned}
        \bar{\bm{R}}_t
        \left(
        \bm{Q}_t^\top \bm{Q}_t
        \right)^{-1}
        \bm{Q}_t^\top
        \VUoutcome{}{}{t}
        &=
        \big(
        \bm{R}_t - \frac{\mu}{N} \ones{N\times N} \bm{Q}_t
        \big)
        \left(
        \bm{Q}_t^\top \bm{Q}_t
        \right)^{-1}
        \bm{Q}_t^\top
        \VUoutcome{\parallel}{}{t}
        =
        \bm{R}_t \Vec{\alpha}_t
        -
        \frac{\mu}{N} \ones{N\times N} \bm{Q}_t \Vec{\alpha}_t.
    \end{aligned}
\end{equation}
where in the last equality we used
$\VUoutcome{\parallel}{}{t} = \bm{Q}_t \Vec{\alpha_t}$. Considering \eqref{eq:CD_proof_nonsym_1} and \eqref{eq:CD_proof_nonsym_2} together, we have
\begin{equation*}
    \begin{aligned}
        \Voutcome{}{}{t+1}\big|_{\Gc_t}
        &\eqd
        \frac{\mu}{N} \ones{N\times N}
        \VUoutcome{\parallel}{}{t}
        +
        \frac{\mu}{N} \ones{N\times N}
        \VUoutcome{\perp}{}{t}
        + \bm{R}_t \Vec{\alpha}_t
        -
        \frac{\mu}{N} \ones{N\times N} \bm{Q}_t \Vec{\alpha}_t
        +
        \widetilde{\IM}_0 P^\perp
        \VUoutcome{}{}{t}
        +
        \IMatT{t} \VUoutcome{}{}{t}
        +
        \Vnoise{}{t}
        \\
        &\eqd
        \frac{\mu}{N} \ones{N\times N}
        \VUoutcome{\perp}{}{t}
        + \bm{R}_t \Vec{\alpha}_t
        +
        \widetilde{\IM}_0 
        \VUoutcome{\perp}{}{t}
        +
        \IMatT{t} \VUoutcome{}{}{t}
        +
        \Vnoise{}{t}
        \\
        &\eqd
        \widetilde{\IM} 
        \VUoutcome{\perp}{}{t}
        + \bm{R}_t \Vec{\alpha}_t
        +
        \IMatT{t} \VUoutcome{}{}{t}
        +
        \Vnoise{}{t},
    \end{aligned}
\end{equation*}
where in the last equality we used $\widetilde{\IM} \eqd \frac{\mu}{N} \ones{N\times N} + \widetilde{\IM}_0$ that concludes the proof.\ep

\subsection{Proof of Theorem~\ref{lm:Big lemma}}
For each $t$, we can assume, without loss of generality, that the function $\outcome{}{}{} \mapsto \outcomeg{t}{}(\outcome{}{}{},\Vtreatment{}{}, \Vcovar{})$ is non-constant with a positive probability with respect to $(\Vtreatment{}{},\Vcovar{}) \sim \Pi \x p_{\covar}$. The case where $\outcome{}{}{} \mapsto \outcomeg{t}{}(\outcome{}{}{},\Vtreatment{}{}, \Vcovar{})$ is almost surely constant is trivial and does not require further analysis. We use induction on $t$. 

\textbf{Step 1.} Let $t=0$ and note that $\bm Q_0$ and $\bm R_0$ are empty matrices and the $\sigma$-algebra $\Gc_0$ is generated by $\Voutcome{}{}{0}$, $\Mtreatment{}{}$, and $\covar$. We prove Parts \ref{part:BL-a} and \ref{part:BL-b} for $t=0$ and Part \ref{part:BL-c} for $t = 1$ as the base case.
\begin{enumerate}[label=(\alph*)]
    \item \label{item:BL-average limit} By Assumption~\ref{asmp:Gaussian Interference Matrice} and \eqref{eq:conditional dist of outcome_nonsym}, conditioning on the values of $\Voutcome{}{}{0}$, $\Mtreatment{}{}$, as well as $\covar$ and so on the value of $\VUoutcome{}{}{0} = \outcomeg{0}{}\big(\Voutcome{}{}{0},\Mtreatment{}{},\covar\big)$, the elements of $\Voutcome{}{}{1}$ are i.i.d. Gaussian random variables with mean $\nu_{1N}$ and variance~$\rho_{1N}^2$:
    \begin{equation}
        \label{eq:BL-a0-Y1 stat}
        \begin{aligned}
            \nu_{1N}
            &:=
            \E
            \left[
            \outcome{}{n}{1}
            \Big|
            \VUoutcome{}{}{0}
            \right]
            =
            \E
            \left[
            [
            \IM \VUoutcome{}{}{0}
            +
            \IMatT{0} \VUoutcome{}{}{0}
            +
            \Vnoise{}{0}
            ]_n
            \Big|
            \VUoutcome{}{}{0}
            \right]
            =
            \frac{\mu + \mu_0}{N}
            \sum_{n=1}^N
            \Uoutcome{n}{}{0},
            \\
            \rho_{1N}^2
            &:=
            \Var
            \left[
            \outcome{}{n}{1}
            \Big|
            \VUoutcome{}{}{0}
            \right]
            =
            \frac{\sigma^2+\sigma_0^2}{N}
            \sum_{n=1}^N
            \left(\Uoutcome{n}{}{0}\right)^2
            +
            \sigma_e^2,
        \end{aligned}
    \end{equation}
    where $\Uoutcome{n}{}{0} = \outcomeg{0}{} \big(\outcome{}{n}{0},\Vtreatment{n}{},\Vcovar{n}\big)$ is the $n^{th}$ element of the column vector $\VUoutcome{}{}{0}$. By Assumption~\ref{asmp:BL}-\ref{asmp:BL-bound on initials}, both $\nu_{1N}$ and $\rho_{1N}^2$ are bounded. Now, if we let $Z$ be a standard Normal random variable, by inequality $(a+b)^l \leq 2^{l-1} (a^l+b^l)$, for $l \geq 1$, that is a straightforward result of Jensen's inequality, we get
    \begin{equation}
    \begin{aligned}
    \label{eq:BL-proof-a0-1}
        \E
        \left[
        \big|\outcome{}{n}{1}\big|^l
        \big|
        \VUoutcome{}{}{0}
        \right]
        &=
        \E
        \left[
        \big|\nu_{1N} + \rho_{1N} Z\big|^l
        \big|
        \VUoutcome{}{}{0}
        \right]
        \leq
        2^{l-1}
        \E
        \left[
        \big|\nu_{1N}\big|^l + \big|\rho_{1N}\big|^l \big|Z\big|^l
        \big|
        \VUoutcome{}{}{0}
        \right] \leq c.
    \end{aligned}
    \end{equation}
    Here, $c$ is a constant independent of $N$ and might alter in different lines. Reusing Jensen's inequality multiple times and the fact that $\psi \in \poly{k}$, for $\kappa > 0$, we get 
    \begin{equation}
    \label{eq:BL-proof-a0-2}
        \begin{aligned}
            &\; \E
            \left[
            \Big|
            \psi\big(
            \outcome{}{n}{1}
            ,
            \Vtreatment{n}{},\Vcovar{n}
            \big)
            -
            \E_{\IMG_0,\Vnoise{}{0}}
            \left[
            \psi\big(
            \outcome{}{n}{1}
            ,
            \Vtreatment{n}{},\Vcovar{n}
            \big)
            \right]
            \Big|^{2+\kappa}
            \right]
            \\
            \leq
            &\; \E
            \left[
            \Big(
            \big|
            \psi\big(
            \outcome{}{n}{1}
            ,
            \Vtreatment{n}{},\Vcovar{n}
            \big)
            \big|
            +
            \E_{\IMG_0,\Vnoise{}{0}}
            \left[
            \big|
            \psi\big(
            \outcome{}{n}{1}
            ,
            \Vtreatment{n}{},\Vcovar{n}
            \big)
            \big|
            \right]
            \Big)^{2+\kappa}
            \right]
            \\
            \leq
            &\; \E
            \left[
            2^{1+\kappa}
            \Big(
            \big|
            \psi\big(
            \outcome{}{n}{1}
            ,
            \Vtreatment{n}{},\Vcovar{n}
            \big)
            \big|^{2+\kappa}
            +
            \E_{\IMG_0,\Vnoise{}{0}}
            \left[
            \big|
            \psi\big(
            \outcome{}{n}{1}
            ,
            \Vtreatment{n}{},\Vcovar{n}
            \big)
            \big|^{2+\kappa}
            \right]
            \Big)
            \right]
            \\
            \leq
            &\;c
            \E
            \left[
            \Big(
            1 + 
            \norm{
            \big(
            \outcome{}{n}{1}
            ,
            \Vtreatment{n}{},\Vcovar{n}
            \big)
            }^k
            \Big)^{2+\kappa}
            +
            \E_{\IMG_0,\Vnoise{}{0}}
            \left[
            \Big(
            1+
            \norm{
            \big(
            \outcome{}{n}{1}
            ,
            \Vtreatment{n}{},\Vcovar{n}
            \big)
            }^k
            \Big)^{2+\kappa}
            \right]
            \right]
            \\
            \leq
            &\;c 
            \E
            \left[
            2
            +
            \Big(
            \big(
            \outcome{}{n}{1}
            \big)^2
            +
            \norm{
            \big(
            \Vtreatment{n}{},\Vcovar{n}
            \big)
            }^2
            \Big)^{k+\frac{k\kappa}{2}}
            +
            \E_{\IMG_0,\Vnoise{}{0}}
            \left[
            \Big(
            \big(
            \outcome{}{n}{1}
            \big)^2
            +
            \norm{
            \big(
            \Vtreatment{n}{},\Vcovar{n}
            \big)
            }^2
            \Big)^{k+\frac{k\kappa}{2}}
            \right]
            \right]
            \\
            \leq
            &\;c 
            \E
            \left[
            2
            +
            \big(
            \outcome{}{n}{1}
            \big)^{2k+k\kappa}
            +
            \norm{
            \big(
            \Vtreatment{n}{},\Vcovar{n}
            \big)
            }^{2k+k\kappa}
            +
            \E_{\IMG_0,\Vnoise{}{0}}
            \left[
            \big(
            \outcome{}{n}{1}
            \big)^{2k+k\kappa}
            +
            \norm{
            \big(
            \Vtreatment{n}{},\Vcovar{n}
            \big)
            }^{2k+k\kappa}
            \right]
            \right],
        \end{aligned}
    \end{equation}
    where $\E_{\IMG_0,\Vnoise{}{0}}$ is the expectation with respect to the randomness of the interference matrix $\IMG_0 = \IM + \IMatT{0}$ and observation noise $\Vnoise{}{0}$.
    In \eqref{eq:BL-proof-a0-2}, because $k \geq 2$ and $\kappa>0$, we get $2+\kappa \geq 1$ and $k+\frac{k\kappa}{2}\geq 1$, and we are allowed to use the inequality $(v_1+v_2)^l \leq 2^{l-1} (v_1^l+v_2^l),\; v_1,v_2 \geq 0$.
    Therefore, by~\eqref{eq:BL-proof-a0-1}, we have
    \begin{align*}
        \frac{1}{N} \sum_{n=1}^N
        \E
        \left[
        \Big|
        \psi\big(
        \outcome{}{n}{1}
        ,
        \Vtreatment{n}{}, \Vcovar{n}
        \big)
        -
        \E_{\IMG_0,\Vnoise{}{0}}
        \left[
        \psi\big(
        \outcome{}{n}{1}
        ,
        \Vtreatment{n}{}, \Vcovar{n}
        \big)
        \right]
        \Big|^{2+\kappa}
        \right]
        \leq
        c N^{\kappa/2},
    \end{align*}
    where $c$ is a constant independent of $N$. Applying the Strong Law of Large Numbers (SLLN) for triangular arrays in Theorem~\ref{thm:SLLN}, we obtain
    \begin{align}
        \label{eq:BL-b0-SLLN}
        \lim_{N \rightarrow \infty}
        \frac{1}{N} \sum_{n=1}^N
        \Big(
        \psi\big(
        \outcome{}{n}{1}
        ,
        \Vtreatment{n}{}, \Vcovar{n}
        \big)
        -
        \E_{\IMG_0,\Vnoise{}{0}}
        \left[
        \psi\big(
        \outcome{}{n}{1}
        ,
        \Vtreatment{n}{}, \Vcovar{n}
        \big)
        \right]
        \Big)
        \eqas
        0.
    \end{align}
    By Assumption~\ref{asmp:BL}-\ref{asmp:BL-bound on initials} and \eqref{eq:BL-a0-Y1 stat}, we can write
    \begin{equation*}
        \begin{aligned}
            \lim_{N\rightarrow \infty} \nu_{1N}
            &=
            \lim_{N\rightarrow \infty}
            \frac{\mu+\mu_0}{N}
            \sum_{n=1}^N
            \outcomeg{0}{}\big(\outcome{}{n}{0},\Vtreatment{n}{}, \Vcovar{n}\big)
            =
            \AVO{}{}{1}
            \\
            \lim_{N\rightarrow \infty} \rho_{1N}^2
            &=
            \lim_{N\rightarrow \infty}
            \frac{\sigma^2+\sigma_0^2}{N}
            \sum_{n=1}^N
            \outcomeg{0}{}\big(\outcome{}{n}{0},\Vtreatment{n}{}, \Vcovar{n}\big)^2
            +
            \sigma_e^2
            =
            \VVO{}{2}{1}.
        \end{aligned}
    \end{equation*}
    Now, we use Theorem~\ref{thm:SLLN-2}  for $f(\Vtreatment{n}{}, \Vcovar{n}) = \E_{\IMG_0,\Vnoise{}{0}}\left[\psi\big(\outcome{}{n}{1},\Vtreatment{n}{}, \Vcovar{n}\big)\right]$. By \eqref{eq:BL-b0-SLLN}, we can write
    \begin{equation}
        \label{eq::BL-proof-a0-dynamics}
        \begin{aligned}
            \lim_{N \rightarrow \infty}
            \frac{1}{N} \sum_{n=1}^N
            \psi\big(
            \outcome{}{n}{1}
            ,
            \Vtreatment{n}{}, \Vcovar{n}
            \big)
            &\eqas
            \lim_{N \rightarrow \infty}
            \frac{1}{N} \sum_{n=1}^N
            \E_{\IMG_0,\Vnoise{}{0}}
            \left[
            \psi\big(
            \outcome{}{n}{1}
            ,
            \Vtreatment{n}{}, \Vcovar{n}
            \big)
            \right]
            \\
            &\eqas
            \lim_{N \rightarrow \infty}
            \E
            \Big[
            \psi
            \big(
            \nu_{1N}
            +
            \rho_{1N} Z
            ,
            \Vtreatment{}{},\Vcovar{}
            \big)
            \Big]
            \eqas
            \E
            \Big[
            \psi
            \big(
            \AVO{}{}{1}
            +
            \VVO{}{}{1} Z
            ,
            \Vtreatment{}{},\Vcovar{}
            \big)
            \Big].
        \end{aligned}
    \end{equation}
    Note that $\E_{\IMG_0,\Vnoise{}{0}}\left[\psi\big(\outcome{}{n}{1},\Vtreatment{n}{},\Vcovar{n}\big)\right] \in \poly{k}$, since $\psi \in \poly{k}$. In the last equality in \eqref{eq::BL-proof-a0-dynamics}, we used the Dominated Convergence Theorem (DCT), see e.g., Theorem 16.4 in \cite{billingsley2008probability}, which allows us to interchange the limit and the expectation. Additionally, we utilized the continuous mapping theorem, stated in Theorem 2.3 in \cite{van2000asymptotic}, to pass the limit through the function. It is important to note that $Z$ is independent of both $\Vtreatment{}{}$ and $\Vcovar{}$ since its randomness arises from the interference matrix $\IMG_0$ and observation noise $\Vnoise{}{0}$, which are independent of $\Vtreatment{}{}$ and $\Vcovar{}$.
    
    In the second step of the induction, we need two more results. The first result is given in \eqref{eq::BL-proof-a0-dynamics-with-eps} and we can derive it by following the same procedure as above.
    \begin{equation}
        \label{eq::BL-proof-a0-dynamics-with-eps}
        \begin{aligned}
            \lim_{N \rightarrow \infty}
            \frac{1}{N} \sum_{n=1}^N
            \psi\big(
            \outcome{}{n}{1}
            ,
            \outcome{}{n}{1}
            - \IMatTv{n\cdot}{0} \VUoutcome{}{}{0}
            - \noise{n}{0}
            ,
            \Vtreatment{n}{},\Vcovar{n}
            \big)
            \eqas
            \E
            \Big[
            \psi
            \big(
            \AVO{}{}{1}
            +
            \VVO{}{}{1} Z
            ,
            \BAVO{}{}{1}
            +
            \BVVO{}{}{1} Z'
            ,
            \Vtreatment{}{},\Vcovar{}
            \big)
            \Big],
        \end{aligned}
    \end{equation}
    where we assume that $\psi:\R^{2+T+M}\mapsto \R$ is a $\poly{k}$ function and $\IMatTv{n\cdot}{0}$ is the $n^{th}$ row the matrix $\IMatT{0}$. For the second result, consider the function $ \outcomeg{0}{}\big(\outcome{}{n}{0},\Vtreatment{n}{},\Vcovar{n} \big)\phi\big(\outcome{}{n}{1},\outcome{}{n}{1}- \IMatTv{n\cdot}{0} \VUoutcome{}{}{0}-\noise{n}{0},\Vtreatment{n}{},\Vcovar{n} \big)$ that $\phi \in \poly{\frac{k}{2}}$ is arbitrary. By Assumption~\ref{asmp:BL}-\ref{asmp:BL-pl functions}, we know that this function lies within $\poly{k}$. Hence, following the same procedure as in \eqref{eq:BL-proof-a0-2}, we can check the conditions of Theorem~\ref{thm:SLLN}. This implies,
    \begin{equation}
    \begin{aligned}
        \label{eq:BL-a0-third result-1}
        &\lim_{N \rightarrow \infty}
        \frac{1}{N} \sum_{n=1}^N
        \Big(
        \outcomeg{0}{}\big(\outcome{}{n}{0},\Vtreatment{n}{},\Vcovar{n} \big)
        \phi\big(\outcome{}{n}{1}, \outcome{}{n}{1}- \IMatTv{n\cdot}{0} \VUoutcome{}{}{0}-\noise{n}{0}, \Vtreatment{n}{},\Vcovar{n} \big)\Big)
        \\
        \eqas
        &\lim_{N \rightarrow \infty}
        \frac{1}{N} \sum_{n=1}^N
        \Big(
        \E_{\IMG_0,\Vnoise{}{0}}
        \left[
        \outcomeg{0}{}\big(\outcome{}{n}{0},\Vtreatment{n}{},\Vcovar{n} \big)
        \phi\big(\outcome{}{n}{1}, \outcome{}{n}{1}- \IMatTv{n\cdot}{0} \VUoutcome{}{}{0}-\noise{n}{0}, \Vtreatment{n}{},\Vcovar{n} \big)
        \right]
        \Big).
    \end{aligned}
    \end{equation}
    Note that in \eqref{eq:BL-a0-third result-1}, the expectation is with respect to the randomness of the interference matrix and observation noise. Thus, letting $\varphi(\Vtreatment{n}{},\Vcovar{n} ) = \E_{\IMG_0,\Vnoise{}{0}} \left[ \phi\big(\outcome{}{n}{1},\outcome{}{n}{1}- \IMatTv{n\cdot}{0} \VUoutcome{}{}{0}-\noise{n}{0},\Vtreatment{n}{},\Vcovar{n} \big) \right]$, by Assumption~\ref{asmp:BL}-\ref{asmp:BL-empirical dist of inits} and the DCT, we have
    \begin{equation*}
        \begin{aligned}
            &\lim_{N \rightarrow \infty}
            \frac{1}{N} \sum_{n=1}^N
            \Big(
            \outcomeg{0}{}\big(\outcome{}{n}{0},\Vtreatment{n}{},\Vcovar{n} \big)
            \phi\big(\outcome{}{n}{1},\outcome{}{n}{1}- \IMatTv{n\cdot}{0} \VUoutcome{}{}{0}-\noise{n}{0},\Vtreatment{n}{},\Vcovar{n} \big)
            \Big)
            \\
            \eqas
            &\lim_{N \rightarrow \infty}
            \frac{1}{N} \sum_{n=1}^N
            \Big(
            \E_{\IMG_0,\Vnoise{}{0}}
            \left[
            \outcomeg{0}{}\big(\outcome{}{n}{0},\Vtreatment{n}{},\Vcovar{n} \big)
            \phi\big(\outcome{}{n}{1}, \outcome{}{n}{1}- \IMatTv{n\cdot}{0} \VUoutcome{}{}{0}-\noise{n}{0},\Vtreatment{n}{},\Vcovar{n} \big)
            \right] \Big)
            \\
            \eqas
            &\lim_{N \rightarrow \infty}
            \E
            \big[
            \bar{g}_0(\Vtreatment{}{},\Vcovar{})
            \phi(\nu_{1N} + \rho_{1N} Z, \bar{\nu}_{1N} + \bar{\rho}_{1N} Z',\Vtreatment{}{},\Vcovar{})
            \big].
        \end{aligned}
    \end{equation*}
    Similar to \eqref{eq::BL-proof-a0-dynamics}, we obtain the desired result as follows
    \begin{equation*}
        \begin{aligned}
            &\lim_{N \rightarrow \infty}
            \frac{1}{N} \sum_{n=1}^N
            \Big(
            \outcomeg{0}{}\big(\outcome{}{n}{0},\Vtreatment{n}{},\Vcovar{n} \big)
            \phi\big(\outcome{}{n}{1},\outcome{}{n}{1}- \IMatTv{n\cdot}{0} \VUoutcome{}{}{0}-\noise{n}{0},\Vtreatment{n}{},\Vcovar{n} \big)
            \Big)
            \\
            \eqas
            &\;\E
            \big[
            \bar{g}_0(\Vtreatment{}{},\Vcovar{})
            \phi\big(
            \AVO{}{}{1}
            +
            \VVO{}{}{1} Z
            ,
            \BAVO{}{}{1}
            +
            \BVVO{}{}{1} Z'
            ,
            \Vtreatment{}{},\Vcovar{}
            \big)
            \big].
        \end{aligned}
    \end{equation*}

    \item By Assumption~\ref{asmp:BL}-\ref{asmp:BL-bound on initials} as well as \eqref{eq::BL-proof-a0-dynamics} for $\psi(y,\cdot,\cdot)=y$ and $\psi(y,\cdot,\cdot)=y^2$, we have
    \begin{equation}
        \label{eq:BL-proof-b0-1}
        \begin{aligned}
            \lim_{N \rightarrow \infty}
            \frac{1}{N}
            \sum_{n=1}^N
            \outcome{}{n}{1}
            &\eqas
            \AVO{}{}{1}
            = 
            \lim_{N \rightarrow \infty}
            \frac{\mu+\mu_0}{N}
            \sum_{n=1}^N
            \Uoutcome{n}{}{0},
            \\
            \lim_{N \rightarrow \infty}
            \frac{1}{N}
            \sum_{n=1}^N
            \big(\outcome{}{n}{1}
            \big)^2
            &\eqas
            \AVO{}{2}{1} + \VVO{}{2}{1}
            =
            \AVO{}{2}{1}
            +
            \lim_{N \rightarrow \infty}
            \frac{\sigma^2+\sigma_0^2}{N}
            \sum_{n=1}^N
            \left(\Uoutcome{n}{}{0}\right)^2
            +
            \sigma_e^2.
        \end{aligned}
    \end{equation}
    Likewise, by Assumption~\ref{asmp:BL}-\ref{asmp:BL-bound on initials} as well as \eqref{eq::BL-proof-a0-dynamics-with-eps} for functions $\psi(y,\bar{y},\cdot,\cdot)=\bar{y}$ and $\psi(y,\bar{y},\cdot,\cdot)=\bar{y}^2$, we have
    \begin{equation*}
        \begin{aligned}
            \lim_{N \rightarrow \infty}
            \frac{1}{N}
            \sum_{n=1}^N
            \big(\outcome{}{n}{1}
            - \IMatTv{n\cdot}{0} \VUoutcome{}{}{0}
            -
            \noise{n}{0}
            \big)
            &\eqas
            \BAVO{}{}{1}
            =
            \lim_{N \rightarrow \infty}
            \frac{\mu}{N}
            \sum_{n=1}^N
            \Uoutcome{n}{}{0},
            \\
            \lim_{N \rightarrow \infty}
            \frac{1}{N}
            \sum_{n=1}^N
            \big(\outcome{}{n}{1}
            - \IMatTv{n\cdot}{0} \VUoutcome{}{}{0}
            -
            \noise{n}{0}
            \big)
            \big(\outcome{}{n}{1}
            - \IMatTv{n\cdot}{0} \VUoutcome{}{}{0}
            -
            \noise{n}{0}
            \big)
            &\eqas
            \BAVO{}{2}{1}
            +
            \lim_{N \rightarrow \infty}
            \frac{\sigma^2}{N}
            \sum_{n=1}^N
            \Uoutcome{n}{}{0}^2.
        \end{aligned}
    \end{equation*}
    Finally, by applying Theorem~\ref{thm:SLLN} and considering the fact that elements of $\Vnoise{}{0}$ are zero-mean random variables independent of everything, we can write
    \begin{equation*}
        \begin{aligned}
            \lim_{N \rightarrow \infty}
            \frac{1}{N}
            \sum_{n=1}^N
            \outcome{}{n}{1}
            \big(\outcome{}{n}{1}
            - \IMatTv{n\cdot}{0} \VUoutcome{}{}{0}
            -
            \noise{n}{0}
            \big)
            &=
            \lim_{N \rightarrow \infty}
            \frac{1}{N}
            \sum_{n=1}^N
            \Big(
            \big(\IMatv{n\cdot}+\IMatTv{n\cdot}{0}\big) \VUoutcome{}{}{0}
            +
            \noise{n}{0}
            \Big)
            (\IMatv{n\cdot} \VUoutcome{}{}{0})
            \\
            &\eqas
            \lim_{N \rightarrow \infty}
            \frac{1}{N}
            \sum_{n=1}^N
            (\IMatv{n\cdot} \VUoutcome{}{}{0})^2
            +
            \lim_{N \rightarrow \infty}
            \frac{1}{N}
            \sum_{n=1}^N
            (\IMatv{n\cdot} \VUoutcome{}{}{0})
            (\IMatTv{n\cdot}{0} \VUoutcome{}{}{0}),
        \end{aligned}
    \end{equation*}
    where $\IMatv{n\cdot}$ and $\IMatTv{n\cdot}{0}$ denote row $n$ of $\IM$ and $\IMatT{0}$, respectively. Note that for all $n\in [N]$, random variables $\IMatv{n\cdot} \VUoutcome{}{}{0}$ and $\IMatTv{n\cdot}{0} \VUoutcome{}{}{0}$ are i.i.d. and
    \begin{align*}
        \IMatv{n\cdot} \VUoutcome{}{}{0} \sim \Nc\left( \frac{\mu}{N} \sum_{i=1}^N \Uoutcome{i}{}{0}, \frac{\sigma^2}{N} \sum_{i=1}^N (\Uoutcome{i}{}{0})^2\right),
        \quad\quad
        \IMatTv{n\cdot}{0} \VUoutcome{}{}{0} \sim \Nc\left( \frac{\mu_0}{N} \sum_{i=1}^N \Uoutcome{i}{}{0}, \frac{\sigma_0^2}{N} \sum_{i=1}^N (\Uoutcome{i}{}{0})^2\right).
    \end{align*}
    Therefore, we get
    \begin{equation*}
        \begin{aligned}
            \lim_{N \rightarrow \infty}
            \frac{1}{N}
            \sum_{n=1}^N
            \outcome{}{n}{1}
            \big(\outcome{}{n}{1}
            - \IMatTv{n\cdot}{0} \VUoutcome{}{}{0}
            -
            \noise{n}{0}
            \big)
            &\eqas
            \lim_{N \rightarrow \infty}
            \frac{\mu(\mu+\mu_0)}{N} \left(\sum_{i=1}^N \Uoutcome{i}{}{0}\right)^2
            +
            \lim_{N \rightarrow \infty}
            \frac{\sigma^2}{N} \sum_{i=1}^N (\Uoutcome{i}{}{0})^2,
        \end{aligned}
    \end{equation*}
    where the limits exist based on Assumption~\ref{asmp:BL}-\ref{asmp:BL-bound on initials}.

    \item For $t = 1$, the matrix $\bm{Q}_1$ is equal to the vector $\VUoutcome{}{}{0}$ and $\bm{V}_1$ is equal to the vector $\Voutcome{}{}{1}$. By Assumption~\ref{asmp:BL}-\ref{asmp:BL-bound on initials}, we have
    \begin{align*}
        \lim_{N \rightarrow \infty} \frac{\bm{Q}_1^\top \bm{Q}_1}{N}
        =
        \lim_{N \rightarrow \infty}
        \pdot{\VUoutcome{}{}{0}}{\VUoutcome{}{}{0}} =
        \lim_{N \rightarrow \infty}
        \frac{1}{N}
        \sum_{i=1}^N \left(\Uoutcome{n}{}{0}\right)^2 > 0,     
    \end{align*}
    as well as
    \begin{align*}
        \lim_{N \rightarrow \infty} \frac{\bm{V}_1^\top \bm{V}_1}{N} 
        -
        \lim_{N \rightarrow \infty} \frac{\bm{V}_1^\top\ones{N\times 1}}{N}
        \lim_{N \rightarrow \infty} \frac{\ones{1\times N}\bm{V}_1}{N}
        &=
        \\
        \lim_{N \rightarrow \infty} \pdot{\Voutcome{}{}{1}}{\Voutcome{}{}{1}}
        -
        \lim_{N \rightarrow \infty} \frac{\Voutcome{}{}{1}^\top \ones{N\times 1}}{N}
        \lim_{N \rightarrow \infty} \frac{\ones{1\times N} \Voutcome{}{}{1}}{N}
        &\eqas
        \VVO{}{2}{1} > 0,
    \end{align*}
    where we also used the result of Step 1-\ref{part:BL-b}.
    \end{enumerate}

    \textbf{Step 2.} Assuming that \eqref{eq:BL-average limit}, \eqref{eq::BL-at-dynamics-with-eps}, and \eqref{eq::BL-at-third result} hold for $s=0,1,\ldots,t-1$, \eqref{eq:BL-b} for $0\leq r,s <t$, and \eqref{eq:BL-lower bound for perps} for $s=1,\ldots,t-1$, we show that they also hold for $t$.
    \begin{equation}
        \label{eq::BL-at-dynamics-with-eps}
        \begin{aligned}
            &\lim_{N \rightarrow \infty}
            \frac{1}{N} \sum_{n=1}^N
            \psi\big(
            \outcome{}{n}{1}
            ,
            \outcome{}{n}{1}-\IMatTv{n\cdot}{0}\VUoutcome{}{}{0} - \noise{n}{0}
            ,
            \ldots
            ,
            \outcome{}{n}{s+1}
            ,
            \outcome{}{n}{s+1}-\IMatTv{n\cdot}{s}\VUoutcome{}{}{s} - \noise{n}{s}
            ,
            \Vtreatment{n}{},\Vcovar{n}
            \big)
            \\
            \eqas
            &\;\E
            \Big[
            \psi
            \big(
            \AVO{}{}{1}
            +
            \VVO{}{}{1} Z_1
            ,
            \BAVO{}{}{1}
            +
            \BVVO{}{}{1} Z'_1
            ,
            \ldots,
            \AVO{}{}{s+1}
            +
            \VVO{}{}{s+1} Z_{s+1}
            ,
            \BAVO{}{}{s+1}
            +
            \BVVO{}{}{s+1} Z'_{s+1}
            ,
            \Vtreatment{}{},\Vcovar{}
            \big)
            \Big],
        \end{aligned}
    \end{equation}
    where $\psi: \R^{2(s+1)+T+M} \mapsto \R$ is within $\poly{k}$. Further, we write
    \begin{equation}
        \label{eq::BL-at-third result}
        \begin{aligned}
            &\lim_{N \rightarrow \infty}
            \frac{1}{N} \sum_{n=1}^N
            \Big(
            \outcomeg{0}{}\big(\outcome{}{n}{0},\Vtreatment{n}{},\Vcovar{n}\big)
            \phi\big(\outcome{}{n}{1}, \outcome{}{n}{1}-\IMatTv{n\cdot}{0}\VUoutcome{}{}{0} - \noise{n}{0},\ldots,
            \\
            &\quad\quad\quad\quad\quad\quad\quad\quad\quad\quad\quad\quad\quad\quad\quad
            \outcome{}{n}{s+1},\outcome{}{n}{s+1}-\IMatTv{n\cdot}{s}\VUoutcome{}{}{s}-\noise{n}{s},\Vtreatment{n}{},\Vcovar{n}\big)
            \Big)
            \\
            \eqas
            &\;\E
            \Big[
            \bar{g}_0(\Vtreatment{}{},\Vcovar{})
            \phi\big(
            \AVO{}{}{1}
            +
            \VVO{}{}{1} Z_1
            ,
            \BAVO{}{}{1}
            +
            \BVVO{}{}{1} Z'_1
            ,
            \ldots,
            \AVO{}{}{s+1}
            +
            \VVO{}{}{s+1} Z_{s+1}
            ,
            \BAVO{}{}{s+1}
            +
            \BVVO{}{}{s+1} Z'_{s+1}
            ,
            \Vtreatment{}{},\Vcovar{}
            \big)
            \Big],
        \end{aligned}
    \end{equation}
    where $\phi: \R^{2(s+1)+T+M} \mapsto \R$ is within $\poly{\frac{k}{2}}$.
    Below, we first prove \eqref{eq:BL-lower bound for perps}.
    \begin{enumerate}[label=(\alph*)]
        \item[(c)] Defining the function $\psi = g_s\big(\outcome{}{n}{s},\Vtreatment{n}{},\Vcovar{n}\big) g_r\big(\outcome{}{n}{r},\Vtreatment{n}{},\Vcovar{n}\big)$, by the induction hypothesis \ref{part:BL-a}, for  $1\leq r,s \leq t$, we have
        \begin{equation}
            \label{eq:BL-proof-ct-1}
            \begin{aligned}            
            \lim_{N \rightarrow \infty}
            \frac{1}{N}
            \sum_{n=1}^N
            \Uoutcome{n}{}{0} \Uoutcome{n}{}{s}
            &=
            \lim_{N \rightarrow \infty}
            \pdot{\VUoutcome{}{}{0}}{\VUoutcome{}{}{s}}
            \eqas
            \E\left[
            \bar{g}_0(\Vtreatment{}{},\Vcovar{}) \outcomeg{s}{}\big(\AVO{}{}{s}+\VVO{}{}{s} Z_s,\Vtreatment{}{},\Vcovar{}\big)
            \right]
            \\
            \lim_{N \rightarrow \infty}
            \frac{1}{N}
            \sum_{n=1}^N
            \Uoutcome{n}{}{s} \Uoutcome{n}{}{r}
            &=
            \lim_{N \rightarrow \infty}
            \pdot{\VUoutcome{}{}{s}}{\VUoutcome{}{}{r}}
            \eqas
            \E\left[
            \outcomeg{s}{}\big(\AVO{}{}{s}+\VVO{}{}{s} Z_s,\Vtreatment{}{},\Vcovar{}\big) \outcomeg{r}{}\big(\AVO{}{}{r}+\VVO{}{}{r} Z_r,\Vtreatment{}{},\Vcovar{}\big)
            \right].
            \end{aligned}
        \end{equation}
        Now, let $\Vec{u} = (u_1,\ldots,u_t)^\top \in \R^t$ be a non-zero vector. By \eqref{eq:BL-proof-ct-1}, we have
        \begin{equation}
            \label{eq:BL-proof-ct-2}
            \begin{aligned}
                \Vec{u}^\top \left(\lim_{N \rightarrow \infty} \frac{\bm{Q}_t^\top \bm{Q}_t}{N}\right) \Vec{u}
                &=
                \lim_{N \rightarrow \infty} \Vec{u}^\top \frac{\bm{Q}_t^\top \bm{Q}_t}{N} \Vec{u}
                \\
                &\eqas
                \E\left[
                \left(
                u_1
                \bar{g}_{0}(\Vtreatment{}{},\Vcovar{})
                +
                \sum_{s=2}^t
                u_s 
                \outcomeg{s-1}{}\big(\AVO{}{}{s-1}+\VVO{}{}{s-1} Z_{s-1},\Vtreatment{}{},\Vcovar{}\big)
                \right)^2
                \right]
                \\
                &\quad+
                u_1^2\left(\frac{\VVO{}{2}{1}-\sigma_e^2}{\sigma^2+\sigma_0^2}-\E\left[ \bar{g}_{0}(\Vtreatment{}{},\Vcovar{})^2 \right]\right).
            \end{aligned}
        \end{equation}
        By Assumption~\ref{asmp:BL}-\ref{asmp:BL-empirical dist of inits}, the last term in \eqref{eq:BL-proof-ct-2} is non-negative. Now, if $u_1 \neq 0 = u_2 = \ldots = u_t$, then the result is immediate by Assumption~\ref{asmp:BL}-\ref{asmp:BL-bound on initials}. Otherwise, there is some $2\leq i \leq t$ such that $u_i \neq 0$. Recalling that $y \mapsto \outcomeg{i-1}{}(y,\Vtreatment{}{},\Vcovar{})$ is a non-constant function with a positive probability with respect to $(\Vtreatment{}{},\Vcovar{}) \sim \pi \times p$, the mapping $(y_1,\ldots,y_t) \mapsto \sum_{s=1}^t u_s \outcomeg{s-1}{}\big(y_s,\Vtreatment{}{},\Vcovar{}\big)$ is a non-constant function. But, by the induction hypothesis \ref{part:BL-b}, it is straightforward to show that 
        \begin{equation*}
            \begin{aligned}
                \Cov\left[\left( \AVO{}{}{1}+\VVO{}{}{1}Z_1,\ldots, \AVO{}{}{t-1}+\VVO{}{}{t-1} Z_{t-1} \right)\right]
                \eqas
                \lim_{N \rightarrow \infty} \frac{V_{t-1}^\top V_{t-1}}{N} 
                -
                \lim_{N \rightarrow \infty} \frac{V_{t-1}^\top \ones{N\times 1}}{N}
                \lim_{N \rightarrow \infty} \frac{\ones{1\times N}V_{t-1}}{N},
            \end{aligned}
        \end{equation*}
        which is by the induction hypothesis positive definite. This implies that the random variable $u_1 \bar{g}_{0}(\Vtreatment{}{},\Vcovar{}) + \sum_{s=2}^t u_s \outcomeg{s-1}{}\big(\AVO{}{}{s-1}+\VVO{}{}{s-1} Z_{s-1},\Vtreatment{}{},\Vcovar{}\big)$ has a non-degenerate distribution. That means,
        \begin{align*}
            \forall \Vec{u}\neq 0,\;\; \Vec{u}^\top \left(\lim_{N \rightarrow \infty} \frac{\bm{Q}_t^\top \bm{Q}_t}{N}\right) \Vec{u} > 0
            \quad\quad
            \implies
            \quad\quad
            \lim_{N \rightarrow \infty} \frac{\bm{Q}_t^\top \bm{Q}_t}{N} \succ 0.
        \end{align*}
        For the second part, for $1\leq r,s \leq t$, let us denote
        \begin{align*}
            v_{r,s}
            :=
            \left[
            \frac{\bm{V}_t^\top \bm{V}_t}{N} 
            -
            \frac{\bm{V}_t^\top\ones{N\times 1}}{N}
            \frac{\ones{1\times N}\bm{V}_t}{N}
            \right]_{r,s}
            =
            \frac{\Voutcome{}{}{r}^\top \Voutcome{}{}{s}}{N} 
            -
            \frac{\Voutcome{}{}{r}^\top\ones{N\times 1}}{N}
            \frac{\ones{1\times N}\Voutcome{}{}{s}}{N}.
        \end{align*}
        By the induction hypothesis~\ref{part:BL-b}, if $r\neq s$, we have
        \begin{align*}
            \lim_{N \rightarrow \infty}
            v_{r,s}
            \eqas
            \AVO{}{}{r} \AVO{}{}{s} +
            \lim_{N \rightarrow \infty}
            \frac{\sigma^2}{N} \sum_{n=1}^N \Uoutcome{}{}{(r-1)n}\Uoutcome{}{}{(s-1)n}
            - \AVO{}{}{r} \AVO{}{}{s}
            = \lim_{N \rightarrow \infty}
            \frac{\sigma^2}{N} \sum_{n=1}^N \Uoutcome{}{}{(r-1)n}\Uoutcome{}{}{(s-1)n},
        \end{align*}
        and if $r=s$, we have
        \begin{align*}
            \lim_{N \rightarrow \infty}
            v_{r,r}
            \eqas
            \AVO{}{2}{r} +
            \lim_{N \rightarrow \infty}
            \frac{\sigma^2+\sigma_r^2}{N} \sum_{n=1}^N \Uoutcome{}{}{(r-1)n}\Uoutcome{}{}{(r-1)n} + \sigma_e^2
            - \AVO{}{2}{r}
            = \lim_{N \rightarrow \infty}
            \frac{\sigma^2+\sigma_r^2}{N} \sum_{n=1}^N \Uoutcome{2}{}{(r-1)n} + \sigma_e^2.
        \end{align*}
        Then, the result is straightforward as $\bm{Q}_t$ is positive definite.
        \begin{corollary}
            \label{cl:alpha is bounded}
            The vector $\Vec{\alpha}$ defined in \eqref{eq:projection coefficients} has a finite limit as $N \rightarrow \infty$.
        \end{corollary}
        Proof. By \eqref{eq:projection coefficients}, we can write
        \begin{align}
            \label{eq:cl-alpha is finite}
            \lim_{N \rightarrow \infty} \Vec{\alpha}_t
            =
            \lim_{N \rightarrow \infty} \left(
            \bm{Q}_t^\top \bm{Q}_t
            \right)^{-1}
            \bm{Q}_t^\top \VUoutcome{}{}{t}
            =
            \lim_{N \rightarrow \infty} \left(
            \frac{\bm{Q}_t^\top \bm{Q}_t}{N}
            \right)^{-1}
            \lim_{N \rightarrow \infty}
            \frac{\bm{Q}_t^\top \VUoutcome{}{}{t}}{N}.
        \end{align}    
        Using the result of part~\ref{part:BL-c}, for large values of $N$, the matrix $\frac{\bm{Q}_t^\top \bm{Q}_t}{N} \in \R^{t\times t}$ is positive definite (this is true because the eigenvalues of a matrix vary continuously with respect to its entries). Then, note that the mapping $\bm{M} \mapsto \bm{M}^{-1}$ is continuous for invertible matrices $\bm{M}$. As a result, we get
        \begin{align*}
            \lim_{N \rightarrow \infty} \left(\frac{\bm{Q}_t^\top \bm{Q}_t}{N}\right)^{-1}
            =
            \left(\lim_{N \rightarrow \infty}\frac{\bm{Q}_t^\top \bm{Q}_t}{N}\right)^{-1}.
        \end{align*}
        Since the matrix $\lim_{N \rightarrow \infty} \frac{\bm{Q}_t^\top \bm{Q}_t}{N}$ is positive definite, the first term in the RHS of \eqref{eq:cl-alpha is finite} is well-defined and finite. The finiteness of the other term is the consequence of \eqref{eq:BL-proof-ct-1}. \ep

        \item[(b)] Below, we first derive several minor results; then, we use them to show \eqref{eq:BL-b} holds true for $0\leq r,s \leq t$. Here, we use the SSLN given in Theorem~\ref{thm:SLLN} multiple times without checking the conditions as they are straightforward.

        Denote by $\IMatvnew{n\cdot}$ and $\IMatTv{n\cdot}{t}$ the $n^{th}$ rows of $\widetilde{\IM}$ and $\IMatT{t}$. Recalling Lemma~\ref{lm:conditional dist of outcome}, note that $\IMatvnew{n\cdot} \VUoutcome{\perp}{}{t}$ and $\IMatTv{n\cdot}{t} \VUoutcome{}{}{t}$ are Gaussian i.i.d. random variables:
        \begin{equation}
            \label{eq:BL-proof-bt-simple-1}
            \begin{aligned}
                \IMatvnew{n\cdot} \VUoutcome{\perp}{}{t} \sim \Nc\left( \frac{\mu}{N} \sum_{i=1}^N \Uoutcome{\perp i}{}{t}, \frac{\sigma^2}{N} \sum_{i=1}^N (\Uoutcome{\perp i}{}{t})^2\right),
            \quad\quad
                \IMatTv{n\cdot}{t} \VUoutcome{}{}{t} \sim \Nc\left( \frac{\mu_t}{N} \sum_{i=1}^N \Uoutcome{i}{}{t}, \frac{\sigma_t^2}{N} \sum_{i=1}^N (\Uoutcome{i}{}{t})^2\right),
            \end{aligned}
        \end{equation}
        where $\Uoutcome{n}{}{t} = \outcomeg{t}{}\big(\outcome{}{n}{t},\Vtreatment{n}{},\Vcovar{n}\big)$ is the $n^{th}$ element of the vector $\VUoutcome{}{}{t}$ and $\Uoutcome{\perp n}{}{t}$ is the $n^{th}$ element of the vector $\VUoutcome{\perp}{}{t}$. Applying Theorem~\ref{thm:SLLN}, we get
        \begin{align}
            \label{eq:BL-proof-bt-simple-2}
            \lim_{N \rightarrow \infty}
            \frac{1}{N}
            \sum_{n=1}^N
            \IMatvnew{n\cdot} 
            \VUoutcome{\perp}{}{t}
            \eqas
            \lim_{N \rightarrow \infty}
            \frac{\mu}{N}
            \sum_{n=1}^N
            \Uoutcome{\perp n}{}{t}, 
        \end{align}
        and similarly
        \begin{align}
            \label{eq:BL-proof-bt-simple-3}
            \lim_{N \rightarrow \infty}
            \frac{1}{N}
            \sum_{n=1}^N
            \IMatTv{n\cdot}{t} 
            \VUoutcome{}{}{t}
            \eqas
            \lim_{N \rightarrow \infty}
            \frac{\mu_t}{N}
            \sum_{n=1}^N
            \Uoutcome{n}{}{t}.
        \end{align}
        Similarly, \eqref{eq:BL-proof-bt-simple-1} implies that:
        \begin{equation}
            \label{eq:BL-proof-bt-simple-4}
            \begin{aligned}
                \lim_{N \rightarrow \infty}
                \frac{1}{N}
                \sum_{n=1}^N
                \left(
                \IMatvnew{n\cdot} 
                \VUoutcome{\perp}{}{t}
                \right)^2
                &\eqas
                \lim_{N \rightarrow \infty}
                \left( \frac{\mu}{N}
                \sum_{n=1}^N
                \Uoutcome{\perp n}{}{t}\right)^2 + 
                \lim_{N \rightarrow \infty} \frac{\sigma^2}{N}
                \sum_{n=1}^N
                \left(\Uoutcome{\perp n}{}{t}\right)^2,
            \end{aligned}
        \end{equation}
        as well as
        \begin{equation}
            \label{eq:BL-proof-bt-simple-5}
            \begin{aligned}
                \lim_{N \rightarrow \infty}
                \frac{1}{N}
                \sum_{n=1}^N
                \left(
                \IMatTv{n\cdot}{t}
                \VUoutcome{}{}{t}
                \right)^2
                &\eqas
                \lim_{N \rightarrow \infty}
                \left( \frac{\mu_t}{N}
                \sum_{n=1}^N
                \Uoutcome{n}{}{t}\right)^2 + 
                \lim_{N \rightarrow \infty} \frac{\sigma_t^2}{N}
                \sum_{n=1}^N
                \left(\Uoutcome{n}{}{t}\right)^2.
            \end{aligned}
        \end{equation}
        Next, by \eqref{eq:Q and R}, the induction hypothesis and Theorem~\ref{thm:SLLN}, we have
        \begin{equation}
        \begin{aligned}
            \label{eq:BL-proof-bt-simple-6}
            \lim_{N \rightarrow \infty}
            \frac{1}{N}
            \sum_{n=1}^N
            \left[
            \bm{R}_t
            \Vec{\alpha}_t
            \right]_n
            &=
            \lim_{N \rightarrow \infty}
            \frac{1}{N}
            \sum_{n=1}^N
            \sum_{s=0}^{t-1}
            \alpha_s
            \big(
            \outcome{}{n}{s+1}
            - \IMatTv{n\cdot}{s} \Uoutcome{n}{}{s}
            - \noise{n}{s}
            \big)
            \\
            &=
            \sum_{s=0}^{t-1}
            \alpha_s
            \left(
            \lim_{N \rightarrow \infty}
            \frac{1}{N}
            \sum_{n=1}^N
            \left(
            \outcome{}{n}{s+1}
            - \IMatTv{n\cdot}{s} \Uoutcome{n}{}{s}
            - \noise{n}{s}
            \right)
            \right)
            \\
            &\eqas
            \sum_{s=0}^{t-1}
            \alpha_s
            \left(
            \lim_{N \rightarrow \infty}
            \frac{\mu}{N}
            \sum_{n=1}^N
            \Uoutcome{n}{}{s}
            \right)
            \\
            &=
            \lim_{N \rightarrow \infty}
            \frac{\mu}{N}
            \sum_{n=1}^N
            \Uoutcome{\parallel n}{}{t},
        \end{aligned}
        \end{equation}
        where in the last equality we used \eqref{eq:projection sum}. 
Considering \eqref{eq:Q and R} and the induction hypothesis \eqref{eq:BL-b-5}, it yields
        \begin{equation}
        \label{eq:BL-proof-bt-simple-7}
            \begin{aligned}
                &\lim_{N \rightarrow \infty}
                \frac{1}{N}
                \sum_{n=1}^N
                \left(
                \left[
                \bm{R}_t
                \Vec{\alpha}_t
                \right]_n
                \right)^2
                \\
                =
                &\lim_{N \rightarrow \infty}
                \frac{1}{N}
                \sum_{n=1}^N
                \left(
                \sum_{s=0}^{t-1}
                \alpha_s
                \big(
                \outcome{}{n}{s+1}
                - \IMatTv{n\cdot}{s} \VUoutcome{}{}{s}
                - \noise{n}{s}
                \big)
                \right)^2
                \\
                =
                &\lim_{N \rightarrow \infty}
                \frac{1}{N}
                \sum_{n=1}^N
                \sum_{0\leq s,r < t}
                \alpha_s
                \alpha_r
                \big(
                \outcome{}{n}{s+1} - \IMatTv{n\cdot}{s} \VUoutcome{}{}{s} - \noise{n}{s}
                \big)
                \big(
                \outcome{}{n}{r+1} - \IMatTv{n\cdot}{r} \VUoutcome{}{}{r} - \noise{n}{r}
                \big)
                \\
                =
                &\sum_{0\leq s,r < t}
                \alpha_s
                \alpha_r
                \left(
                \lim_{N \rightarrow \infty}
                \frac{1}{N}
                \sum_{n=1}^N
                \big(
                \outcome{}{n}{s+1} - \IMatTv{n\cdot}{s} \VUoutcome{}{}{s} - \noise{n}{s}
                \big)
                \big(
                \outcome{}{n}{r+1} - \IMatTv{n\cdot}{r} \VUoutcome{}{}{r} - \noise{n}{r}
                \big)\right)
                \\
                \eqas
                &\lim_{N \rightarrow \infty} \frac{\mu^2}{N^2}
                \left(
                \sum_{n=1}^N
                \sum_{0\leq s <t}
                \alpha_s
                \Uoutcome{n}{}{s}
                \right)
                \left(
                \sum_{n=1}^N
                \sum_{0\leq r <t}
                \alpha_r
                \Uoutcome{n}{}{r}
                \right)
                +
                \lim_{N \rightarrow \infty}
                \frac{\sigma^2}{N}
                \sum_{n=1}^N
                \sum_{0\leq s,r < t}
                \alpha_s
                \alpha_r
                \Uoutcome{n}{}{s} \Uoutcome{n}{}{r}
                \\
                =
                &\lim_{N \rightarrow \infty}
                \left(\frac{\mu}{N} \sum_{n=1}^{N}
                \Uoutcome{\parallel n}{}{t} \right)^2
                +
                \lim_{N \rightarrow \infty}
                \frac{\sigma^2}{N}
                \sum_{n=1}^N
                \left(\Uoutcome{\parallel n}{}{t}\right)^2,
            \end{aligned}
        \end{equation}
        where in the last line we used \eqref{eq:projection sum}. Further, because the elements of $\Vnoise{n}{t}$ are independent zero mean Gaussian random variables, a straightforward application of Theorem~\ref{thm:SLLN} implies that for any term $\delta$ in this proof, we have:
        \begin{align}
            \label{eq:BL-proof-bt-simple-eps1}
            \lim_{N \rightarrow \infty}
                \frac{1}{N}
                \sum_{n=1}^N
                \delta \noise{n}{t}
                \eqas
                0,
        \end{align}
        and
        \begin{align}
            \label{eq:BL-proof-bt-simple-eps2}
            \lim_{N \rightarrow \infty}
                \frac{1}{N}
                \sum_{n=1}^N
                \big(\noise{n}{t}\big)^2
                \eqas
                \sigma_e^2.
        \end{align}

        Now, we first obtain \eqref{eq:BL-b-1} for $s=t$.  Based on \eqref{eq:conditional dist of outcome_nonsym}, we can write
        \begin{equation}
            \label{eq:BL-proof-bt-1}
            \begin{aligned}
                \lim_{N \rightarrow \infty}
                \frac{1}{N}
                \sum_{n=1}^N
                \outcome{}{n}{t+1}
                &\eqas
                \lim_{N \rightarrow \infty}
                \frac{1}{N}
                \sum_{n=1}^N
                \left(
                \IMatvnew{n\cdot} 
                \VUoutcome{\perp}{}{t}
                +
                \left[
                \bm{R}_t
                \Vec{\alpha}_t
                \right]_n
                +
                \IMatTv{n\cdot}{t} \VUoutcome{}{}{t}
                +
                \noise{n}{t}
                \right)
                \\
                &\eqas
                \lim_{N \rightarrow \infty}
                \frac{\mu+\mu_t}{N}
                \sum_{n=1}^N
                \outcomeg{t}{}\big(\outcome{}{n}{t},\Vtreatment{n}{},\Vcovar{n}\big)
                \\
                &\eqas
                (\mu+\mu_t)
                \E
                \left[
                \outcomeg{t}{}\big(\AVO{}{}{t} + \VVO{}{}{t} Z, \Vtreatment{}{},\Vcovar{}\big)
                \right] = \AVO{}{}{t+1},
        \end{aligned}
        \end{equation}
        where we used the induction hypothesis~\ref{part:BL-a}, \eqref{eq:BL-proof-bt-simple-2}, \eqref{eq:BL-proof-bt-simple-3}, \eqref{eq:BL-proof-bt-simple-6}, and \eqref{eq:BL-proof-bt-simple-eps1}. The derivation of \eqref{eq:BL-b-2} is similar and is removed to avoid repetition.

        We next obtain \eqref{eq:BL-b-3} for $s=r=t$. By \eqref{eq:conditional dist of outcome_nonsym}, \eqref{eq:BL-proof-bt-simple-4}, \eqref{eq:BL-proof-bt-simple-5}, \eqref{eq:BL-proof-bt-simple-7}, \eqref{eq:BL-proof-bt-simple-eps1}, and \eqref{eq:BL-proof-bt-simple-eps2}, we can write
        \begin{equation}
            \label{eq:BL-proof-bt-2-1}
            \begin{aligned}
                \lim_{N \rightarrow \infty}
                \frac{1}{N}
                \sum_{n=1}^N
                \big(
                \outcome{}{n}{t+1}
                \big)^2
                &\eqas
                \lim_{N \rightarrow \infty}
                \frac{1}{N}
                \sum_{n=1}^N
                \left(
                \IMatvnew{n\cdot} 
                \VUoutcome{\perp}{}{t}
                +
                \left[
                \bm{R}_t
                \Vec{\alpha}_t
                \right]_n
                +
                \IMatTv{n\cdot}{t} \VUoutcome{}{}{t}
                +
                \noise{n}{t}
                \right)^2
                \\
                &\eqas
                \lim_{N \rightarrow \infty}
                \left( \frac{\mu}{N}
                \sum_{n=1}^N
                \Uoutcome{\perp n}{}{t}\right)^2 + 
                \lim_{N \rightarrow \infty} \frac{\sigma^2}{N}
                \sum_{n=1}^N
                \left(\Uoutcome{\perp n}{}{t}\right)^2,
                \\
                &\quad+
                \lim_{N \rightarrow \infty}
                \left(\frac{\mu}{N} \sum_{n=1}^{N}
                \Uoutcome{\parallel n}{}{t} \right)^2
                +
                \lim_{N \rightarrow \infty}
                \frac{\sigma^2}{N}
                \sum_{n=1}^N
                \left(\Uoutcome{\parallel n}{}{t}\right)^2
                \\
                &\quad +
                \lim_{N \rightarrow \infty}
                \left( \frac{\mu_t}{N}
                \sum_{n=1}^N
                \Uoutcome{n}{}{t}\right)^2 + 
                \lim_{N \rightarrow \infty} \frac{\sigma_t^2}{N}
                \sum_{n=1}^N
                \left(\Uoutcome{n}{}{t}\right)^2
                +
                \sigma_e^2
                \\
                &\quad
                +
                \lim_{N \rightarrow \infty}
                \frac{2}{N}
                \sum_{n=1}^N
                \left(\IMatvnew{n\cdot} 
                \VUoutcome{\perp}{}{t} \left[
                \bm{R}_t
                \Vec{\alpha}_t
                \right]_n \right)
                +
                \lim_{N \rightarrow \infty}
                \frac{2}{N}
                \sum_{n=1}^N
                \left(\IMatvnew{n\cdot} 
                \VUoutcome{\perp}{}{t} \IMatTv{n\cdot}{t} \VUoutcome{}{}{t} \right)
                \\
                &\quad
                +
                \lim_{N \rightarrow \infty}
                \frac{2}{N}
                \sum_{n=1}^N
                \left( \left[
                \bm{R}_t
                \Vec{\alpha}_t
                \right]_n \IMatTv{n\cdot}{t}\VUoutcome{}{}{t} \right).
            \end{aligned}
        \end{equation}
        Note that the only random objects in the right-hand side of \eqref{eq:BL-proof-bt-2-1} are $\IMatvnew{n\cdot} $ and $\IMatTv{n\cdot}{t}$. Thus, by \eqref{eq:Q and R}, an argument similar to \eqref{eq:BL-proof-bt-simple-1}, and applying Theorem~\ref{thm:SLLN}, we can write
        \begin{equation}
        \label{eq:BL-proof-bt-2-AR term}
            \begin{aligned}
                \lim_{N \rightarrow \infty}
                \frac{2}{N}
                \sum_{n=1}^N
                \left(\IMatvnew{n\cdot} 
                \VUoutcome{\perp}{}{t} \left[
                \bm{R}_t
                \Vec{\alpha}_t
                \right]_n \right)
                &=
                \lim_{N \rightarrow \infty}
                \frac{2}{N}
                \sum_{n=1}^N
                \sum_{s=0}^{t-1}
                \alpha_s
                \IMatvnew{n\cdot} 
                \VUoutcome{\perp}{}{t}
                \big( \outcome{}{n}{s+1}
                - \IMatTv{n\cdot}{s} \VUoutcome{}{}{s}
                - \noise{n}{s} \big)
                \\
                &\eqas
                \lim_{N \rightarrow \infty}
                \frac{2}{N}
                \sum_{n=1}^N
                \frac{\mu}{N}
                \sum_{m=1}^N
                \Uoutcome{\perp m}{}{t}
                \left(
                \sum_{s=0}^{t-1}
                \alpha_s
                \big( \outcome{}{n}{s+1}
                - \IMatTv{n\cdot}{s} \VUoutcome{}{}{s}
                - \noise{n}{s} \big)\right)
                \\
                &\eqas
                \lim_{N \rightarrow \infty}
                \frac{2\mu}{N}
                \left(
                \sum_{n=1}^N
                \Uoutcome{\perp n}{}{t}
                \right)
                \lim_{N \rightarrow \infty}
                \frac{\mu}{N}
                \left(
                \sum_{n=1}^N
                \Uoutcome{\parallel n}{}{t}
                \right),
            \end{aligned}
        \end{equation}
        where in the last line we used the induction hypothesis \eqref{eq:BL-b-2} and \eqref{eq:projection sum}. Likewise, we get
        \begin{equation}
        \label{eq:BL-proof-bt-2-BR term}
            \begin{aligned}
                \lim_{N \rightarrow \infty}
                \frac{2}{N}
                \sum_{n=1}^N
                \left(\IMatTv{n\cdot}{t} 
                \VUoutcome{}{}{t} \left[
                \bm{R}_t
                \Vec{\alpha}_t
                \right]_n \right)
                \eqas
                \lim_{N \rightarrow \infty}
                \frac{2\mu_t}{N}
                \left(
                \sum_{n=1}^N
                \Uoutcome{n}{}{t}
                \right)
                \lim_{N \rightarrow \infty}
                \frac{\mu}{N}
                \left(
                \sum_{n=1}^N
                \Uoutcome{\parallel n}{}{t}
                \right),
            \end{aligned}
        \end{equation}
        We continue the proof by calculating the next term using \eqref{eq:BL-proof-bt-simple-1} and Theorem~\ref{thm:SLLN}:
        \begin{equation}
            \label{eq:BL-proof-bt-2-AB term}
            \begin{aligned}
                \lim_{N \rightarrow \infty}
                \frac{2}{N}
                \sum_{n=1}^N
                \left(\IMatvnew{n\cdot} 
                \VUoutcome{\perp}{}{t} \IMatTv{n\cdot}{t} \VUoutcome{}{}{t} \right)
                \eqas
                \lim_{N \rightarrow \infty}
                \frac{2\mu \mu_t}{N^2}
                \sum_{n=1}^N
                \Uoutcome{\perp n}{}{t}
                \sum_{n=1}^N
                \Uoutcome{n}{}{t}.
            \end{aligned}
        \end{equation}
        Combining \eqref{eq:BL-proof-bt-2-1}-\eqref{eq:BL-proof-bt-2-AB term} yields:
        \begin{align*}
            \lim_{N \rightarrow \infty}
            \frac{1}{N}
            \sum_{n=1}^N
            \big(
            \outcome{}{n}{t+1}
            \big)^2
            &\eqas
            \lim_{N \rightarrow \infty}
            \left( \frac{\mu+\mu_t}{N}
            \sum_{n=1}^N
            \Uoutcome{n}{}{t}\right)^2 + 
            \lim_{N \rightarrow \infty} \frac{\sigma^2+\sigma_t^2}{N}
            \sum_{n=1}^N
            \left(\Uoutcome{n}{}{t}\right)^2 + \sigma_e^2.
        \end{align*}
        The desired result is immediate by applying the induction hypothesis~\ref{part:BL-a}.
        We next derive \eqref{eq:BL-b-3-rs} for $r=t$ and $0\leq s \leq t-1$. Considering \eqref{eq:conditional dist of outcome_nonsym} and \eqref{eq:BL-proof-bt-simple-eps1}, we can write
        \begin{equation}
        \label{eq:BL-proof-bt-3-1}
            \begin{aligned}
                &\lim_{N \rightarrow \infty}
                \frac{1}{N}
                \sum_{n=1}^N
                \outcome{}{n}{s+1}
                \outcome{}{n}{t+1}
                \\
                \eqas
                &\lim_{N \rightarrow \infty}
                \frac{1}{N}
                \sum_{n=1}^N
                \outcome{}{n}{s+1}
                \left(
                \IMatvnew{n\cdot} 
                \VUoutcome{\perp}{}{t}
                +
                \left[
                \bm{R}_t
                \Vec{\alpha}_t
                \right]_n
                + \IMatTv{n\cdot}{t} \VUoutcome{}{}{t}
                + \noise{n}{t}
                \right)
                \\
                \eqas
                &\lim_{N \rightarrow \infty}
                \frac{1}{N}
                \sum_{n=1}^N
                \left(
                \left(
                \IMatvnew{n\cdot} 
                \VUoutcome{\perp}{}{t}
                \right)\outcome{}{n}{s+1}
                +
                \left(
                \left[
                \bm{R}_t
                \Vec{\alpha}_t
                \right]_n
                \right)\outcome{}{n}{s+1}
                +
                \left(
                \IMatTv{n\cdot}{t} 
                \VUoutcome{}{}{t}
                \right)\outcome{}{n}{s+1}
                \right).
            \end{aligned}
        \end{equation}
        Note that after conditioning on $\Gc_t$, the quantity $\outcome{}{n}{s+1}$ is known. Then, applying Theorem~\ref{thm:SLLN} and considering \eqref{eq:BL-proof-bt-simple-1}, we obtain
        \begin{equation}
            \label{eq:BL-proof-bt-3-2}
            \begin{aligned}
                \lim_{N \rightarrow \infty}
                \frac{1}{N}
                \sum_{n=1}^N
                \left(
                \IMatvnew{n\cdot} 
                \VUoutcome{\perp}{}{t}
                \right)\outcome{}{n}{s+1}
                &\eqas
                \lim_{N \rightarrow \infty}
                \frac{\mu}{N^2}
                \sum_{n=1}^N
                \left(
                \outcome{}{n}{s+1}
                \sum_{m=1}^N
                \Uoutcome{\perp m}{}{t}
                \right)
                \\
                &=
                \lim_{N \rightarrow \infty}
                \frac{\mu}{N^2}
                \left(
                \sum_{n=1}^N
                \outcome{}{n}{s+1}
                \sum_{n=1}^N
                \Uoutcome{\perp n}{}{t}
                \right)
                \\
                &\eqas
                \lim_{N \rightarrow \infty}
                \frac{\mu (\mu + \mu_s)}{N^2}
                \left(
                \sum_{n=1}^N
                \Uoutcome{n}{}{s}
                \sum_{n=1}^N
                \Uoutcome{\perp n}{}{t}
                \right).
            \end{aligned}
        \end{equation}
        In a similar fashion, we can see that
        \begin{equation}
            \label{eq:BL-proof-bt-3-3}
            \begin{aligned}
                \lim_{N \rightarrow \infty}
                \frac{1}{N}
                \sum_{n=1}^N
                \left(
                \IMatTv{n\cdot}{t} 
                \VUoutcome{}{}{t}
                \right)\outcome{}{n}{s+1}
                &\eqas
                \lim_{N \rightarrow \infty}
                \frac{\mu_t}{N^2}
                \sum_{n=1}^N
                \left(
                \outcome{}{n}{s+1}
                \sum_{m=1}^N
                \Uoutcome{m}{}{t}
                \right)
                \\
                &=
                \lim_{N \rightarrow \infty}
                \frac{\mu_t}{N^2}
                \left(
                \sum_{n=1}^N
                \outcome{}{n}{s+1}
                \sum_{n=1}^N
                \Uoutcome{n}{}{t}
                \right)
                \\
                &\eqas
                \lim_{N \rightarrow \infty}
                \frac{\mu_t (\mu + \mu_s)}{N^2}
                \left(
                \sum_{n=1}^N
                \Uoutcome{n}{}{s}
                \sum_{n=1}^N
                \Uoutcome{n}{}{t}
                \right).
            \end{aligned}
        \end{equation}
        Lastly, considering \eqref{eq:Q and R}, by the induction hypothesis \eqref{eq:BL-b-4}, we get
        \begin{equation}
            \label{eq:BL-proof-bt-3-4}
            \begin{aligned}
                \lim_{N \rightarrow \infty}
                \frac{1}{N}
                \sum_{n=1}^N
                \left(
                \left[
                \bm{R}_t
                \Vec{\alpha}_t
                \right]_n
                \right)\outcome{}{n}{s+1}
                &=
                \lim_{N \rightarrow \infty}
                \frac{1}{N}
                \sum_{n=1}^N
                \left(
                \sum_{r=0}^{t-1}
                \alpha_r
                \big(
                \outcome{}{n}{r+1}
                - \IMatTv{n\cdot}{r} \VUoutcome{}{}{r}
                - \noise{n}{r}
                \big)
                \outcome{}{n}{s+1}
                \right)
                \\
                &\eqas
                \sum_{r=0}^{t-1}
                \alpha_r 
                \left(
                \lim_{N \rightarrow \infty}
                \frac{\mu(\mu+\mu_s)}{N^2}
                \sum_{n=1}^N \Uoutcome{n}{}{r}
                \sum_{n=1}^N \Uoutcome{n}{}{s}
                +
                \lim_{N \rightarrow \infty}
                \frac{\sigma^2}{N} \sum_{n=1}^N \Uoutcome{n}{}{r}\Uoutcome{n}{}{s}
                \right)
                \\
                &\eqas
                \lim_{N \rightarrow \infty}\hspace{-1mm}
                \left(
                \frac{\mu(\mu+\mu_s)}{N^2} \sum_{n=1}^N
                \sum_{r=0}^{t-1}
                \alpha_r \Uoutcome{n}{}{r}
                \sum_{n=1}^N \Uoutcome{n}{}{s}
                +
                \frac{\sigma^2}{N} \sum_{n=1}^N \sum_{r=0}^{t-1}
                \alpha_r
                \Uoutcome{n}{}{r}\Uoutcome{n}{}{s}\right)
                \\
                &=
                \lim_{N \rightarrow \infty}
                \frac{\mu(\mu+\mu_s)}{N^2}
                \left(
                \sum_{n=1}^N
                \Uoutcome{n}{}{s}
                \sum_{n=1}^N
                \Uoutcome{\parallel n}{}{t}
                \right)
                +
                \lim_{N \rightarrow \infty}
                \frac{\sigma^2}{N}
                \sum_{n=1}^N
                \left(
                \Uoutcome{\parallel n}{}{t}
                \Uoutcome{n}{}{s}
                \right)
                \\
                &=
                \lim_{N \rightarrow \infty}
                \frac{\mu(\mu+\mu_s)}{N^2}
                \left(
                \sum_{n=1}^N
                \Uoutcome{n}{}{s}
                \sum_{n=1}^N
                \Uoutcome{\parallel n}{}{t}
                \right)
                +
                \lim_{N \rightarrow \infty}
                \frac{\sigma^2}{N}
                \sum_{n=1}^N
                \left(
                \Uoutcome{n}{}{t}
                \Uoutcome{n}{}{s}
                \right).
            \end{aligned}
        \end{equation}
        In the last line of \eqref{eq:BL-proof-bt-3-4}, we used the fact that $\pdot{\VUoutcome{}{}{t}}{\VUoutcome{}{}{s}} = \pdot{\VUoutcome{\parallel}{}{t}}{\VUoutcome{}{}{s}}$ as $\VUoutcome{\perp}{}{t} \perp \VUoutcome{}{}{s}$. Considering \eqref{eq:BL-proof-bt-3-1}-\eqref{eq:BL-proof-bt-3-4} together concludes the proof of \eqref{eq:BL-b-3-rs}.

        We follow a similar approach as above to obtain \eqref{eq:BL-b-4}. Fixing $0 \leq r \leq t-1$ and letting $s=t$, by \eqref{eq:conditional dist of outcome_nonsym}, \eqref{eq:BL-proof-bt-3-2}, and \eqref{eq:BL-proof-bt-3-4}, we can write
        \begin{equation*}
            \begin{aligned}
                &\lim_{N \rightarrow \infty}
                \frac{1}{N}
                \sum_{n=1}^N
                \big(
                \outcome{}{n}{t+1}
                - \IMatTv{n\cdot}{t} \VUoutcome{}{}{t}
                - \noise{n}{t}
                \big)
                \outcome{}{n}{r+1}
                \\
                \eqas
                &\lim_{N \rightarrow \infty}
                \frac{1}{N}
                \sum_{n=1}^N
                \left(
                \IMatvnew{n\cdot} 
                \VUoutcome{\perp}{}{t}
                +
                \left[
                \bm{R}_t
                \Vec{\alpha}_t
                \right]_n
                \right)\outcome{}{n}{r+1}
                \\
                \eqas
                &\lim_{N \rightarrow \infty}
                \frac{\mu (\mu + \mu_r)}{N^2}
                \left(
                \sum_{n=1}^N
                \Uoutcome{n}{}{r}
                \sum_{n=1}^N
                \Uoutcome{\perp n}{}{t}
                \right)
                \\
                &+
                \lim_{N \rightarrow \infty}
                \frac{\mu(\mu+\mu_r)}{N^2}
                \left(
                \sum_{n=1}^N
                \Uoutcome{n}{}{r}
                \sum_{n=1}^N
                \Uoutcome{\parallel n}{}{t}
                \right)
                +
                \lim_{N \rightarrow \infty}
                \frac{\sigma^2}{N}
                \sum_{n=1}^N
                \left(
                \Uoutcome{n}{}{t}
                \Uoutcome{n}{}{r}
                \right)
            \end{aligned}
        \end{equation*}
        Likewise, we can show the result for the case that $r=t$ and $0\leq s \leq t-1$:
        \begin{equation}
            \label{eq:BL-proof-bt-4-1}
            \begin{aligned}
                &\lim_{N \rightarrow \infty}
                \frac{1}{N}
                \sum_{n=1}^N
                \big(
                \outcome{}{n}{s+1}
                - \IMatTv{n\cdot}{s} \VUoutcome{}{}{s}
                - \noise{n}{s}
                \big)
                \outcome{}{n}{t+1}
                \\
                \eqas
                &\lim_{N \rightarrow \infty}
                \frac{1}{N}
                \sum_{n=1}^N
                \big(
                \outcome{}{n}{s+1}
                - \IMatTv{n\cdot}{s} \VUoutcome{}{}{s}
                - \noise{n}{s}
                \big)
                \left(
                \IMatvnew{n\cdot} 
                \VUoutcome{\perp}{}{t}
                +
                \left[
                \bm{R}_t
                \Vec{\alpha}_t
                \right]_n
                + \IMatTv{n\cdot}{t} \VUoutcome{}{}{t}
                + \noise{n}{t}
                \right)
                \\
                =
                &\lim_{N \rightarrow \infty}
                \frac{1}{N}
                \sum_{n=1}^N
                \left(
                \IMatvnew{n\cdot} 
                \VUoutcome{\perp}{}{t}
                \right)
                \big(
                \outcome{}{n}{s+1}
                - \IMatTv{n\cdot}{s} \VUoutcome{}{}{s}
                - \noise{n}{s}
                \big)
                \\
                &+
                \lim_{N \rightarrow \infty}
                \frac{1}{N}
                \sum_{n=1}^N
                \left(
                \left[
                \bm{R}_t
                \Vec{\alpha}_t
                \right]_n
                \right)
                \big(
                \outcome{}{n}{s+1}
                - \IMatTv{n\cdot}{s} \VUoutcome{}{}{s}
                - \noise{n}{s}
                \big)
                \\
                &+
                \lim_{N \rightarrow \infty}
                \frac{1}{N}
                \sum_{n=1}^N
                \left(
                \IMatTv{n\cdot}{t} \VUoutcome{}{}{t}
                \right)
                \big(
                \outcome{}{n}{s+1}
                - \IMatTv{n\cdot}{s} \VUoutcome{}{}{s}
                - \noise{n}{s}
                \big),
            \end{aligned}
        \end{equation}
        where we also used \eqref{eq:BL-proof-bt-simple-eps1}. By Theorem~\ref{thm:SLLN} and the induction hypothesis \eqref{eq:BL-b-2}, we have
        \begin{equation}
            \label{eq:BL-proof-bt-4-2}
            \begin{aligned}
                \lim_{N \rightarrow \infty}
                \frac{1}{N}
                \sum_{n=1}^N
                \left(
                \IMatvnew{n\cdot} 
                \VUoutcome{\perp}{}{t}
                \right)
                \big(
                \outcome{}{n}{s+1}
                - \IMatTv{n\cdot}{s} \VUoutcome{}{}{s}
                - \noise{n}{s}
                \big)
                \eqas
                \lim_{N \rightarrow \infty}
                \frac{\mu^2}{N^2}
                \sum_{n=1}^N
                \Uoutcome{\perp n}{}{t}
                \sum_{n=1}^N
                \Uoutcome{n}{}{s},
            \end{aligned}
        \end{equation}
        as well as,
        \begin{equation}
            \label{eq:BL-proof-bt-4-3}
            \begin{aligned}
                \lim_{N \rightarrow \infty}
                \frac{1}{N}
                \sum_{n=1}^N
                \left(
                \IMatTv{n\cdot}{t} 
                \VUoutcome{}{}{t}
                \right)
                \big(
                \outcome{}{n}{s+1}
                - \IMatTv{n\cdot}{s} \VUoutcome{}{}{s}
                - \noise{n}{s}
                \big)
                \eqas
                \lim_{N \rightarrow \infty}
                \frac{\mu\mu_t}{N^2}
                \sum_{n=1}^N
                \Uoutcome{n}{}{t}
                \sum_{n=1}^N
                \Uoutcome{n}{}{s}.
            \end{aligned}
        \end{equation}
        Finally, \eqref{eq:Q and R}, \eqref{eq:projection sum}, and the induction hypothesis \eqref{eq:BL-b-5} imply that
        \begin{equation}
        \label{eq:BL-proof-bt-4-4}
            \begin{aligned}
                &\lim_{N \rightarrow \infty}
                \frac{1}{N}
                \sum_{n=1}^N
                \left(
                \left[
                \bm{R}_t
                \Vec{\alpha}_t
                \right]_n
                \right)
                \big(
                \outcome{}{n}{s+1}
                - \IMatTv{n\cdot}{s} \VUoutcome{}{}{s}
                - \noise{n}{s}
                \big)
                \\
                =
                &\lim_{N \rightarrow \infty}
                \frac{1}{N}
                \sum_{n=1}^N
                \sum_{r=0}^{t-1}
                \alpha_r
                \big(
                \outcome{}{n}{r+1}
                - \IMatTv{n\cdot}{r} \VUoutcome{}{}{r}
                - \noise{n}{r}
                \big)
                \big(
                \outcome{}{n}{s+1}
                - \IMatTv{n\cdot}{s} \VUoutcome{}{}{s}
                - \noise{n}{s}
                \big)
                \\
                \eqas
                &\lim_{N \rightarrow \infty}
                \frac{\mu^2}{N^2} \left(\sum_{n=1}^N \Uoutcome{n}{}{s}\right)
                \left(\sum_{n=1}^N \Uoutcome{\parallel n}{}{t}\right)
                +
                \lim_{N \rightarrow \infty}
                \frac{\sigma^2}{N}
                \sum_{n=1}^N
                \Uoutcome{n}{}{s} \Uoutcome{n}{}{t},
            \end{aligned}
        \end{equation}
        where in the last line, we used $\pdot{\VUoutcome{}{}{t}}{\VUoutcome{}{}{s}} = \pdot{\VUoutcome{\parallel}{}{t}}{\VUoutcome{}{}{s}}$. The desired result follows by aggregating \eqref{eq:BL-proof-bt-4-1}-\eqref{eq:BL-proof-bt-4-4}.

        To conclude the proof of part~\ref{part:BL-b}, we need to show \eqref{eq:BL-b-5} for $r=s=t$ as well as $s=t,\; 0\leq r<t-1$. If $r=s=t$, then considering \eqref{eq:conditional dist of outcome_nonsym}, the result immediately follows from \eqref{eq:BL-proof-bt-simple-4}, \eqref{eq:BL-proof-bt-simple-7}, and \eqref{eq:BL-proof-bt-2-AR term}. For the case that $s=t,\; 0\leq r<t-1$, by \eqref{eq:conditional dist of outcome_nonsym}, we have
        \begin{equation}
        \label{eq:BL-proof-bt-5}
            \begin{aligned}
                &\lim_{N \rightarrow \infty}
                \frac{1}{N}
                \sum_{n=1}^N
                \big(
                \outcome{}{n}{t+1}
                - \IMatTv{n\cdot}{t} \VUoutcome{}{}{t}
                -
                \noise{n}{t} \big)
                \big(
                \outcome{}{n}{r+1}
                - \IMatTv{n\cdot}{r} \VUoutcome{}{}{r}
                -
                \noise{n}{r} \big)
                \\
                =
                &\lim_{N \rightarrow \infty}
                \frac{1}{N}
                \sum_{n=1}^N
                \big(
                \IMatvnew{n\cdot} \Uoutcome{\perp}{}{t} 
                +
                \left[\bm{R}_t\Vec{\alpha}_t\right]_n
                \big)
                \big(
                \outcome{}{n}{r+1}
                - \IMatTv{n\cdot}{r} \VUoutcome{}{}{r}
                -
                \noise{n}{r} \big)
                \\
                \eqas
                &\lim_{N \rightarrow \infty}
                \frac{\mu^2}{N^2}
                \sum_{n=1}^N
                \Uoutcome{\perp n}{}{t}
                \sum_{n=1}^N
                \Uoutcome{n}{}{r}
                +
                \lim_{N \rightarrow \infty}
                \frac{\mu^2}{N^2} \left(\sum_{n=1}^N \Uoutcome{n}{}{r}\right)
                \left(\sum_{n=1}^N \Uoutcome{\parallel n}{}{t}\right)
                +
                \lim_{N \rightarrow \infty}
                \frac{\sigma^2}{N}
                \sum_{n=1}^N
                \Uoutcome{n}{}{r} \Uoutcome{n}{}{t}
                \\
                =
                &\lim_{N \rightarrow \infty}
                \frac{\mu^2}{N^2} \left(\sum_{n=1}^N \Uoutcome{n}{}{r}\right)
                \left(\sum_{n=1}^N \Uoutcome{n}{}{t}\right)
                +
                \lim_{N \rightarrow \infty}
                \frac{\sigma^2}{N}
                \sum_{n=1}^N
                \Uoutcome{n}{}{r} \Uoutcome{n}{}{t},
            \end{aligned}
        \end{equation}
        where we used \eqref{eq:BL-proof-bt-4-2} and \eqref{eq:BL-proof-bt-4-4}. The induction hypothesis \ref{part:BL-a} implies the last equality in \eqref{eq:BL-b-5}.

        \item Given, $\Voutcome{N}{}{1}, \ldots, \Voutcome{N}{}{t}$, $\Mtreatment{}{N}$, and $\covar(N)$, define
        \begin{equation*}
            \begin{aligned}
                \Psi_{n}(N)
                &:=
                \psi\big(
                \outcome{N}{n}{1}
                , \ldots,
                \outcome{N}{n}{t}
                ,
                \outcome{N}{n}{t+1}
                ,
                \Vtreatment{n}{}(N)
                ,
                \Vcovar{n}(N)
                \big).
            \end{aligned}
        \end{equation*}
        Dropping the notation $N$ in the right-hand side, based on \eqref{eq:conditional dist of outcome_nonsym}, we can write
        \begin{equation*}
            \begin{aligned}
                \Psi_{n}(N)
                \Big|_{\Gc_t}
                \eqd
                \psi\left(
                \outcome{}{n}{1}
                , \ldots,
                \outcome{}{n}{t}
                ,
                \left[
                \widetilde{\IM}
                \VUoutcome{\perp}{}{t}
                + \bm{R}_t 
                \Vec{\alpha}_t
                +
                \IMatT{t} \VUoutcome{}{}{t}
                +
                \Vnoise{}{t}
                \right]_n
                ,
                \Vtreatment{n}{}
                ,
                \Vcovar{n}
                \right),
            \end{aligned}
        \end{equation*}
        where $\widetilde{\IM}$ has the same distribution as $\IM$ independent of everything else. We also let
        \begin{align*}
            \widetilde{\Psi}_{n}(N) = \Psi_{n}(N) - \E_{\IMG_t,\Vnoise{}{t}}[\Psi_{n}(N)].
        \end{align*}
        where $\E_{\IMG_t,\Vnoise{}{t}}$ denotes the expectation with respect to the randomness of the  matrices $\widetilde{\IM}$, $\IMatT{t}$, as well as the noise vector $\Vnoise{}{t}$. We follow the same approach as Step~1-\ref{item:BL-average limit}. Note that given $\Gc_t$, the elements of $\widetilde{\IM} \VUoutcome{\perp}{}{t} + \IMatT{t} \VUoutcome{}{}{t} + \Vnoise{}{t}$ are i.i.d. Gaussian random variables with mean $\tilde{\nu}_{tN}$ and variance~$\tilde{\rho}_{tN}^2$:
        \begin{equation}
            \label{eq:BL-bt-Yt stat}
            \begin{aligned}
                \tilde{\nu}_{tN}
                &:=
                \E
                \left[
                [
                \widetilde{\IM} \VUoutcome{\perp}{}{t} + \IMatT{t} \VUoutcome{}{}{t} + \Vnoise{}{t}
                ]_n
                \Big|
                \VUoutcome{}{}{t}
                \right]
                =
                \frac{\mu}{N}
                \sum_{n=1}^N
                \Uoutcome{\perp n}{}{t}
                +
                \frac{\mu_t}{N}
                \sum_{n=1}^N
                \Uoutcome{n}{}{t},
                \\
                \tilde{\rho}_{tN}^2
                &:=
                \Var
                \left[
                [
                \widetilde{\IM} \VUoutcome{\perp}{}{t} + \IMatT{t} \VUoutcome{}{}{t} + \Vnoise{}{t}
                ]_n
                \Big|
                \VUoutcome{}{}{t}
                \right]
                =
                \frac{\sigma^2}{N}
                \sum_{n=1}^N
                \left(\Uoutcome{\perp n}{}{t}\right)^2
                +
                \frac{\sigma_t^2}{N}
                \sum_{n=1}^N
                \left(\Uoutcome{n}{}{t}\right)^2
                +
                \sigma_e^2,
            \end{aligned}
        \end{equation}
        where $\Uoutcome{n}{}{t} = \outcomeg{t}{}\big(\outcome{}{n}{t},\Vtreatment{n}{},\Vcovar{n}\big)$ is the $n^{th}$ element of the column vector $\VUoutcome{}{}{t}$ and let
        \begin{align}
            \label{eq:BL-bt-(-1)}
            \tilde{\nu}_t = \lim_{N\rightarrow \infty }\tilde{\nu}_{tN},
            \quad\quad\quad
            \tilde{\rho}^2_t = \lim_{N\rightarrow \infty }\tilde{\rho}_{tN}^2.
        \end{align}
        We show that $\tilde{\nu}_t$ and $\tilde{\rho}_t^2$ are almost surely finite. That is, with a probability of 1, we have
        \begin{align}
            \label{eq:BL-bt-1}
            \lim_{N \rightarrow \infty}
            \frac{1}{N}
            \sum_{n=1}^N
            \Uoutcome{\perp n}{}{t} < \infty,
            \;\;\;
            \lim_{N \rightarrow \infty}
            \frac{1}{N}
            \sum_{n=1}^N
            \Uoutcome{n}{}{t} < \infty,
            \;\;\;
            \lim_{N \rightarrow \infty}
            \frac{1}{N}
            \sum_{n=1}^N
            \left(\Uoutcome{\perp n}{}{t}\right)^2 < \infty,
            \;\;\;
            \lim_{N \rightarrow \infty}
            \frac{1}{N}
            \sum_{n=1}^N
            (\Uoutcome{n}{}{t})^2 < \infty.
        \end{align}
        By definition, we can write
        \begin{equation}
        \label{eq:BL-bt-2}
        \begin{aligned}
            \frac{1}{N}
            \sum_{n=1}^N
            \Uoutcome{\perp n}{}{t}
            &=
            \frac{1}{N}
            \sum_{n=1}^N
            \Uoutcome{n}{}{t}
            -
            \frac{1}{N}
            \sum_{n=1}^N
            \Uoutcome{\parallel n}{}{t},
            \\
            \frac{1}{N}
            \sum_{n=1}^N
            \left(\Uoutcome{\perp n}{}{t}\right)^2
            =
            \pdot{\VUoutcome{\perp}{}{t}}{\VUoutcome{\perp}{}{t}}
            &=
            \pdot{\VUoutcome{}{}{t}}{\VUoutcome{}{}{t}}
            -
            \pdot{\VUoutcome{\parallel}{}{t}}{\VUoutcome{\parallel}{}{t}}
            =
            \frac{1}{N}
            \sum_{n=1}^N
            \left(\Uoutcome{n}{}{t}\right)^2
            -
            \frac{1}{N}
            \sum_{n=1}^N
            \left(\Uoutcome{\parallel n}{}{t}\right)^2.
        \end{aligned}
        \end{equation}
        Then, by the induction hypothesis and Assumption~\ref{asmp:BL}-\ref{asmp:BL-pl functions} for functions $\psi = \outcomeg{t}{}\big(\outcome{}{n}{t},\Vtreatment{n}{},\Vcovar{n}\big)$ and $\psi = \outcomeg{t}{}\big(\outcome{}{n}{t},\Vtreatment{n}{},\Vcovar{n}\big)^2$, we get
        \begin{equation}
        \label{eq:BL-bt-3}
        \begin{aligned}
            \lim_{N \rightarrow \infty}
            \frac{1}{N}
            \sum_{n=1}^N
            \Uoutcome{n}{}{t}
            &=
            \lim_{N \rightarrow \infty}
            \frac{1}{N}
            \sum_{n=1}^N
            \outcomeg{t}{} \big(\outcome{}{n}{t},\Vtreatment{n}{},\Vcovar{n}\big)
            \eqas
            \E\left[
            \outcomeg{t}{} \big(\AVO{}{}{t} + \VVO{}{}{t} Z,\Vtreatment{}{},\Vcovar{}\big)
            \right]  < \infty
            \\
            \lim_{N \rightarrow \infty}
            \frac{1}{N}
            \sum_{n=1}^N
            \left(\Uoutcome{n}{}{t}\right)^2
            &=
            \lim_{N \rightarrow \infty}
            \frac{1}{N}
            \sum_{n=1}^N
            \outcomeg{t}{} \big(\outcome{}{n}{t},\Vtreatment{n}{},\Vcovar{n}\big)^2
            \eqas
            \E\left[
            \outcomeg{t}{} \big(\AVO{}{}{t} + \VVO{}{}{t} Z,\Vtreatment{}{},\Vcovar{}\big)^2
            \right]  < \infty,
        \end{aligned}
        \end{equation}
        where $Z \sim \Nc(0,1)$. Further, by \eqref{eq:projection sum}, we have
        \begin{align*}
            \frac{1}{N}
            \sum_{n=1}^N
            \Uoutcome{\parallel n}{}{t}
            &=
            \frac{1}{N}
            \sum_{n=1}^N
            \sum_{s=0}^{t-1} \alpha_s \Uoutcome{n}{}{s}
            =
            \sum_{s=0}^{t-1}
            \frac{\alpha_s}{N}
            \sum_{n=1}^N
            \Uoutcome{n}{}{s}
            \\
            \frac{1}{N}
            \sum_{n=1}^N
            \left(\Uoutcome{\parallel n}{}{t}\right)^2
            &=
            \frac{1}{N}
            \sum_{n=1}^N
            \left(
            \sum_{s=0}^{t-1} \alpha_s \Uoutcome{n}{}{s}
            \right)^2
            =
            \sum_{r,s=0}^{t-1} \alpha_r \alpha_s \pdot{\VUoutcome{}{}{r}}{\VUoutcome{}{}{s}}.
        \end{align*}
        Considering Corollary~\ref{cl:alpha is bounded}, the vector $\Vec{\alpha}$ has a finite limit as $N\rightarrow \infty$. Similar to \eqref{eq:BL-bt-3}, the induction hypothesis for functions $\psi = \outcomeg{s}{}\big(\outcome{}{n}{s},\Vtreatment{n}{},\Vcovar{n}\big)$ and $\psi = \outcomeg{r}{}\big(\outcome{}{n}{r},\Vtreatment{n}{},\Vcovar{n}\big) \outcomeg{s}{}\big(\outcome{}{n}{s},\Vtreatment{n}{},\Vcovar{n}\big)$ implies that almost surely
        \begin{align}
            \label{eq:BL-bt-4}
            \lim_{N\rightarrow \infty}
            \frac{1}{N}
            \sum_{n=1}^N
            \Uoutcome{\parallel n}{}{t}
            < \infty
            ,
            \quad\quad
            \lim_{N\rightarrow \infty}
            \frac{1}{N}
            \sum_{n=1}^N
            \left(\Uoutcome{\parallel n}{}{t}\right)^2
            =
            \lim_{N\rightarrow \infty}
            \sum_{r,s=0}^{t-1} \alpha_r \alpha_s \pdot{\VUoutcome{}{}{r}}{\VUoutcome{}{}{s}}
            < \infty.
        \end{align}
        Consequently, by \eqref{eq:BL-bt-2}-\eqref{eq:BL-bt-4}, we get the result in \eqref{eq:BL-bt-1}. This also implies that $\tilde{\nu}_{tN}$ and $\tilde{\rho}_{tN}^2$, in~\eqref{eq:BL-bt-Yt stat}, are almost surely bounded for all values of $N$. As an immediate result, for $l\geq 1$, we get
        \begin{align}
            \label{eq:BL-proof-t-2}
            \E
            \left[
            \left|
                \big[
                \widetilde{\IM} \VUoutcome{\perp}{}{t} + \IMatT{t} \VUoutcome{}{}{t} + \Vnoise{}{t}
                \big]_n
                + 
                \left[\bm{R}_t 
                \Vec{\alpha}_t
                \right]_n
            \right|^{l}
            \right]
            \leq
            2^{l-1}
            \E
            \left[
            \left|
                \big[
                \widetilde{\IM} \VUoutcome{\perp}{}{t} + \IMatT{t} \VUoutcome{}{}{t} + \Vnoise{}{t}
                \big]_n
                \right|^{l}
                +
                \left|
                \left[\bm{R}_t 
                \Vec{\alpha}_t
                \right]_n
            \right|^{l}
            \right]
            \leq c,
        \end{align}
        where $c$ is a constant independent of $N$ and we used the inequality $(v_1+v_2)^l \leq 2^{l-1} (v_1^l+v_2^l),\; v_1,v_2 \geq 0$. Note that in \eqref{eq:BL-proof-t-2}, given $\Gc_t$, the term $\bm{R}_t  \Vec{\alpha}_t$ is deterministic and bounded in view of Corollary~\ref{cl:alpha is bounded}. Now, fixing $0< \kappa < 1$ and using the fact that $\psi \in \poly{k}$ and so $|\psi(\Vec{\omega})|\leq c (1 + \norm{\Vec{\omega}}^k)$, similar to \eqref{eq:BL-proof-a0-2}, we get that
        \begin{equation}
            \label{eq:BL-proof-t-1}
            \begin{aligned}
                \frac{1}{N}
                \sum_{n=1}^N
                \E\left[\left| \widetilde{\Psi}_{n}(N)\right|^{2+\kappa} \right]
                \leq c N^{\kappa/2}.
            \end{aligned}
        \end{equation}
        In order to obtain \eqref{eq:BL-proof-t-1}, the only difference, compared to \eqref{eq:BL-proof-a0-2}, is that we need the following inequality for $l\geq 1$ as a straightforward application of Jensen's inequality:
        \begin{align*}
            \left(\frac{v_1+\ldots+v_j}{l}\right)^l
            \leq
            \frac{v_1^l+\ldots+v_j^l}{l},\quad\quad
            v_i \geq 0.
        \end{align*}
        Therefore, we can apply the SLLN for triangular arrays, Theorem~\ref{thm:SLLN}. We get
        \begin{equation}
        \label{eq:BL-proof-t-3}
        \begin{aligned}
            \lim_{N \rightarrow \infty}
            \frac{1}{N}
            \sum_{n=1}^N
            \psi&
            \big(
            \outcome{}{n}{1}
            , \ldots,
            \outcome{}{n}{t}
            ,
            \outcome{}{n}{t+1}
            ,
            \Vtreatment{n}{}
            ,
            \Vcovar{n}
            \big)
            \\
            \eqas
            \;\lim_{N \rightarrow \infty}
            \frac{1}{N}
            \sum_{n=1}^N
            \E_{\IMG_t,\Vnoise{}{t}}
            \Big[
            \psi&\Big(
                \outcome{}{n}{1}
                , \ldots,
                \outcome{}{n}{t}
                ,
                \left[
                \widetilde{\IM}
                \VUoutcome{\perp}{}{t}
                + \bm{R}_t 
                \Vec{\alpha}_t
                +
                \IMatT{t} \VUoutcome{}{}{t}
                +
                \Vnoise{}{t}
                \right]_n
                ,
                \Vtreatment{n}{}
                ,
                \Vcovar{n}
                \Big)
            \Big].
        \end{aligned}
        \end{equation}
        Note that for any Borel measurable function $\phi$ and $Z \sim \Nc(0,1)$, the following random variables have the same distribution:
        \begin{align*}
            \phi\left(\left[\widetilde{\IM} \VUoutcome{\perp}{}{t} + \bm{R}_t \Vec{\alpha}_t + \IMatT{t} \VUoutcome{}{}{t} + \Vnoise{}{t} \right]_n\right),
            \quad\quad
            \phi\left(\tilde{\nu}_{tN} + \tilde{\rho}_{tN} Z + \sum_{s=0}^{t-1} \alpha_s \big( \outcome{}{n}{s+1} - \IMatTv{n\cdot}{s}\VUoutcome{}{}{s} - \noise{n}{s} \big)\right).
        \end{align*}
        Thus, we define
        \begin{equation*}
        \begin{aligned}
            \widehat{\psi}&
            \big(
            \outcome{}{n}{1}
            ,
            \outcome{}{n}{1} - \IMatTv{n\cdot}{0}\VUoutcome{}{}{0} - \noise{n}{0}
            , \ldots,
            \outcome{}{n}{t},
            \outcome{}{n}{t} - \IMatTv{n\cdot}{t-1}\VUoutcome{}{}{t-1} - \noise{n}{t-1}
            ,
            \Vtreatment{n}{}
            ,
            \Vcovar{n}
            \big)
            \\
            :=
            \E_{Z}
            \Big[
            \psi&\Big(
                \outcome{}{n}{1}
                , \ldots,
                \outcome{}{n}{t}
                ,
                \tilde{\nu}_{tN}
                +
                \tilde{\rho}_{tN} Z
                +
                \sum_{s=0}^{t-1}
                \alpha_s
                \big(
                \outcome{}{n}{s+1}
                - \IMatTv{n\cdot}{s}\VUoutcome{}{}{s}
                - \noise{n}{s}
                \big)
                ,
                \Vtreatment{n}{}
                ,
                \Vcovar{n}
                \Big)
            \Big].
        \end{aligned}
        \end{equation*}
        Using the induction hypothesis in \eqref{eq::BL-at-dynamics-with-eps}, for the function $\widehat{\psi}$, by \eqref{eq:BL-proof-t-3}, we have
        \begin{equation}
        \label{eq:BL-proof-t-new function}
        \begin{aligned}
            &\lim_{N \rightarrow \infty}
            \frac{1}{N}
            \sum_{n=1}^N
            \psi
            \big(
            \outcome{}{n}{1}
            , \ldots,
            \outcome{}{n}{t}
            ,
            \outcome{}{n}{t+1}
            ,
            \Vtreatment{n}{}
            ,\Vcovar{n}
            \big)
            \\
            \eqas
            &\lim_{N \rightarrow \infty}
            \frac{1}{N}
            \sum_{n=1}^N
            \widehat{\psi}
            \big(
            \outcome{}{n}{1}
            ,
            \outcome{}{n}{1}- \IMatTv{n\cdot}{0}\VUoutcome{}{}{0} - \noise{n}{0}
            , \ldots,
            \outcome{}{n}{t},
            \outcome{}{n}{t}- \IMatTv{n\cdot}{t-1}\VUoutcome{}{}{t-1} - \noise{n}{t-1}
            ,
            \Vtreatment{n}{},\Vcovar{n}
            \big)
            \\
            \eqas
            &\;
            \E
            \left[
            \widehat{\psi}
            \big(
            \AVO{}{}{1}
            +
            \VVO{}{}{1} Z_1
            ,
            \BAVO{}{}{1}
            +
            \BVVO{}{}{1} Z'_1
            , \ldots,
            \AVO{}{}{t}
            +
            \VVO{}{}{t} Z_t
            ,
            \BAVO{}{}{t}
            +
            \BVVO{}{}{t} Z'_t
            ,
            \Vtreatment{}{},\Vcovar{}
            \big)
            \right]
            \\
            =
            &\;
            \E
            \E_{Z}
            \bigg[
            \psi\bigg(
                \AVO{}{}{1}
                +
                \VVO{}{}{1} Z_1
                , \ldots,
                \AVO{}{}{t}
                +
                \VVO{}{}{t} Z_t,
                \tilde{\nu}_{t}
                +
                \tilde{\rho}_{t} Z
                +
                \sum_{s=0}^{t-1}
                \alpha_s
                \big(
                \BAVO{}{}{s+1}
                +
                \BVVO{}{}{s+1}
                Z'_{s+1}
                \big)
                ,
                \Vtreatment{}{},\Vcovar{}
                \bigg)
            \bigg],
        \end{aligned}
        \end{equation}
        where $Z$ is an independent Normal random variable. Similar to \eqref{eq:BL-proof-a0-1}, we used the DCT and the continuous mapping theorem to interchange the limit and the expectation and then pass the limit through the function in \eqref{eq:BL-proof-t-new function}.
        Now, we need to show that
        \begin{equation}
            \label{eq:BL-proof-bt-dynamics-1}
            \begin{aligned}
                \E
                \left[
                \tilde{\nu}_{t}
                +
                \tilde{\rho}_{t} Z
                +
                \sum_{s=0}^{t-1}
                \alpha_s
                \big(
                \BAVO{}{}{s+1}
                +
                \BVVO{}{}{s+1}
                Z'_{s+1}
                \big)
                \right]
                &=
                \AVO{}{}{t+1},
                \\
                \Var
                \left[
                \tilde{\nu}_{t}
                +
                \tilde{\rho}_{t} Z
                +
                \sum_{s=0}^{t-1}
                \alpha_s
                \big(
                \BAVO{}{}{s+1}
                +
                \BVVO{}{}{s+1}
                Z'_{s+1}
                \big)
                \right]
                &=
                \VVO{}{2}{t+1}.
            \end{aligned}
        \end{equation}
        Then, the proof is complete because $(Z'_1,\ldots,Z'_t)$ has a joint Normal distribution independent of $Z$ and so the random variable $\tilde{\nu}_{t} + \tilde{\rho}_{t} Z + \sum_{s=0}^{t-1} \alpha_s \big( \BAVO{}{}{s+1} + \BVVO{}{}{s+1} Z'_{s+1})$ is Gaussian as well. To obtain \eqref{eq:BL-proof-bt-dynamics-1}, we let $\psi(y_1,\ldots,y_{t+1},w,\Vcovar{}) = y_{t+1}$ and $\psi(y_1,\ldots,y_{t+1},w,\Vcovar{}) = y_{t+1}^2$ in \eqref{eq:BL-proof-t-new function}. We get
        \begin{equation}
            \label{eq:BL-proof-bt-dynamics-2}
            \begin{aligned}
                \lim_{N \rightarrow \infty}
                \frac{1}{N}
                \sum_{n=1}^N
                \outcome{}{n}{t+1}
                &=
                \E
                \left[
                \tilde{\nu}_{t}
                +
                \tilde{\rho}_{t} Z
                +
                \sum_{s=0}^{t-1}
                \alpha_s
                \big(
                \BAVO{}{}{s+1}
                +
                \BVVO{}{}{s+1}
                Z_{s+1}
                \big)
                \right],
                \\
                \lim_{N \rightarrow \infty}
                \frac{1}{N}
                \sum_{n=1}^N
                \big(
                \outcome{}{n}{t+1}\big)^2
                &=
                \E
                \left[
                \left(
                \tilde{\nu}_{t}
                +
                \tilde{\rho}_{t} Z
                +
                \sum_{s=0}^{t-1}
                \alpha_s
                \big(
                \BAVO{}{}{s+1}
                +
                \BVVO{}{}{s+1}
                Z_{s+1}
                \big)
                \right)^2
                \right].
            \end{aligned}
        \end{equation}
        But, by part~\ref{part:BL-b}, we have
        \begin{align*}
            \lim_{N \rightarrow \infty}
            \frac{1}{N}
            \sum_{n=1}^N
            \outcome{}{n}{t+1}
            \eqas \AVO{}{}{t+1},
            \quad\quad
            \lim_{N \rightarrow \infty}
            \frac{1}{N}
            \sum_{n=1}^N
            \outcome{}{n}{t+1}^2
            \eqas \AVO{}{2}{t+1} + \VVO{}{2}{t+1}.
        \end{align*}
        That yields the desired result in \eqref{eq:BL-proof-bt-dynamics-1}.
        
        To conclude the proof, we need to show that the relations in \eqref{eq::BL-at-dynamics-with-eps} and \eqref{eq::BL-at-third result} hold true for~$s=t$.
        Following a similar argument as above, the former result is immediate. Here, we show that \eqref{eq::BL-at-third result} is true for~$s=t$; that is, for $\phi: \R^{2(t+1)+T+M} \mapsto \R$ within $\poly{\frac{k}{2}}$, we claim that
        \begin{equation}
            \label{eq::BL-at-third result-proof}
            \begin{aligned}
                &\lim_{N \rightarrow \infty}
                \frac{1}{N} \sum_{n=1}^N
                \Big(
                \outcomeg{0}{}\big(\outcome{}{n}{0},\Vtreatment{n}{},\Vcovar{n}\big)
                \phi\big(\outcome{}{n}{1}, \outcome{}{n}{1} - \IMatTv{n\cdot}{0}\VUoutcome{}{}{0} - \noise{n}{0},\ldots,
                \\
                &\quad\quad\quad\quad\quad\quad
                \outcome{}{n}{t+1},\outcome{}{n}{t+1}-\IMatTv{n\cdot}{t}\VUoutcome{}{}{t}-\noise{n}{t}, \Vtreatment{n}{}, \Vcovar{n}\big)
                \Big)
                \\
                \eqas
                &\;\E
                \Big[
                \bar{g}_0(\Vtreatment{}{},\Vcovar{})
                \phi\big(\AVO{}{}{1} + \VVO{}{}{1} Z_1, \BAVO{}{}{1} + \BVVO{}{}{1} Z'_1,\ldots,\AVO{}{}{t+1} + \rho_{t+1} Z_{t+1}, \BAVO{}{}{t+1} + \BVVO{}{}{t+1} Z'_{t+1},\Vtreatment{}{},\Vcovar{}\big)
                \Big],
            \end{aligned}
        \end{equation}
        Note that the vector $\Voutcome{}{}{0}$ is given and both $\outcomeg{0}{}$ and $\phi$ are $\poly{\frac{k}{2}}$ functions. Therefore, we can verify the conditions of Theorem~\ref{thm:SLLN} by following the same argument as the one resulting in \eqref{eq:BL-proof-t-1}. Then, applying the SLLN for the triangular arrays, similar to \eqref{eq:BL-proof-t-new function}, we get
        \begin{equation*}
            \begin{aligned}
                &\lim_{N \rightarrow \infty}
                \frac{1}{N} \sum_{n=1}^N
                \Big(
                \outcomeg{0}{}\big(\outcome{}{n}{0},\Vtreatment{n}{},\Vcovar{n}\big)
                \phi\big(\outcome{}{n}{1}, \outcome{}{n}{1} - \IMatTv{n\cdot}{0}\VUoutcome{}{}{0} - \noise{n}{0},\ldots,
                \\
                &\quad\quad\quad\quad\quad\quad
                \outcome{}{n}{t+1},\outcome{}{n}{t+1}-\IMatTv{n\cdot}{t}\VUoutcome{}{}{t}-\noise{n}{t}, \Vtreatment{n}{}, \Vcovar{n}\big)
                \Big)
                \\
                \eqas
                &\lim_{N \rightarrow \infty}
                \frac{1}{N} \sum_{n=1}^N
                \E_{\IMG_t,\Vnoise{}{t}}
                \bigg[
                \outcomeg{0}{}\big(\outcome{}{n}{0}, \Vtreatment{n}{},\Vcovar{n}\big)
                \phi\big(\outcome{}{n}{1},\outcome{}{n}{1}-\IMatTv{n\cdot}{0}\VUoutcome{}{}{0} -\noise{n}{0},\ldots,
                \\
                &\quad\quad\quad\quad\quad\quad\quad\outcome{}{n}{t},\outcome{}{n}{t}-\IMatTv{n\cdot}{t-1}\VUoutcome{}{}{t-1}-\noise{n}{t-1},
                \\
                &\quad\quad\quad\quad\quad\quad\quad\left[
                \widetilde{\IM}
                \VUoutcome{\perp}{}{t}
                +
                \bm{R}_t 
                \Vec{\alpha}_t
                +
                \IMatTv{n\cdot}{t}\VUoutcome{}{}{t}
                +
                \Vnoise{}{t}
                \right]_n,
                \left[
                \widetilde{\IM}
                \VUoutcome{\perp}{}{t}
                +
                \bm{R}_t 
                \Vec{\alpha}_t
                \right]_n, \Vtreatment{n}{}, \Vcovar{n}\big)
                \bigg]
                \\
                \eqas
                &\;\E
                \E_{Z}
                \bigg[
                \bar{g}_0(\Vtreatment{}{},\Vcovar{})
                \phi\big(\AVO{}{}{1}+\VVO{}{}{1} Z_1, \BAVO{}{}{1}+\BVVO{}{}{1} Z'_1,\ldots,\AVO{}{}{t}+\VVO{}{}{t} Z_t, \BAVO{}{}{t}+\BVVO{}{}{t} Z'_t,
                \tilde{\nu}_{t}
                +
                \tilde{\rho}_{t} Z
                +
                \sum_{s=0}^{t-1}
                \alpha_s
                \big(
                \BAVO{}{}{s+1}
                +
                \BVVO{}{}{s+1}
                Z'_{s+1}
                \big),
                \\
                &\quad\quad\quad\quad\quad\quad\quad
                \BAVO{}{}{t}
                +
                \BVVO{}{}{t} Z'
                +
                \sum_{s=0}^{t-1}
                \alpha_s
                \big(
                \BAVO{}{}{s+1}
                +
                \BVVO{}{}{s+1}
                Z'_{s+1}
                \big), \Vtreatment{}{},\Vcovar{}\big)
                \bigg],
            \end{aligned}
        \end{equation*}
        where in the last equality we used the induction hypothesis stated in \eqref{eq::BL-at-third result}. An argument similar to \eqref{eq:BL-proof-bt-dynamics-1} besides \eqref{eq:BL-b-2} and \eqref{eq:BL-b-5} concludes the proof. \ep
    \end{enumerate}

\subsection{Proof of Theorem~\ref{thm:consistency}: Strong consistency of the estimator}
Without loss of generality, and by relabeling, we assume that Assumptions~\ref{asmp:BL}-\ref{asmp:BL-bound on initials} and \ref{asmp:BL-empirical dist of inits} hold true for $t=-1$. This allows us to consider the state evolution equations in \eqref{eq:state evolution} for $t\geq 0$. Now, let $f$ be the function corresponding to Algorithm~\ref{alg:causal-mp-original}. That is, 
\begin{align}
    \label{eq:TTE_algorithm_function}
    f: \R^{T+1} \mapsto \R^{T+1},
    \quad\quad
    \left(\ETTE{0}{\desired,0}, \ldots, \ETTE{T}{\desired,0}\right) = f\big(\HAVO{}{}{0}, \ldots, \HAVO{}{}{T}\big)
\end{align}
Here, the function $f$ maps the sample means of the observed outcomes over time to the output of Algorithm~\ref{alg:causal-mp-original}. Consequently, $f$ implicitly depends on the other inputs of the algorithm, including the desired treatment level denoted by $\desired$. Below, we first demonstrate that $f$ is a continuous function in its arguments. In the second step, we show that $\big(\TTE{0}{\desired,0}, \ldots, \TTE{T}{\desired,0}\big) = f\big(\AVO{\expd}{}{0}, \ldots, \AVO{\expd}{}{T}\big)$. Then, considering Theorem~\ref{thm:Big theorem}, we know that $\AVO{\expd}{}{t} \eqas \lim_{N \rightarrow \infty} \HAVO{}{}{t},\; t\in[T]_0$, which implies the following and concludes the proof:
\begin{align*}
    \lim_{N \rightarrow \infty} \left(\ETTE{0}{\desired,0}, \ldots, \ETTE{T}{\desired,0}\right)
    &\;=
    \lim_{N \rightarrow \infty} f\big(\HAVO{}{}{0}, \ldots, \HAVO{}{}{T}\big)
    \\
    &\eqas
    f\big(\AVO{\expd}{}{0}, \ldots, \AVO{\expd}{}{T}\big)
    \\
    &\;=
    \left(\TTE{0}{\desired,0}, \ldots, \TTE{T}{\desired,0}\right),
\end{align*}
where the second line holds due to the continuous mapping theorem, Theorem 2.3 in \cite{van2000asymptotic}.

\textbf{Step 1.} In order to verify the continuity of $f$, note that regression coefficients $a_1,\;b_1,\;a_2,$ and $b_2$ are continuous in the input data $\big(\HAVO{}{}{0}, \ldots, \HAVO{}{}{T}\big)$. As a result, the coefficients $(\widehat{\PE},\widehat{\ME},\widehat{\DE})$ can also be understood as continuous functions of the input data. Now, we use an induction on $t$ to show that $\desiredHAVO{}{}{t}$ and $\ETTE{t}{\desired,0}$ are continuous in $\big((\HAVO{}{}{0}, \ldots, \HAVO{}{}{T}\big)$, as well. The result for $t=0$ is immediate by the definition. For the induction step, considering $\desiredHAVO{}{}{t+1} = \HAVO{}{}{t+1} + \widehat{\PE}\left(\desiredHAVO{}{}{t}-\HAVO{}{}{t}\right) + \widehat{\DE} (\desired-\pi_j) + \widehat{\ME} \left(\desired\desiredHAVO{}{}{t}-\pi_j\HAVO{}{}{t}\right),\; j =1,2,$ the right-hand side is a continuous function of the data, as the sum and product of a finite number of continuous functions is a continuous function. The same result is valid for $\ETTE{t+1}{\desired,0} = \widehat{\PE} \ETTE{t}{\desired,0} + \widehat{\DE} \desired + \widehat{\ME} \desired \desiredHAVO{}{}{t}$ and the induction is complete. Finally, since the composition of continuous functions is continuous, the whole procedure in Steps 2 and 3 of Algorithm~\ref{alg:causal-mp-original} is continuous in $\big(\HAVO{}{}{0}, \ldots, \HAVO{}{}{T}\big)$, implying the continuity of $f$.

\textbf{Step 2.} Note that by Theorem~\ref{thm:Big theorem}, there exists a set $\Cset \subset \Omega$ with $\P(\Cset) = 1$ such that $\HAVO{\expd}{}{t} = \frac{1}{N} \sum_{n=1}^N \outcome{}{n}{t}$ converges point-wise to $\AVO{\expd}{}{t}$ over $\Cset$ as $N \rightarrow \infty$, where $t\geq 0$ (note that a countable union of zero-measure sets also has measure zero). Henceforth, we focus on the set $\Cset$ and do all the computations over that. By \eqref{eq:function_structure} and \eqref{eq:state evolution}, we get
\begin{equation}
\label{eq:consistency_proof_SE}
\begin{aligned}
    \AVO{\expd}{}{t+1}
    &=
    \ACE + \APE \AVO{\expd}{}{t} + \ADE \pi_1 + \AME \AVO{\expd}{}{t} \pi_1 + \ACC^{\;\top} \Acovar,
    \quad\quad\quad
    &&t = 0,\ldots,T_1-1,
    \\
    \AVO{\expd}{}{t+1}
    &=
    \ACE + \APE \AVO{\expd}{}{t} + \ADE \pi_2 + \AME \AVO{\expd}{}{t} \pi_2 + \ACC^{\;\top} \Acovar,
    \quad\quad\quad
    &&t = T_1,\ldots,T_1+T_2-1.
\end{aligned}
\end{equation}
Above, we utilized the following notations:
\begin{equation*}
    \begin{aligned}
        \ACE = \E[\CE^n],\quad\quad\quad
        \APE = \E[\PE^n],\quad\quad\quad
        \ADE = \E[\DE^n],\quad\quad\quad
        \AME = \E[\ME^n],\quad\quad\quad
        \ACC = \E[\CC^n],\quad\quad\quad
        \Acovar = \E[\covar],
    \end{aligned}
\end{equation*}
where $\ACC$ and $\Acovar$ are the vectors representing the mean of the columns of matrices $\CCM$ and $\covar$, respectively. Specifically, in \eqref{eq:consistency_proof_SE}, we handle the randomness of $\CE^n,\PE^n,\DE^n,\ME^n,$ and $\CC^n$ by extending the matrix $\covar$; that is, we incorporate these coefficients as components of the individual's covariate vector within an extended matrix $\widetilde\covar$. Note that Assumption~\ref{asmp:BL} holds for the upgraded model, because of the independence and bounded moment assumption of $\CE^n,\PE^n,\DE^n,\ME^n,$ and $\CC^n$.

Then, we can rewrite \eqref{eq:consistency_proof_SE} as follows:
\begin{equation}
\label{eq:consistency_proof_SE_2}
\begin{aligned}
    \AVO{\expd}{}{t+1}
    &=
    \left(\APE  + \AME  \pi_1
    \right) \AVO{\expd}{}{t}
    +
    \ACE + \ADE \pi_1 + \ACC^{\;\top} \Acovar,
    \quad\quad\quad
    &&t = 0,\ldots,T_1-1,
    \\
    \AVO{\expd}{}{t+1}
    &=
    \left(\APE  + \AME  \pi_2
    \right) \AVO{\expd}{}{t}
    +
    \ACE + \ADE \pi_2 + \ACC^{\;\top} \Acovar,
    \quad\quad\quad
    &&t = T_1,\ldots,T_1+T_2-1.
\end{aligned}
\end{equation}
Therefore, regressing $\big(\AVO{\expd}{}{1},\ldots, \AVO{\expd}{}{T_1}\big)^\top$ on
$\big(\AVO{\expd}{}{0},\ldots, \AVO{\expd}{}{T_1-1}\big)^\top$ is exact and we get the coefficient $b_1 = \APE  + \AME  \pi_1$ and intercept $a_1 = \ACE + \ADE \pi_1 + \ACC^{\;\top} \Acovar$. Likewise, we obtain
$b_2 = \APE  + \AME  \pi_2$ and $a_2 = \ACE + \ADE \pi_2 + \ACC^{\;\top} \Acovar$. This implies that
\begin{align}
    \label{eq:consistency_proof_coeffs}
    \widehat{\Xi} = \APE,\quad\quad\quad\quad\quad
    \widehat{\ME} = \AME, \quad\quad\quad\quad\quad
    \widehat{\DE} = \ADE.
\end{align}
On the other hand, by Theorem~\ref{thm:Big theorem} and \eqref{eq:consistency_proof_SE}, for $t=0,\ldots,T_1-1$, we can write
\begin{equation}
    \label{eq:consistency_proof_counterfactual_1}
\begin{aligned}
    \AVO{\desired}{}{t+1}
    &=
    \ACE + \APE \AVO{\desired}{}{t} + \ADE \desired + \AME \desired \AVO{\desired}{}{t} + \ACC^{\;\top} \Acovar
    \\
    &=
    \ACE + \APE \AVO{\expd}{}{t} + \ADE\pi_1 + \AME \pi_1 \AVO{\expd}{}{t} + \ACC^{\;\top} \Acovar
    +
    \APE
    \left(
    \AVO{\desired}{}{t} - \AVO{\expd}{}{t}
    \right)
    +
    \AME \left(
    \desired \AVO{\desired}{}{t}
    -
    \pi_1 \AVO{\expd}{}{t}
    \right)
    +
    \ADE (\desired-\pi_1)
    \\
    &=
    \AVO{\expd}{}{t+1}
    +
    \APE
    \left(
    \AVO{\desired}{}{t} - \AVO{\expd}{}{t}
    \right)
    +
    \AME \left(
    \desired \AVO{\desired}{}{t}
    -
    \pi_1 \AVO{\expd}{}{t}
    \right)
    +
    \ADE (\desired-\pi_1),
\end{aligned}
\end{equation}
where we set $\AVO{\desired}{}{0} = \AVO{\expd}{}{0}$. Here, with a slight abuse of notation, we used $\AVO{\desired}{}{t}$ since the corresponding experimental design has only one element, which is $\desired$. Likewise, for $t=T_1,\ldots,T_1+T_2-1$, we can obtain
\begin{equation}
    \label{eq:consistency_proof_counterfactual_2}
\begin{aligned}
    \AVO{\desired}{}{t+1}
    &=
    \AVO{\expd}{}{t+1}
    +
    \APE
    \left(
    \AVO{\desired}{}{t} - \AVO{\expd}{}{t}
    \right)
    +
    \AME (\desired \AVO{\desired}{}{t} - \pi_2 \AVO{\expd}{}{t})
    +
    \ADE (\desired-\pi_2).
\end{aligned}
\end{equation}
Also, it is straightforward to check that
\begin{equation}
\label{eq:consistency_proof_counterfactual_3}
\begin{aligned}
    \AVO{0}{}{t+1}
    &=
    \AVO{\expd}{}{t+1}
    +
    \APE
    \left(
    \AVO{0}{}{t} - \AVO{\expd}{}{t}
    \right)
    -
    \AME \pi_1 \AVO{\expd}{}{t}
    -
    \ADE \pi_1,
    \quad\quad\quad
    &&t = 0,\ldots,T_1-1,
    \\
    \AVO{0}{}{t+1}
    &=
    \AVO{\expd}{}{t+1}
    +
    \APE
    \left(
    \AVO{0}{}{t} - \AVO{\expd}{}{t}
    \right)
    -
    \AME \pi_2 \AVO{\expd}{}{t}
    -
    \ADE \pi_2,
    \quad\quad\quad
    &&t = T_1,\ldots,T_1+T_2-1,
\end{aligned}
\end{equation}
where $\AVO{0}{}{0} = \AVO{\expd}{}{0}$.

Considering \eqref{eq:consistency_proof_coeffs}-\eqref{eq:consistency_proof_counterfactual_3}, by the definition of the total treatment effect in \eqref{eq:TTE_def_fixed} and Theorem~\ref{thm:Big theorem}, over the set~$\Cset$, we have
\begin{align}
    \label{eq:consistency_proof_11}
    \TTE{t+1}{\desired,0}
    =
    \AVO{\desired}{}{t+1} - \AVO{0}{}{t+1}
    =
    \APE
    \TTE{t}{\desired,0}
    +
    \AME \desired \AVO{\desired}{}{t}
    +
    \ADE \desired.
\end{align}
Recalling the fact that $\P(\Cset) = 1$ concludes the proof. \ep

\subsection{Proof of Theorem~\ref{thm:Estimator at EQ}: Analysis of the estimator at equilibrium}
The proof technique is similar to the proof of Theorem~\ref{thm:consistency}. Based on \eqref{eq:consistency_proof_11}, it is straightforward to obtain the following result: 
\begin{align}
    \label{eq:E_at_Eq_proof_1}
    \TTE{}{1,0} \eqas \frac{\AME \AVO{1}{}{} + \ADE}{1-\APE}.
\end{align}
But, we have
\begin{equation}
\label{eq:E_at_Eq_proof_2}
\begin{aligned}
    \AVO{\pi_1}{}{}
    &=
    \left(\APE  + \AME  \pi_1
    \right) \AVO{\pi_1}{}{}
    +
    \ACE + \ADE \pi_1 + \ACC^{\;\top} \Acovar,
    \\
    \AVO{\pi_2}{}{}
    &=
    \left(\APE  + \AME  \pi_2
    \right) \AVO{\pi_2}{}{}
    +
    \ACE + \ADE \pi_2 + \ACC^{\;\top} \Acovar.
\end{aligned}
\end{equation}
On the other hand, by \eqref{eq:Estimator at EQ} and Theorem~\ref{thm:Big theorem}, we have
\begin{equation*}
\begin{aligned}
    \lim_{N \rightarrow \infty}
    \ETTE{}{1,0}
    &=
    \frac{1}{\pi_2-\pi_1} 
    \lim_{N \rightarrow \infty}
    \sum_{n=1}^N 
    \frac{\left(
    \outcome{\pi_2}{n}{}
    -
    \outcome{\pi_1}{n}{}
    \right)}{N}
    \\
    &\eqas
    \frac{\AVO{\pi_2}{}{}-\AVO{\pi_1}{}{}}{\pi_2-\pi_1}
    \\
    &=
    \frac{\APE (\AVO{\pi_2}{}{}-\AVO{\pi_1}{}{}) + \AME (\pi_2 \AVO{\pi_2}{}{} - \pi_1 \AVO{\pi_1}{}{})}{\pi_2-\pi_1}
    +
    \ADE
    \\
    &=
    \frac{\APE (\AVO{\pi_2}{}{}-\AVO{\pi_1}{}{})}{\pi_2-\pi_1}
    +
    \frac{\AME (\pi_2 \AVO{\pi_2}{}{} - \pi_1 \AVO{\pi_1}{}{})}{\pi_2-\pi_1}
    +
    \ADE
    \\
    &\eqas
    \APE
    \lim_{N \rightarrow \infty}
    \ETTE{}{1,0}
    +
    \frac{\AME (\pi_2 \AVO{\pi_2}{}{} - \pi_1 \AVO{\pi_1}{}{})}{\pi_2-\pi_1}
    +
    \ADE,
\end{aligned}    
\end{equation*}
that implies
\begin{align}
    \label{eq:E_at_Eq_proof_3}
    \lim_{N \rightarrow \infty}
    \ETTE{}{1,0} \eqas
    \frac{1}{1-\APE}
    \left(
    \frac{\AME (\pi_2 \AVO{\pi_2}{}{} - \pi_1 \AVO{\pi_1}{}{})}{\pi_2-\pi_1}
    +
    \ADE
    \right)
\end{align}
Then, by \eqref{eq:E_at_Eq_proof_1}-\eqref{eq:E_at_Eq_proof_3}, we get
\begin{align*}
    \lim_{N \rightarrow \infty}
    \ETTE{}{1,0} - \TTE{}{1,0} \eqas
    \frac{\AME}{1-\APE}
    \left(
    \frac{\pi_2 \AVO{\pi_2}{}{} - \pi_1 \AVO{\pi_1}{}{}}{\pi_2-\pi_1}
    -
    \AVO{1}{}{}
    \right).
\end{align*}
But, by Theorem~\ref{thm:Big theorem}, we know that $\lim_{N \rightarrow \infty} \HAVO{\expd}{}{} \eqas \AVO{\expd}{}{}$, and the proof is complete. \ep

\subsection{Two Versions of Strong Law of Large Numbers}
We need the following strong law of large numbers (SLLN) for triangular arrays of independent but not identically distributed random variables. The form stated below is Theorem~3 in \cite{bayati2011dynamics} that is adapted from Theorem~2.1 in \cite{hu1997strong}.
\begin{theorem}[SLLN]
    \label{thm:SLLN}
    Let $\left\{X_{n,i}:1\leq i \leq n,\; n \geq 1\right\}$ be a triangular array of random variables such that $(X_{n,1},\ldots,X_{n,n})$ are mutually independent with a mean equal to zero for each $n$ and $\frac{1}{n} \sum_{i=1}^n E\left[|X_{n,i}|^{2+\kappa}\right] \leq c n^{\kappa/2}$ for some $0 < \kappa < 1$ and $c < \infty$. Then, we have
    \begin{align}
        \label{eq:SLLN}
        \lim_{N \rightarrow \infty}
        \frac{1}{n} \sum_{i=1}^n X_{n,i} \eqas 0.
    \end{align}
\end{theorem}
We also need the following form of the law of large numbers which is an extension of Lemma~4 in \cite{bayati2011dynamics}.
\begin{theorem}
    \label{thm:SLLN-2}
    Fix $k\geq 2$ and an integer $l$ and let $\left\{\bm v(N)\right\}_{N \geq 1}$ be a sequence of vectors that $\bm v(N) \in \R^{N\x l}$. That means, $\bm v(N)$ is a matrix with $N$ rows and $l$ columns. Assume that the empirical distribution of $\bm v(N)$, denoted by $\hat{p}_{N}$, converges weakly to a probability measure $p_v$ on $\R^l$ such that $\E_{p_v}\left[\norm{\Vec{V}}^k\right] < \infty$ and $\E_{\hat{p}_{N}}\left[\norm{\Vec{V}}^k\right] \rightarrow \E_{p_v}\left[\norm{\Vec{V}}^k\right]$ as $N \rightarrow \infty$. Then, for any continuous function $f:\R^l \mapsto \R$ with at most polynomial growth of order $k$, we have
    \begin{align}
        \label{eq:SLLN-2}
        \lim_{N \rightarrow \infty}
        \frac{1}{N} \sum_{n=1}^N f\big(\bm v_n(N)\big)
        \eqas \E_{p_v} [f(\Vec{V})].
    \end{align}
\end{theorem}
Proof. We use the same truncation technique as Lemma~4 in \cite{bayati2011dynamics}. For a positive integer $h$, we define
\begin{align}
    \label{eq:SLLN-2-proof-truncated function}
    f_{h}(\Vec{\omega}) :=
    \begin{cases}
    h &f(\Vec{\omega}) > h,
    \\
    f(\Vec{\omega}) \quad\quad &|f(\Vec{\omega})| \leq h,
    \\
    -h &f(\Vec{\omega}) < h,
    \end{cases}
\end{align}
and write $\tilde{f}_h(\Vec{\omega}) := f(\Vec{\omega}) - f_h(\Vec{\omega})$. Then, by definition of empirical measure, we have
\begin{align*}
    \frac{1}{N} \sum_{n=1}^N f\big(\bm v_n(N)\big)
    =
    \E_{\hat{p}_{N}}[f(\Vec{V})]
    =
    \E_{\hat{p}_{N}}[\tilde{f}_h(\Vec{V})]
    +
    \E_{\hat{p}_{N}}[f_h(\Vec{V})].
\end{align*}
Also, we can write
\begin{equation}
    \label{eq:SLLN-2-proof-1}
    \begin{aligned}
        \liminf_{N \rightarrow \infty}\left(
        \E_{\hat{p}_{N}}[\tilde{f}_h(\Vec{V})]
        +
        \E_{\hat{p}_{N}}[f_h(\Vec{V})]\right)
        &=
        \liminf_{N \rightarrow \infty} \E_{\hat{p}_{N}}[f(\Vec{V})]
        \\
        &\leq
        \limsup_{N \rightarrow \infty} \E_{\hat{p}_{N}}[f(\Vec{V})]
        =
        \limsup_{N \rightarrow \infty}\left(
        \E_{\hat{p}_{N}}[\tilde{f}_h(\Vec{V})]
        +
        \E_{\hat{p}_{N}}[f_h(\Vec{V})]\right).
    \end{aligned}
\end{equation}
On the other hand, because $\hat{p}_{N}$ converges weakly to $p_{v}$, for the bounded continuous function $f_h$ (see, e.g., Section~2 of \cite{billingsley2013convergence}), we have
\begin{align}
    \label{eq:SLLN-2-proof-2}
    \lim_{N \rightarrow \infty}
    \E_{\hat{p}_{N}}[f_h(\Vec{V})] = \E_{p_{v}}[f_h(\Vec{V})].
\end{align}
Considering that $f$ has at most polynomial growth of order $k$, we can write
\begin{align}
    \label{eq:SLLN-2-proof-3}
    \left|\tilde{f}_h(\Vec{V})\right|
    \leq
    \left|f(\Vec{V})\right| \1_{\left\{\left|f(\Vec{V})\right|>h\right\}}
    \leq
    c\left(1+\norm{\Vec{V}}^k\right) \1_{\left\{\left|f(\Vec{V})\right|>h\right\}}
    \leq
    c\left(1+\norm{\Vec{V}}^k\right) \1_{\left\{\frac{h}{c}-1 < \norm{\Vec{V}}^k\right\}}.
\end{align}
Putting \eqref{eq:SLLN-2-proof-1}-\eqref{eq:SLLN-2-proof-3} together, we get
\begin{align*}
    &\E_{p_v}[f_h(\Vec{V})]
    -
    \limsup_{N \rightarrow \infty}
    \E_{\hat{p}_{N}}\left[c\left(1+\norm{\Vec{V}}^k\right) \1_{\left\{\frac{h}{c}-1 < \norm{\Vec{V}}^k\right\}}\right]
    \\
    &\leq
    \liminf_{N \rightarrow \infty} \E_{\hat{p}_{N}}[f(\Vec{V})]
    \\
    &\leq
    \limsup_{N \rightarrow \infty} \E_{\hat{p}_{N}}[f(\Vec{V})]
    \\
    &\leq
    \E_{p_v}[f_h(\Vec{V})]
    +
    \limsup_{N \rightarrow \infty}
    \E_{\hat{p}_{N}}\left[c\left(1+\norm{\Vec{V}}^k\right) \1_{\left\{\frac{h}{c}-1 < \norm{\Vec{V}}^k\right\}}\right].
\end{align*}
Now, based on the weak convergence of $\hat{p}_{N}$ to $p_v$, we can write
\begin{align*}
    \lim_{N \rightarrow \infty}
    \E_{\hat{p}_{N}}\left[c\left(1+\norm{\Vec{V}}^k\right) \1_{\left\{\frac{h}{c}-1 \geq \norm{\Vec{V}}^k\right\}}\right]
    =
    \E_{p_v}\left[c\left(1+\norm{\Vec{V}}^k\right) \1_{\left\{\frac{h}{c}-1 \geq \norm{\Vec{V}}^k\right\}}\right].
\end{align*}
Recalling the assumption $\lim_{N \rightarrow \infty} \E_{\hat{p}_{N}}\left[\norm{\Vec{V}}^k\right] = \E_{p_v}\left[\norm{\Vec{V}}^k\right]$, this implies that
\begin{align*}
    \limsup_{N \rightarrow \infty}
    \E_{\hat{p}_{N}}\left[c\left(1+\norm{\Vec{V}}^k\right) \1_{\left\{\frac{h}{c}-1 < \norm{\Vec{V}}^k\right\}}\right]
    &=
    \lim_{N \rightarrow \infty}
    \E_{\hat{p}_{N}}\left[c\left(1+\norm{\Vec{V}}^k\right) \1_{\left\{\frac{h}{c}-1 < \norm{\Vec{V}}^k\right\}}\right]
    \\
    &=
    \E_{p_v}\left[c\left(1+\norm{\Vec{V}}^k\right) \1_{\left\{\frac{h}{c}-1 < \norm{\Vec{V}}^k\right\}}\right].
\end{align*}
But, applying the dominated convergence theorem (e.g., Theorem 16.4 in \cite{billingsley2008probability}), we get
\begin{align*}
    \lim_{h \rightarrow \infty}
    \E_{p_v}\left[c\left(1+\norm{\Vec{V}}^k\right) \1_{\left\{\frac{h}{c}-1 < \norm{\Vec{V}}^k\right\}}\right] = 0,
\end{align*}
where we used $\E_{p_v}\left[\norm{\Vec{V}}^k\right] < \infty$. To conclude the proof, note that $\E_{p_v}[f_h(\Vec{V})] \rightarrow \E_{p_v}[f(\Vec{V})]$ as $h \rightarrow \infty$ by reusing the dominated convergence theorem. \ep

\end{document}